\documentclass{elsart}
\newcommand{\mass}{m}
\newcommand{\vmass}{\vec\mass}
\newcommand{\constsep}{W}
\newcommand{\massp}{{m^{\prime}}}
\newcommand{\rate}{{\overline K}}
\newcommand{\pref}{\Gamma}
\newcommand{\scalmom}[1][\indmom]{\overline M_{#1}}
\newcommand{\expmom}{\alpha}
\newcommand{\indmom}{p}
\newcommand{\mom}[1][\indmom]{M_{#1}}
\newcommand{\taup}{\tau^{\prime}}
\newcommand{\wpr}{w^{\prime}}
\newcommand{\pos}[1][r]{\vec#1}
\newcommand{\const}{B}
\newcommand{\laplace}{{\mathcal L}}
\newcommand{\order}{r}
\newcommand{\lap}{\zeta}
\newcommand{\lapp}{s}
\newcommand{\tres}{\theta}
\newcommand{\explarge}{\vartheta}
\newcommand{\cinf}{c^{(\infty)}}
\newcommand{\sinit}{S_0}
\newcommand{\lapbar}{\overline\lap}
\newcommand{\fbar}{\overline F}
\newcommand{\gbar}{\overline G}
\newcommand{\suma}{\sigma}
\newcommand{\difer}{\delta}

\newcommand{\gp}[1][0]{g^\prime(#1)}
\newcommand{\gpp}[1][0]{g^{\prime\prime}(#1)}

\newcommand{\Kfin}[1][N]{K^{(#1)}}
\newcommand{\Kfinbar}[1][N]{\overline{K}^{(#1)}}
\newcommand{\Ei}{\mathop {\rm Ei}\nolimits}
\newcommand{\dens}{{\rho_0}}
\newcommand{\trans}{{\mathcal T}}
\newcommand{\transit}{{\mathcal T}_\infty}
\newcommand{\ptil}{{\tilde p}}
\newcommand{\gtil}{{\tilde G}}
\newcommand{\laptil}{{\tilde\lap}}
\newcommand{\tpinf}{{\tilde p_\infty}}
\newcommand{\tD}{{\tilde D}}
\newcommand{\tP}{{\tilde P}}
\newcommand{\tp}{{\tilde p}}
\newcommand{\ztil}{{\tilde z}}

\usepackage{graphicx}

\begin{document}
\begin{frontmatter}
\title{Scaling Theory and Exactly Solved Models In the Kinetics of
Irreversible Aggregation}
\author{F. Leyvraz}
\address{Centro de Ciencias F\'\i sicas, Universidad Nacional
Aut\'onoma de M\'exico, Avenida Universidad s/n, Colonia
Chamilpa, 62251 Cuernavaca, Morelos, Mexico}
\ead{leyvraz@fis.unam.mx}
\begin{abstract}
The scaling theory of irreversible aggregation is discussed in some
detail. First, we review the general theory in the simplest case
of binary reactions. We then extend consideration to
ternary reactions, multispecies aggregation, inhomogeneous situations
with arbitrary size dependent diffusion constants as well as arbitrary
production terms. A precise formulation of the scaling hypothesis
is given as well as a general theory of crossover phenomena.
The consequences of this definition are described at length.
The specific issues arising in the case in which an
infinite cluster forms at finite times (the so-called
gelling case) are discussed, in order to address
discrepancies between theory and recent numerical work. Finally, a large
number of exactly solved models are reviewed extensively with a view
to pointing out precisely in which sense the scaling hypothesis holds
in these various models. It is shown that the specific definition given
here will give good results for almost all cases. On the other hand, we
show that it is usually possible to find counterexamples to stronger
formulations of the scaling hypothesis.
\end{abstract}
\tableofcontents
\end{frontmatter}
\section{Introduction}
\label{sec:intro}
\setcounter{equation}0
\subsection{What is this all about}
What do a glass of fresh milk, a planetary system in formation and
Los Angeles on a smoggy day all have in common? At first sight,
not much, even from the slightly skewed viewpoint of a physicist.
The various phenomena involved in these systems have quite different
physics and the interesting questions are wholly
unrelated. Nevertheless, there is one aspect which these
three (and a large number of other) systems share:
in all of them one finds some process of irreversible aggregation, the
characterization of which is of some importance. To be specific, in
the case of milk, we are dealing with small globules
driven by buoyancy which coalesce
irreversibly to form cream; in a planetary system, on the other
hand, one has planetesimals colliding inelastically to
form ever larger blocks, eventually leading to planet formation.
Finally, in aeorosols such as those found in smog, we have
airborne particles so small that Brownian motion is (frequently)
an important transport mechanism. There the
aggregation mechanism is due to the
presence of van der Waals forces between the particles, which
are considerably stronger than the effects of thermal agitation.

Let us therefore review what these systems have in common:
in all of them, we have identifiable aggregates of particles
playing an important role. These aggregates grow by sticking
to each other. Specifically, they are usually not limited to
growing by accretion of the very smallest clusters, which
is another topic altogether. One has, therefore, a situation
in which  there is originally quite a large number of small
particles, and these then coalesce as time goes on. No allowance
is made for the possibility of break-up. One is thus not concerned
with a problem of approach to equilibrium, but with an
entirely {\em dynamical\/} problem\footnote{It is argued by
some that this is unrealistic, since on fundamental grounds
backward reactions {\em always\/} exist. To this the short answer
is that I  shall only be interested in time scales for which
the backward processes can indeed be neglected.}.

Aggregation phenomena are, as hinted at above, quite common in
nature. Apart from the examples mentioned above, we may mention
astrophysics (for further discussion of the subject with references see
e.g. \cite{lee00}), cloud physics
(see \cite{pru78})  and polymer chemistry, where a great deal of the
theoretical
work underpinning the gelation transition actually originated (see
\cite{sto43}).
Even to attempt to give a fair idea of the variety of applications
of aggregation and related concepts, however, is beyond both my
abilities and the scope of this work. The reader specifically interested
in the applications to aerosol physics is referred to the review by Drake
\cite{drake} which addresses a large number of such issues in detail.
Friedlander \cite{fri00} and Hidy \cite{hid70} are also important general
references on the subject.

What, then, are the issues we want to address when dealing with
such a system?  Overall, the questions
most frequently asked fall into two categories:
the first concerns the sizes and the second the morphology
of the aggregates. Let us deal shortly with the
latter first. It is, of course, quite obvious that if
two liquid droplets coalesce, they will, under the influence
of surface tension, rapidly  relax to a spherical shape.
Slightly more complex liquid systems may behave in a less
obvious way (think for example of a water drop on an inclined
glass plane, where surface tension and gravity act together),
but the geometry still remains fairly straightforward.
Matters change dramatically, however, when solid
particles stick together. In this case, depending on the degree
of rigidity of the bonds, the aggregates may or may not
rearrange. The  simplest case, because it is quite well-defined,
is the one in which there is no rearrangement whatever.
Then, as is well-known, the aggregates grow
fractal shapes and their morphology can only be characterized
at the statistical level. The main tool used
in order to characterize different morphologies is the
so-called mass-radius relationship
\begin{equation}
R(\mass)=const.\cdot\mass^{1/D_f},
\label{eqi:1}
\end{equation}
for large values of the mass $\mass$.
Here $R(\mass)$ is some average measure of the radius of the aggregate
as a function of its mass $\mass$ and $D_f$ is an exponent known
as the fractal dimension. A considerable amount of work has gone
into the study of such systems (for a review and for
an introduction to the extensive literature, see \cite{jul87,mea98}).
The upshot is that
a fractal dimension can indeed be defined, but that
it depends, although not very sensitively, on some of the
details of the transport mechanism. In particular, ballistic
transport yields more compact structures
(and hence larger values of the fractal dimension $D_f$)
than diffusion, and low sticking probability has a similar effect.

A short summary of the above discussion could therefore be stated
as follows: morphology is extremely important, but it is
strongly dependent on the specific mechanisms involved
in each particular system (many features of general interest
such as, say, the appearance of complex self-similar
structures in the case where no restructuration is present,
occur quite generally, but detailed features,
such as the actual values of the  fractal dimension,
do depend on specifics).

What will principally concern us in this review is the
study of the masses  of the aggregates. Specifically,
since there are always a large variety of masses at
one given time, we need the so-called {\em cluster
size distribution\/} as a function of time. (Note that
in the following, by size, we
shall always mean mass and never some measure of
geometric size such as radius or volume.) By this we mean that we
study the function $c(\mass, t)$ giving the concentration of
aggregates of mass $\mass$ at time $t$.  Roughly speaking,
one uses the following strategy: in many physical situations,
it is possible to express the rate of aggregation as the
encounter probability between aggregates of masses
$\mass$ and $\massp$, multiplied by a reaction rate
$K(\mass, \massp)$ depending in quite  a general way on
the two masses $\mass$ and $\massp$. The possibility of
describing the aggregation process in this very simplified
manner depends on several assumptions, which will be
shortly discussed later. The most important is a mean-field
hypothesis, which states that no spatial correlations develop during
the course of the reaction. From the theoretical point of
view, this is an extremely severe assumption, but in practice
it is often quite well satisfied.

If this can be done, it means that we have summarized all the
information concerning the complicated and specific physics of
the various individual systems in the matrix $K(\mass, \massp)$
characterizing the mass--dependence of the reaction rates.
It should be emphasized that the determination of the rates $K(\mass,
\massp)$ is in general by no means straightforward. However, as it
depends intimately on the specific physics of the system involved, we
shall not consider it here, but rather always assume the rates $K(\mass,
\massp)$ to be externally given.
One is then led to a system of infinitely
many coupled non-linear rate equations. These still present
a formidable problem, which usually cannot be solved by analytic means.
As for numerical studies, while they are certainly very valuable,
they are necessarily limited in their ability to deal
with either large cluster sizes or large times.
In this review, we extensively describe an approach due originally
to Friedlander, and then considerably expanded upon by Ernst
and van Dongen, namely the scaling approach. The crucial idea
consists precisely in limiting oneself to the case of large
masses and large times. One may then, as was shown,
take only the crudest features of $K(\mass, \massp)$ into account and
arrive at quite precise predictions for the overall features of
the cluster size distribution.

Let us illustrate this by a simple and classical example:
consider particles moving with a mass-dependent diffusion
constant $D(\mass)$ having a radius $R(\mass)$ growing as
in (\ref{eqi:1}). It is then reasonable to assume that
the reaction rates are given by
\begin{equation}
K(\mass,\massp)=\big(D(\mass)+D(\massp)\big)\big(
R(\mass)+R(\massp)
\big)^{d-2},
\label{eqi:2}
\end{equation}
where $d$ is the dimension of space. This follows from the
elementary fact that the rate at which a pointlike
particle diffusing with diffusion constant $D$ is
captured by a spherical trap of radius $R$ is of the
order $DR^{d-2}$. In order to proceed, we need to know
the asymptotic behaviour of the diffusion constants
$D(\mass)$. Let us assume, in keeping with tradition,
a fully general dependence of the type
\begin{equation}
D(\mass)=const.\cdot\mass^{-\gamma}.
\label{eqi:3}
\end{equation}
The scaling theory referred to above then states the following
facts concerning the cluster size distribution $c(\mass, t)$, that
is, the number of aaggregates of mass $\mass$ per unit volume
present in the system at time $t$:
\begin{enumerate}
\item The so-called typical cluster size $s(t)$ grows as
$t^z$, where the exponent $z$ is given by
\begin{equation}
z=\frac{D_f}{D_f+\gamma-(d-2)}.
\label{eqi:4}
\end{equation}
This means, as we shall see in far greater detail. that
if we consider an appropriate rescaling of
the cluster size distribution on this scale,
it collapses onto a single time-independent distribution.

\item From the above follows also that
\begin{equation}
\mom(t):=\int_0^\infty \mass^\indmom c(\mass, t)d\mass=const.
\cdot t^{(\indmom-1)z},
\label{eqi:5}
\end{equation}
at least for sufficiently large values of $\indmom$.
Further work shows that if $\gamma>0$, which is certainly
the most common case, (\ref{eqi:5}) actually holds
for {\em all\/} values of $\indmom$. When $\gamma\leq0$,
on the other hand, the critical value for $\indmom$ is not known,
but is known to lie between $0$ and $(d-2)/D_f-\gamma$.

\item Finally, the behaviour of the aggregates that are much
smaller than $s(t)$ can be evaluated. Thus it can be shown that
such aggregates decay as $\exp(-const.\cdot t^{(d-2)/D_f})$
if $\gamma<0$, whereas they decay as a power law otherwise.
\end{enumerate}
The way in which these and similar conclusions are drawn will be
the topic of this paper.

At this stage, the better read may be experiencing some doubts:
all of the above example, in fact, was essentially known
in the sixties, see for example \cite{swi64}.
So why am I talking about such things,
and what is new? The answer comes in two parts.

First, I have attempted to show how scaling theories can
be constructed in a nearly automatic manner for almost any
type of problem involving irreversible aggregation.
By this I mean that I have treated a large number of
extensions (source terms, aggregates with variable
compositions, inhomogeneous situations and, finally, the
effect of three-body interactions). In all of these cases
I have shown a perfectly routine way of obtaining
a scaling theory, which yields sensible results.
The point of this excercise, which may be said to
be the main purpose of this review altogether, is to show
how to compute scaling functions systematically in almost
any reasonable situation by simply turning a crank. In this
respect, the case of three-body interactions is particularly
instructive: indeed, I was just busy working out just how
the crank worked in this particular case, when suddenly it
failed in quite an unexpected way.
What happened is that an exact result, quite
easy to derive, conflicted with the equally evident scaling
theory I had devised. Appendix \ref{app:deriv-scaling-3}
presents my present view of what went wrong, and should
be borne in mind whenever constructing any kind of scaling
theory along these lines. Another rather tricky issue
concerns gelation: in this case, all the original literature
had predicted certain values for various exponents,
which were clearly refuted in the remarkable numerical
simulations of Lee \cite{lee01}. An attempt is made
to show that the previous arguments \cite{ley82,zif82,zif83} were
indeed inconclusive\footnote{or, in plain English, wrong},
to show in detail why and to present
some indication of how a theory of Lee's work might go.

As the title of this work indicates, however, there is yet another
side to this review, which concerns exactly solved models.
Indeed, there exists a considerable body of knowledge concerning
those few models for which the rate equations can be solved
analytically. This is extremely useful in order to
{\em test\/} the scaling approach. The problem is that the
exact nature of the error made when using the
scaling ansatz is not known {\em a priori}.
On the other hand, when one is dealing
with exactly solved models, one has a full overview of the solution's
behaviour, so that a detailed comparison with the scaling predictions
becomes possible. To this end, however, it
is necessary to investigate the exact solutions in the case
of arbitrary initial conditions, which has not, to my knowledge,
been done earlier to any great extent. Here I present such exact
solutions and show that scaling, in a very specific sense
of the word, indeed invariably holds, whereas many of the
consequences which are ordinarily believed to follow directly
from it, actually fail. This may be viewed as a list
of cautionary counterexamples, which show the  need for care
in stating what follows from scaling and what does not (thus,
I have been characteristically careless in asserting, for example,
the behaviour of the small clusters stated in the example above).
The whole issue is rather complex, but a simple summary\footnote{valid at
least for the non-gelling case} may be given
as follows: whenever a given property of the cluster size
distribution involves a finite fraction of the total mass,
one may be reasonably confident to obtain it correctly
via scaling. Whenever it does not, however, counterexamples
are possible, and indeed, in generic cases, rather likely.
\subsection{Outline of the Paper}
Let me here briefly summarize the layout of this paper.
In Section \ref{sec:mft} I state the known results
on the properties of existence,  uniqueness and mass conservation
for the solutions of the rate equations. These are rather
few, but quite instructive. Indeed, they are  probably also
in some sense optimal, according to the results suggested
by the scaling approach. In Section \ref{sec:scaling} we develop
the scaling approach in several stages. First, we show how
scale invariant solutions exist whenever the rate kernel
$K(\mass, \massp)$ is homogeneous in its arguments.
This in itself, however, says nothing about the
general solutions of the equations. To this end, we require the
{\em scaling hypothesis}, which states that all solutions
which start from an specific class of initial conditions
eventually approach this scale invariant solution at
large times. We then go on to discuss the precise sense
in which such convergence should be assumed. This question
may seem to be entirely technical and devoid of interest, but
in fact it lies at the very heart of the issues concerning the
validity of scaling. I shall suggest that the appropriate
choice of mode of convergence is {\em weak convergence},
where suitably normalized measures are defined in terms of the
cluster size distribution $c(\mass, t)$.
To express this in non-technical terms, I will say
that a cluster size distribution $c(\mass, t)$ converges
weakly to a given distribution $\Phi(x)$ with respect
to a typical size $s(t)$ if the expectation value of any function
which varies smoothly on the scale $s(t)$ tends to the corresponding
expectation value as $t\to\infty$. If this specific type of
convergence is used, it can be shown, as will be done extensively in
Section \ref{sec:exact}, that scaling obtains under
all reasonable circumstances. In other words, all the issues
that have been raised concerning so-called ``violations of
scaling'' are seen to arise from the fact that the correct
questions are not being asked, or equivalently, that conclusions
are being drawn from the scaling hypothesis, which are not
legitimate once scaling is defined in this precise manner
\footnote{Before scaling is defined in some way or other, of course,
the whole debate remains void}. I then proceed to derive
a general integral equation for the scaling function $\Phi(x)$.
This is simply a mild generalization of the Ernst--van Dongen equation
and allows in the usual way to derive the standard cases
I, II and III\footnote{These will be described in due time.}
for non-gelling aggregation. A similar equation is also obtained
for the gelling case, and some very
partial results for the behaviour
of the solution at the origin are obtained. These, while still
quite incomplete, clearly show that the original claims
of \cite{ley81} concerning the so-called $\tau$-exponent in the
gelling case were incorrect. The resulting claims are now
more nearly in agreement with the numerical findings of Lee
\cite{lee01}. The theory further suggests a new scaling relation which
should, perhaps, be verified numerically.
After these general remarks, I proceed to show
how the scaling approach can always be obtained in an essentially
mechanical fashion.  In particular, extensions to monomer production,
spatially inhomogeneous systems, aggregation with aggregates of
many species multicomponent aggregation and, finally, three-body
aggregation, are all worked out in some detail. A theory of crossover,
for the case in which one aggregation mechanism is supplanted
by another at large sizes, is also presented. In Section \ref{sec:exact},
on the other hand, we run through most of what was worked out
in Section \ref{sec:scaling}, but now using exactly solved models. Our
main workhorse, of course, is the constant kernel, for which a great
deal is known, and which can be solved exactly in almost
any of the extensions one cares to think about. However, we
also consider the sum and product kernels, and present
some results on the general bilinear kernel, in order to
illustrate the concepts of crossover previously developed.
We also discuss two more recent exactly solved models, both of them
variations on the constant kernel: one is a set of reaction rates
$K(k,l)$ which depend on the parity of the sizes
$k$ and $l$ of the reactants, but on
nothing else; the other is also only defined for discrete values
of the masses and is given by $2-q^k-q^l$, where $q$ is an arbitrary
number between zero and one.  Both of these can be solved exactly
to a large extent, and both display strange effects which have
occasionally been argued to be ``violations of scaling''. Again, a
discussion of these models' behaviour shows the contrary, but also
sheds considerable light on the limits of our definition
of scaling, which has a very hard time making any reasonable
statements about clusters of fixed size. But then, as these models
in fact show, the behaviour of fixed size clusters can be
completely unexpected. Finally, in section \ref{sec:non-mft},
I discuss the issue of models for which the mean-field
theory does not apply. In the cases I discuss, spatial correlations
are built up by the interplay between transport and reaction.
Whenever the reaction is slow, these are negligible, but
as soon as transport  becomes the rate-limiting factor,
spatial correlations invalidate the use of the rate equations.
In this case, no straightforward application of the formalism
developed in the previous sections is, of course, possible.
Scaling concepts, however, remain extremely useful as we shall see, as
they allow to analyze the information one has in a systematic way.
\subsection{Various Topics of Interest not Treated Here}
There are, as always, a considerable number of interesting subjects which
could, and perhaps ought to, have been included in this review, but which
have been left out, for reasons either of space, of limited competence on
my part, or else finally because, although undeniably beautiful, certain
subjects do not fall readily within the scope of this paper as it has been
described in the previous subsections. A necessarily incomplete list might
include the following
\begin{enumerate}
\item The rate equations are not, of course, primary: rather they are
derived, after some approximations, from a master equation for some
appropriate Markov process. Using then the so-called $\Omega$-expansion
devised by van Kampen, one may derive the rate equations. This leads to the
discussion of corrections due to particle number fluctuations, for a
thorough discussion of which the reader is referred to
\cite{lus78,don87a,don87b,don89,buf90}. A considerable literature has also
arisen in the literature on probability concerning such models and their
connection to  the Smolchowski equations. For a review, see \cite{ald99}.
An altogether remarkable claim of convergence of the above stochastic
process to the rate equations treated in this paper is made in
\cite{nor99}.
\item It is possible to go into much greater detail concerning the exact
solution of the full bilinear kernel. I have essentially limited myself to
analyzing various types of crossover and showing that scaling behaviour
arises for arbitrary  initial conditions.
Much more explicit forms can in fact be obtained, and a very interesting
combinatorial interpretation of them can be given. The reader is referred
to \cite{spo83a,spo83b,spo85a} for further discussion of these issues.
\item The list of extensions to the aggregation equations is of necessity
incomplete. Thus one findstreatments of aggregation with replication
or with annihilation \cite{kra96b,ebn95} as well as much more.
The elements presented in
this paper could also be multiplied in various ways. This has not been
attempted. Rather, I have tried to emphasize the applicability of a
general method, which could then be carried over to arbitrary cases as
desired. The systems studied were chosen on
the basis of a wholly subjective feeling
concerning  both their practical importance and the relevance of the
conceptual issues attached to them.
\item As will be seen, there exist several cases in which the
polydispersity exponent $\tau$
cannot be determined via a straightforward study of the integral
equation for the scaling function. For these a very powerful method was
developed by Cueille and Sire \cite{cue97} in order to obtain rigorous
bounds on the $\tau$ exponent. The method turns out to be so powerful that
these bounds are often tantamount to exact evaluations: the upper
and the lower bounds coincide to, say, four or five decimals. They also
give a systematic approach to make ever better bounds. This should
therefore be viewed as a solution of the problem of evaluating $\tau$ for
those particular cases. However, the approach is quite subtle, so that I
must refer the reader to the original paper.
\item My treatment of non mean-field models is completely insufficient.
This, however, may perhaps be excused in the light of the fact that my aim
here is to present the scaling theory of aggregation, not all the beautiful
results in some way connected with irreversible processes and models for
chemical reactions, such a task being quite beyond my abilities. I have
therefore strictly limited myself to such models as were strictly
aggregation models, and have only shown how the scaling theory can be
brought to bear in such case. Even so, I have not, of course, been able to
do the subject justice. The references given in this Section may get the
reader started.
\end{enumerate}
\section{Irreversible Aggregation: The Mean-field Approach}
\setcounter{equation}0
\label{sec:mft}
\subsection{Generalities}
\label{subsec:gen}
As already stated in the Introduction, the phenomenon of
irreversible aggregation is described as follows: Aggregates of
mass $\mass$ and $\massp$ react to form
aggregates of mass $\mass+\massp$. In mean-field theory, one writes
down rate equations for this process, asssuming that
\begin{equation}
A_{\mass}+A_{\massp}\mathop{\longrightarrow}_{K(\mass,\massp)}
A_{\mass+\massp},
\label{eq:2.1}
\end{equation}
where the $K(\mass, \massp)$ are the rates at which the aggregation
process takes
place. The index $\mass$ may range over a discrete or a continuous range
of values. If we then denote by $c(\mass,t)$ the concentration of
aggregates $A_{\mass}$ at time $t$, we obtain the following rate
equations to describe the dynamics of the cluster size distribution
\begin{eqnarray}
\partial_tc(\mass,t)&=&\half\int d\mass_1\,d\mass_2\,
K(\mass_1,\mass_2)\,c(\mass_1, t)
c(\mass_2, t)\times\nonumber\\
&&\times\left[
\delta(\mass_1+\mass_2-\mass)-\delta(\mass_1-\mass)-\delta(\mass_2-\mass)
\right].
\label{eq:2.2}
\end{eqnarray}
Here, as in the rest of this paper, I choose the convention that
$\mass$ runs over continuous values, since  the modifications to
obtain the discrete case can always be viewed as specializations
to singular measures. However, since in many cases it is much more
convenient to solve the case in which only discrete multiples of a
certain mass occur, let me introduce some notations which will recur
throughout this paper: if we have discrete initial
conditions, that is, if there is an $\mass_0$ such that
\begin{equation}
c(\mass,0)=\sum_{k=1}^\infty c_k(0)\delta(\mass-k\mass_0),
\label{eq:2.205}
\end{equation}
then the solution $c(\mass,t)$ at all times has the form
\begin{equation}
c(\mass,t)=\sum_{k=1}^\infty c_k(t)\delta(\mass-k\mass_0).
\label{eq:2.210}
\end{equation}
Here the r.h.s of (\ref{eq:2.210}) {\em defines} the functions
$c_k(t)$. A very important special case is that of so-called
monodisperse initial conditions, for which $c_k(0)=\delta_{k,1}$.

The derivation of (\ref{eq:2.2}) involves certain hypotheses,
which I shortly discuss: First, the effect of
cluster morphology on the rates of aggregation must either be neglected
or be taken into account using some appropriate averaging techniques.
Second, we need to assume that no spatial correlations
between the clusters build up as the reaction proceeds. It is
well-known that such an assumption is by no means trivial. In fact,
one may argue that most realistic systems will,
for sufficiently large times, display
such correlations. However, it can be shown that these
can indeed be neglected when the transport mechanisms
responsible for bringing the clusters to react are significantly faster
than the reaction step. Since this is a fairly common situation
in practice, we will restrict ourselves to this mean-field situation
except in Section \ref{sec:non-mft}.

Equations (\ref{eq:2.2}) can, of course, be generalized in many ways.
One may include, among others, the following additional effects:
\begin{enumerate}
\item Reactions of higher  order than binary. This is certainly
important when the volume fraction of the aggregates grows in
time, which occurs whenever a model displays fractal growth.
In this situation, however, it is quite unlikely that the hypotheses
necessary for mean-field theory to hold remain valid when these
terms become dominant. However, their effect as corrections can be
estimated within this framework, which allows to
decide whether such corrections are likely to be important or not.

\item The parameter $\mass$ can be generalized to be a multicomponent
vector. This describes, for example, situations in which the
aggregates consist of various substances. In this case the scaling
theory must be generalized somewhat to incorporate various different
mass scales which describe different aspects of the cluster size
distribution, namely the {\em average} as well as the
{\em spread} in composition, which both scale differently, due to the
central limit theorem.

\item Spatially inhomogeneous situations can be considered. In this
case, diffusion terms must be explicitly added to the equation. The
diffusion constants involved have, of course, in general a
non-trivial mass-dependence, which leads to quite complicated
equations.

\item Finally, monomer production can be introduced. One may
either consider the case in which this occurs
homogeneously throughout the system, or on the contrary, we may assume
that monomer is being produced at a given point and
diffuses as it aggregates.
\end{enumerate}
As we shall see, the scaling approach can be
extended to give a straightforward
description of these various systems. However, before we proceed,
I would like to review shortly some of the rigorous results known on
the existence and uniqueness properties of solutions of
(\ref{eq:2.2}).
\subsection{Existence and Uniqueness Results}
\label{subsec:ex}
In the following, we shall treat the discrete case of (\ref{eq:2.2})
first: results have usually first been obtained for it, since the
general case involves some additional difficulties.
These equations represent an extremely challenging problem
from the point of view of pure mathematics. The reason is that they
consist of an infinite number of coupled nonlinear ordinary
differential equations. There do not exist standard results
for such cases.

In this case, existing results fall into two classes: First, those
which make an assumption of the type:
\begin{equation}
K(\mass, \massp)\leq C(\mass+\massp).
\label{eq:2.3}
\end{equation}
Here, and throughout the rest of this paper, reaction rates will always
tacitly be assumed to be positive.
{}From such an assumption it can be shown \cite {whi1} that the
$c(\mass, t)$ decay exponentially in $\mass$ for all $t$ for which the
solution exists. From this one shows that the quantity
\begin{equation}
M_1(t)=\sum_{m=1}^\infty mc(m,t)
\label{eq:2.4}
\end{equation}
remains constant as long as the solution is defined. This in turn
allows to prove existence and uniqueness of the solution for all times.
Note that it is formally trivial to prove the constancy of $M_1(t)$,
which physically corresponds to the conservation of the total mass
contained within the aggregating system. However, from a rigorous
point of view, it is necessary that
\begin{equation}
\int_0^\infty \mass
K(\mass,\massp)c(\mass,t)c(\massp,t)d\mass\,d\massp<\infty
\label{eq:2.5}
\end{equation}
for all times.

These results are in a sense optimal: they yield both existence and
uniqueness, as well as mass conservation. Further, it is expected that
mass-conservation will fail whenever the hypothesis
(\ref{eq:2.3}) does.
However, this hypothesis is often too restrictive. Thus, a
reasonable model for aggregation of branched aggregates (without loops)
assumes that the number of reactive sites on each aggregate
grows as $\mass$, from which follows
\begin{equation}
K(\mass, \massp)=C\mass\massp,
\label{eq:2.6}
\end{equation}
which obviously does not satisfy (\ref{eq:2.3}). In this case, it was
first shown \cite{McLeoda,McLeodb}, that no solution satisfying constancy of
$M_1(t)$ could exist beyond a certain time. This negative result was
later put into proper perspective by the discovery \cite{zif80,ley81}
of an exact solution for all times, for which $M_1(t)$ decays after a
finite time. This phenomenon, which is found to be linked to the appearance
of an infinite cluster at finite time, is known as gelation. It is of
considerable importance and we shall discuss it extensively further on in
this  paper.

This led to a different set of existence results
\cite{ley81}, which did not aim to prove either constancy of $M_1(t)$ or
any decay condition leading to (\ref{eq:2.5}). It could
then be shown in \cite{ley81} that, if
\begin{equation}
K(\mass, \massp)\leq r(\mass)r(\massp)\qquad r(\mass)=o(\mass),
\label{eq:2.7}
\end{equation}
then a solution of (\ref{eq:2.2}) with finite and monotonically
decreasing $M_1(t)$ exists for all times, but no
statement was made either concerning the conservation of mass or
uniqueness of the solution. The latter is certainly
unfortunate, but is intimately related to the technique of proof used.
This issue has now essentially been solved in \cite{nor99}, where many
other striking results are shown. The result claimed is that under the
hypothesis (\ref{eq:2.7}) uniqueness holds.
The former is, on the other hand, only to be expected. In fact, as we shall
see, the scaling theory strongly suggests that the bound
(\ref{eq:2.3}) is, in fact, exactly the one that separates gelation at
finite time from regular behaviour at all times. As a further
indication of this fact, the model
\begin{equation}
K(\mass, \massp)=\mass^\alpha\delta_{\mass, \massp}
\label{eq:2.8}
\end{equation}
can be rigorously shown to
violate mass conservation if $\alpha>1$ \cite{ley83}, whereas it
satisfies (\ref{eq:2.3})---and hence mass conservation---if
$\alpha\leq1$.
The marginal case $\alpha=1$ is discussed in greater detail in
\cite{buf89}.

Let me shortly describe the technique involved, since it turns out to
be of fairly general applicability: one first defines a finite,
mass-conserving system, for which standard results can be invoked to
guarantee existence of the solution for all times. A compactness
argument is then used to show that the solutions to the finite
systems have a convergent subsequence. It can then be shown that the
limit of this subsequence satisfies a weak form of (\ref{eq:2.2}),
and finally that it satisfies (\ref{eq:2.2}) in a strong sense as well.
The mode of convergence, however, is not strong enough to conclude
that the mass contained in the limiting solution is the same as that
in the finite approximants, which is what gives to this approach the
requisite generality.
The extension to the case of continuously variable $\mass$ presents
some technical difficulties. These were adressed by Bak \cite{bak}
in the case of a reaction kernel bounded above and below by a constant.
The general case was treated by Ball and Carr in \cite{bal90}.

Another extremely important issue from the point of view of
applications concerns the extension to non-uniform systems This
involves replacing (\ref{eq:2.2}) by a reaction diffusion equation.
This has been treated by Slemrod \cite{slem90},
along lines similar to those sketched
above. Here, however, it should be emphasized that even the
``standard'' results on the finite approximants are highly non-trivial
in this case.
While the results obtained are remarkable indeed, I believe that they
are probably not yet optimal: indeed, the strongest results obtained
concern diffusion constants bounded from below and above by a constant.
In a realistic situation, however, these always decrease with
mass; an extension of these results to more general cases
would certainly be very desirable.

Finally, it should be emphasized that the issues adressed here are not
altogether idle: existence may indeed fail under some circumstances. There
is considerable evidence, for example, that if
$K(\mass_0,\mass)/\mass\to\infty$ as $\mass\to\infty$ for fixed $\mass_0$,
then either no solutions exist or these solutions have the peculiar
property of violating mass conservation from the very start. In this case
it is an open problem whether the solution exists at all for a sufficiently
wide set of initial conditions. For more details on
these fascinating questions, see \cite{don87d,car92,lee00,lee01}.

\section{The Scaling Hypothesis}
\setcounter{equation}0
\label{sec:scaling}
\subsection{Basic Concepts: Moments and Typical Size}
\label{subsec:mom}
The crucial observation underlying the scaling approach is the following:
at large times the value of most quantities of interest---namely those
that result from an average over the whole cluster size distribution---can
be computed using a single ``typical size'' which grows indefinitely with
time. Under these circumstances, we expect that only the coarsest features
of the $K(\mass, \massp)$ will be relevant. Let us therefore assume that
$K(\mass, \massp)$ is asymptotically homogeneous of degree $\lambda$,
that is, that there exist a $\lambda$ and a function $\rate(\mass, \massp)$
such that
\begin{equation}
\rate(a\mass,a\massp)=\lim_{s\to\infty}\left[
s^{-\lambda}K(s\mass, s\massp)
\right]
\label{eq:3.1}
\end{equation}
The function $\rate(\mass, \massp)$ is then clearly homogeneous of
order $\lambda$. To describe $\rate(\mass, \massp)$ in greater
detail, note that it can be written as
\begin{equation}
\rate(\mass, \massp)=\mass^\lambda k\left(
\frac{\massp}{\mass}
\right),
\label{eq:3.105}
\end{equation}
where $k(z)$ is a function subject to the symmetry condition
\begin{equation}
k(z)=z^\lambda k(1/z),
\label{eq:3.110}
\end{equation}
but which is otherwise quite arbitrary. It can therefore be chosen
arbitrarily, say, between 0 and 1. We now define the exponent $\mu$
and the prefactor $\pref$ which describe
the (power law) behaviour of $k(z)$ near the origin.
More specifically:
\begin{equation}
k(z)=\pref z^\mu[1+o(1)]
\label{eq:3.115}
\end{equation}
One additionally defines $\nu$ as the exponent describing the
behaviour of $k(z)$ at infinity. It follows from (\ref{eq:3.110}) that
\begin{equation}
\nu=\lambda-\mu.
\label{eq:3.120}
\end{equation}
It follows from these definitions that there exist reaction rates
corresponding to arbitrary values\footnote{This issue has caused some
confusion in the literature, as a lot of work has gone into the study of
kernels of the form
$\mass^\alpha\massp^{\lambda-\alpha}+\mass^{\lambda-\alpha}\massp^\alpha$,
which do not generate the whole range of possible $\mu$ and $\nu$ values,
as opposed to (\ref{eq:3.125})}
of $\mu$ and $\nu$, as evidenced by
\begin{equation}
K(\mass, \massp)=\left\{
\begin{array}{cc}
\mass^\mu\massp^\nu&\qquad(\mass\leq\massp)\\
\mass^\nu\massp^\mu&\qquad(\mass\geq\massp)
\end{array}
\right.
\label{eq:3.125}
\end{equation}
At this stage, we should rephrase the existence and uniqueness
results described in the previous subsection. Condition (\ref{eq:2.3})
is equivalent to $\lambda\leq1$ and $\nu\leq1$. It is indeed
presumably optimal for the existence of a global mass-conserving solution
of (\ref{eq:2.2}).
On the other hand, condition (\ref{eq:2.7}) which
guarantees existence only, is equivalent to the conditions $\lambda<2$
and $\nu<1$. However, a straightforward extension of the proof given
in \cite{ley81} shows that the former assumption can be dispensed
with. The only essential condition is, therefore, $\nu<1$.\footnote{
The exponents $\lambda$ and $\mu$ above can be defined in a rather
rough manner, except if we consider the limiting cases $\lambda=1$
and $\nu=1$. For these it is essential that there be no logarithmic
corrections to the power law behaviour. see e.g. \cite{buf89} for details.
}

Let us temporarily assume that $K(\mass, \massp)$ is exactly
homogeneous. From this assumption
follows \cite{hen83} that, if $c(m,t)$ is an arbitrary solution of
(\ref{eq:2.2}), then so is
\begin{equation}
\left(T_{a,b}c\right)(\mass, t)=a^{\lambda+1}b^{-1}c(a\mass, bt)
\label{eq:3.2}
\end{equation}
for arbitrary positive $a$ and $b$. The existence of such a symmetry
group motivates the search for solutions which remain invariant under it.
In other words, we ask whether solutions exist, which, as time evolves,
merely transform into similar solutions generated from a constant
cluster size distribution by a variable group transformation of the
form (\ref{eq:3.2}). In the so-called regular case, that is, when
mass conservation holds, see
(\ref{eq:2.3}) , this can only occur if the group transformation
belongs to the subgroup of (\ref{eq:3.2}) which maintains the total
mass in the system invariant, that is $T_{a,a^{\lambda-1}}$. From this
follows that such a solution necessarily has the form
\begin{equation}
c(\mass,t)=\constsep t^{-2/(1-\lambda)}\Phi\left(
m/t^{1/(1-\lambda)}
\right),
\label{eq:3.3}
\end{equation}
where $\constsep$ is a constant factor which will be of use
later\footnote{Specialists will recognize here the separation
constant $w$ of the papers of Ernst and van Dongen}.
The gelling case, as we shall see below, requires more careful
considerations. Quite generally, however, we shall say that a solution
is of scaling form if it can be written in the form
\begin{equation}
c(\mass, t)=\constsep s(t)^{-2}\Phi\left(
\frac{j}{s(t)}
\right),
\label{eq:3.4}
\end{equation}
where $s(t)$ is an arbitrary (growing) function of time, which is
called the typical size. From this follows, as we show in greater
detail later, that if we define the moments of the distribution as
follows
\begin{equation}
\mom(t)=\int_0^\infty \mass^\indmom c(\mass, t)d\mass,
\label{eq:3.5}
\end{equation}
one finds as an estimate for the order of magnitude of $s(t)$
\begin{equation}
s(t)\simeq\frac{\mom[\indmom+1](t)}{\mom(t)},
\label{eq:3.6}
\end{equation}
at least when $\indmom$ is larger than a given $\indmom_0$, the value
of which depends on the details of the system under consideration. In
fact, in many contexts,the typical size is {\em defined} as some such
moment ratio: such are, for example,
the weight average, defined as $M_2(t)/M_1(t)$ and the
$z$-average, given by $M_3(t)/M_2(t)$. The constant $\constsep$ will
allow us to maintain any such definitions, while keeping simple
normalizations for the function $\Phi(x)$.

We shall show in the following sections that, if such a solution
exists, the function $\Phi(x)$ must satisfy a certain integral equation.
Conversely, it is straightforward to see that any solution to the
integral equation also defines via (\ref{eq:3.3}) a self-similar solution
of (\ref{eq:2.2}).

The scaling hypothesis now reduces to the following
statement: every solution starting from a ``narrow'' cluster size
distribution (that is, one that does not have already some kind of
power law distribution at large masses) evolves for large times into
a self-similar solution of the type we have just described
\footnote{In fact, we shall see that, quite often, even initial conditions
with power-law tails will approach the same scaling limit as narrow ones.
The above is merely the most conservative formulation of the hypothesis.}.
We are
therefore assuming that these specific solutions, which can be
described quite accurately in a broad variety of cases, are in fact
the only relevant solutions at large times, as they are approached
from essentially any initial conditions. It should be emphasized that
there is no proof of this so-called ``scaling hypothesis''. nor is there
any idea of how such a proof might go. However, the numerical evidence for
it is good, and we shall see that it is satisfied in all exactly solved
models.
\subsection{Different Kinds of Scaling}
\label{subsec:scal1}
In the following, we define more accurately in which sense we expect
solutions of (\ref{eq:2.2}) to tend towards a scaling solution. This
issue is of fundamental importance, as it will accurately determine
the kind of conclusions one may legitimately draw from scaling
and those one may not.

To justify the definition I shall make, let me first state the
nature of the object we are interested in: this is the cluster size
distribution $c(\mass, t)$. The information it gives us is the
following: how many clusters are there at time $t$ in the mass range
going from $\mass$ to $\mass+d\mass$. As such, the cluster size
distribution is essentially a {\em time-dependent measure}
on the set of all masses. We therefore need a
notion of convergence adapted to measures.

It turns out, as we shall justify in the following, that the
appropriate type of convergence is {\em weak
convergence}. Specifically, we shall say that a cluster size
distribution $c(\mass, t)$ tends to a scaling form if there is a
function $s(t)\to\infty$ as $t\to\infty$ and such that
\begin{equation}
\int_0^\infty \mass c(\mass, t) f[\mass/s(t)]d\mass
\mathop{\longrightarrow}_{t\to\infty}
\constsep\int_0^\infty
x\Phi(x)f(x)dx
\label{eq:3.7}
\end{equation}
for every continuous bounded function $f(x)$ on the positive real
axis. Note that the use of $x\Phi(x)$ in the definition is purely
conventional: it serves to make notation consistent with the previous
subsection, as well as with ordinary usage.

We shall argue in the following that (\ref{eq:3.7}) is the most
appropriate definition for the appproach to scaling.
On the one hand we shall find that the various exactly solved
models which we discuss in Section \ref{sec:exact} yield counterexamples
to many attempts to make sharper statements, whereas (\ref{eq:3.7})
is, as we shall see in some detail, fulfilled in the vast majority
of cases.

Let me here make a technical---but quite important---remark: in order to
prove weak convergence of measures, it is necessary and sufficient,
by a standard result in probability theory \cite{wil}, to show that
the corresponding Fourier transforms converge. Since all the measures
we consider are concentrated on the positive real axis
\footnote{Well, almost all. When we look into the distribution of
compositions in multicomponent aggregation, we shall in fact use Fourier
transforms},
this is
equivalent to convergence of the corresponding Laplace transforms, or
generating functions in the discrete case. In other words, if we define
\begin{equation}
G(\lap; t)=\int_0^\infty c(\mass, t)e^{\lap\mass}d\mass
\label{eq:3.705}
\end{equation}
we obtain the following two equivalent formal expressions for the scaling
function $\Phi(x)$:
\begin{eqnarray}
&&\constsep\int_0^\infty \Phi(x)\left(
e^{-\rho x}-1
\right)=\lim_{t\to\infty}\left\{
s(t)G[-\rho/s(t), t]
\right\}\nonumber\\
&&x\Phi(x)=\constsep^{-1}\laplace^{-1}\left\{
\lim_{t\to\infty}\left[
G_\rho(-\rho/(s(t),t)
\right]
\right\}(x),
\label{eq:3.8}
\label{eq:3.801}
\end{eqnarray}
where $\laplace^{-1}$ denotes the inverse Laplace transform. Since
generating functions such as $G(\lap,t)$ are ubiquitous in exact solutions,
where they arise in a very natural manner, we may already argue from an
aesthetic viewpoint in favour of the above defiinition of convergence to
scaling.

Finally note that there is one element
of arbirariness in the definition (\ref{eq:3.7}) that I have not yet
addressed: it concerns the choice of the measure we decide to
consider. In (\ref{eq:3.7}) we chose $\mass c(\mass, t)$, partly
because it is automatically normalized. It is, however, quite possible
to use, for example, $\mass^nc(\mass, t)/M_n(t)$ in its stead. We
shall say that the $n$'th moment of the cluster size distribution
approaches a scaling form if there is a
function $s(t)$ as above  such that
\begin{equation}
\mom[n](t)^{-1}\int_0^\infty \mass^n c(\mass, t) f[\mass/s(t)]d\mass
\mathop{\longrightarrow}_{t\to\infty}
\constsep\int_0^\infty
x^n\Phi(x)f(x)dx
\label{eq:3.810}
\end{equation}
for every continuous bounded function $f(x)$ on the positive real
axis. Now,
it is straightforward to check that, if the $n$'th
moment of a cluster size distribution
tends to a scaling form, then so do all
higher ones. However, the opposite need not be the
case. Consider for example the exact solution to (\ref{eq:2.2}) for
$K(\mass, \massp)=\mass\massp$ and an initial condition $c(\mass,
0)=\delta(\mass-1)$, which is given by
\begin{equation}
c(\mass, t)=\sum_{k=1}^\infty\frac{k^{k-2}}{k!}\frac{(te^{-t})^k}{t}
\delta(\mass-k),
\label{eq:3.9}
\end{equation}
and is valid for $t\leq1$. In this case, one readily verifies that,
choosing $s(t)=(1-t)^{-2}$, the second
moment of (\ref{eq:3.9}) approaches a scaling form with $\Phi(x)$
given by
$x^{-5/2}e^{-x}$, whereas the first does not tend to a scaling form
at all. Generally speaking, the scaling function to which the $n$'th
moment tends must satisfy
\begin{equation}
\int_0^\infty dx\,x^n\Phi(x)<\infty
\label{eq:3.904}
\end{equation}
so that a singularity of the type $x^{-5/2}$ could never arise as a limit
of first moments.
\subsection{Some Consequences that Follow from Scaling,
and More that Don't}
\label{subsec:scal2}
We now need to understand which consequences follow from the scaling
hypothesis. In many systems, it is possible to measure some of the
moments $\mom(t)$ of the cluster size
distribution $c(\mass, t)$, as defined in (\ref{eq:3.5}).
In fact, it is often, at least in
principle, possible to  reconstruct the full distribution from a
knowledge of its moments. We therefore concentrate first on these.
{}From the definition of convergence to a scaling form (\ref{eq:3.7})
follows
\begin{equation}
\lim_{t\to\infty}\left[
s(t)^{-(\indmom-1)}\mom(t)
\right]
=\constsep\int_0^\infty x^\indmom\Phi(x)dx
=W\scalmom,
\label{eq:3.905}
\end{equation}
whenever the integral on the r.h.s converges. Here the last equality
defines the scaled moments $\scalmom$ which play a
considerable role throughout the theory.
As we shall find out
later, $\Phi(x)$ has no singularities for $x\neq0$ and decays
exponentially as $x\to\infty$.

The only issue is therefore the
behaviour at zero. In many cases, this is given by a power law as follows
\begin{equation}
\Phi(x)\simeq x^{-\tau}\qquad(x\to0).
\label{eq:3.910}
\end{equation}
The meaning of such a behaviour at the origin is the following: it
states that the concentrations averaged in the low end of the cluster
distribution go as $\mass^{-\tau}$. More formally,
\begin{equation}
\lim_{t\to\infty}
\left[s(t)^2
\frac{1}{\epsilon s(t)}\int_{\epsilon s(t)}^{2\epsilon s(t)}c(\massp,
t)d\massp
\right]
=O(\epsilon^{-\tau})
\label{eq:3.911}
\end{equation}
as $\epsilon\to0$. Here the l.h.s. is an averaged concentration
rescaled by $s(t)^2$ so as to cancel the time dependence.

It now follows that a necessary condition for (\ref{eq:3.905}) to hold is
\begin{equation}
\indmom>\tau-1.
\label{eq:3.915}
\end{equation}
(\ref{eq:3.905}) and (\ref{eq:3.915}) therefore confirm
the claim made in (\ref{eq:3.6}) that $s(t)$ can be defined as the
ratio of successive moments, if the order of the moments is
sufficiently large.

Let us now introduce some notation: denote by $z$ the exponent with
which $s(t)$ grows as a function of $t$ and $\expmom_\indmom$ the one
with which $\mom(t)$ grows
\begin{eqnarray}
s(t)&\simeq&t^z\nonumber\\
\mom(t)&\simeq&t^{\expmom_\indmom}
\label{eq:3.920}
\end{eqnarray}
{}From these definitions we obtain the following relationships
\begin{equation}
\expmom_\indmom=(\indmom-1)z\qquad(\indmom>\tau-1).
\label{eq:3.925}
\end{equation}
We may further define $w$ as the exponent with which $c(\mass, t)$
decays when $s(t)$ becomes much larger than $\mass$. Formally once more
\begin{equation}
\lim_{\epsilon\to0}
\left[
\frac{1}{\epsilon s(t)}\int_{\epsilon s(t)}^{2\epsilon s(t)}c(\massp,
t)d\massp
\right]
=O(t^{-w})
\label{eq:3:926}
\end{equation}
{}From (\ref{eq:3.910}) and (\ref{eq:3:926}) one readily shows the
following celebrated identity \cite{vic84}
\begin{equation}
(2-\tau)z=w.
\label{eq:scaling-rel}
\end{equation}
This relation relates large time behaviour of small clusters to the
shape of the cluster size distribution at the low end of the
distribution for fixed large times. Since these two quantities are
not related in a straightforward way, this relation often gives very
useful and non-trivial information.

Note, however, that the above
definitions of $\tau$ and $w$ seem quite artificial. It appears
much more
natural to ask about the large time behaviour of $c(\mass, t)$ at
{\em fixed} $\mass$ and the small $\mass$ behaviour at fixed $t$.
Here, however, it must clearly be stated that these asymptotic
behaviours {\em cannot} be determined from the scaling hypothesis in the
form that we have given it up to now.
Similarly, moments $\mom(t)$ of order lower than $\tau-1$
cannot be described by scaling theory as it stands.
In fact, their behaviour is linked to that of the $c(\mass, t)$
at fixed $\mass$ for large $t$, which
lies beyond the reach of the scaling approach. We may define some
additional exponents: for example, define $\taup$ and $\wpr$ as follows
\begin{eqnarray}
&c&(\mass, t)\simeq \mass^{-\taup}\qquad(1\ll\mass\ll s(t))\nonumber\\
&c&(\mass, t)\simeq t^{-\wpr}\qquad(t\to\infty).
\label{eq:3.927}
\end{eqnarray}
At this stage, however, I should again
remark that exactly solved models
are known (see section \ref{sec:exact}) for which the $\wpr$ exponent
depends on $\mass$, whereas the $\taup$ exponent does not exist.
These, however, satisfy ordinary scaling and have well defined
values of the exponents $\tau$, $z$ and $w$.

Nevertheless, since such systems are not very frequent (they are not,
however, pathological), it makes sense to say that a system satisfies
the {\em strong scaling hypothesis} if $\taup$ and $\wpr$ both exist
and satisfy
\begin{equation}
\taup=\tau\qquad\wpr=w.
\label{eq:3.928}
\end{equation}
Under these circumstances, we can say how the moments $\mom(t)$ of
order $\indmom<\indmom_0$ behave if they are finite initially:
the $\expmom_\indmom$ are given by
\begin{equation}
\expmom_\indmom=\left\{
\begin{array}{ll}
(\indmom-1)z&\qquad(\indmom>\tau-1)\nonumber\\
w&\qquad(\indmom\leq\tau-1).
\end{array}
\right.
\label{eq:3.929}
\end{equation}
Note, however, that if strong scaling is violated, it may well happen
that $\alpha_\indmom\neq(\indmom-1)z$, even when $\indmom>\tau-1$, as we
shall later show in explicit examples. The only generally true statement is
that $\alpha_\indmom=(\indmom-1)z$ if $\indmom>1$.
Since the evaluation of moments is of central importance in
applications of the theory to real systems, this result
is basic. It shows that the large time behaviour of moments can always be
obtained to a knowledge of the
exponents $\tau$ and $z$ for given forms of the reaction kernel
$K(\mass, \massp)$. It is this task to which we now turn.
\subsection{The Scaling Equation and Its Derivation}
\label{subsec:scal3}
The basic kinetic equations (\ref{eq:2.2}) together with the
definition (\ref{eq:3.3}) of the fact that the first moment of
$c(\mass,t)$
approaches a scaling limit, given in (\ref{eq:3.7}) lead to  the
following condition on the scaling function $\Phi(x)$: for all
continuous functions on the positive real axis which vanish
sufficiently fast at infinity, one has, as shown in Appendix
\ref{app:deriv-scaling}:
\begin{equation}
\int_0^\infty dx\,dy\,\rate(x,y)\Phi(x)\Phi(y)x\left[
f(x+y)-f(x)
\right]=
\int_0^\infty dx\,x^2f^\prime(x)\Phi(x).
\label{eq:3.10}
\end{equation}
Here $\rate(x,y)$ is the continuous function defined by
\begin{equation}
\rate(x,y)=\lim_{s\to\infty}\left[
s^{-\lambda}K(s
x,sy).
\right]
\label{eq:3.11}
\end{equation}
As part of the derivation of this equation, one obtains a condition
on $s(t)$, namely
\begin{equation}
\dot s(t)=\constsep s(t)^{\lambda},
\label{eq:3.1105}
\end{equation}
the solution of which is in agreement with the statements obtained on
quite general grounds in (\ref{eq:3.3}). This therefore confirms that
\begin{equation}
z=\frac{1}{1-\lambda}.
\label{eq:3.1110}
\end{equation}
when gelation does not occur.
The derivation of (\ref{eq:3.10}) is somewhat technical and is
given in appendix \ref{app:deriv-scaling}. We present it in some
detail, though, as the scheme is very straightforward and can be extended
in a mechanical way to all the other cases we shall be
considering in this paper.

A particularly useful form
of (\ref{eq:3.10}) is the one in which $f(x)$ is taken as an
arbitrary exponential:
\begin{eqnarray}
I(\rho)&:&=\int_0^\infty dx\,dy\,\rate(x,y)\Phi(x)\Phi(y)xe^{-\rho x}\left[
1-e^{-\rho y}
\right]
\label{eq:3.1199}\\
I(\rho)&=&
\rho\int_0^\infty dx\,x^2e^{-\rho x}\Phi(x).
\label{eq:3.12}
\end{eqnarray}
Here $I(\rho)$ is defined for future reference in (\ref{eq:3.1199})
A very similar equation is central to the work of van Dongen
\cite{don88}. It is obtained from (\ref{eq:3.10}) by setting
$f(x)=\Theta(a-x)$ for arbitrary $a$ and yields the relation
\begin{equation}
a^2\Phi(a)=\int_0^a dx\int_{a-x}^\infty dy\,x\rate(x,y)\Phi(x)\Phi(y)
\label{eq:3.13}
\end{equation}
I present (\ref{eq:3.13}) because of its importance in previous work,
but shall not make much use of it in the sequel. From (\ref{eq:3.10})
also follows immediately the following central expression for the
scaled moments $\scalmom$.
\begin{equation}
\indmom\scalmom[\indmom+1]=\int_0^\infty
dx\,dy\,x\rate(x,y)\Phi(x)\Phi(y)\left[
(x+y)^\indmom-x^\indmom
\right]
\label{eq:3.1302}
\end{equation}
When $\rate(x,y)$ is of the form $x^\alpha y^{\lambda-\alpha}$, then
(\ref{eq:3.1302}) becomes a nonlinear recursion between the scaled moments
whic is frequently useful, see for example subsection
\ref{subsubsec:large}. Also,
as we shall later show, the scaled moments enter in the determination of
$\tau$ for certain cases (see (\ref{eq:3.23}) below),
and (\ref{eq:3.1302}) then turns out to be
very useful to get interesting results from these connections.

Finally, let me make some remarks on uniqueness and normalization of
the solutions of (\ref{eq:3.10}) as well as all its special variants.
It is clear that (\ref{eq:3.10}) inherits the symmetry of the
equations (\ref{eq:2.2}), as stated in (\ref{eq:3.2}). Namely,
whenever $\Phi(x)$ is a solution, so is
\begin{equation}
\Phi_b(x)=b^{1+\lambda}\Phi\left(
\frac{x}{b}
\right)
\label{eq:3.1310}
\end{equation}
for all $b>0$. This ambiguity in the definition of $\Phi(x)$ arises from
the absence
of scale in the equations (\ref{eq:2.2}) at least when aggregates are
large enough and differs from the ambiguity which simply results
from the choice of the scale of $s(t)$, which we have absorbed in the
prefactor $\constsep$ in (\ref{eq:3.7}). It should
generally be checked that final results are not affected by the
transformation (\ref{eq:3.1310}).
To fix the normalization, we use the fact that,
since $x\Phi(x)$ is defined as the weak limit of a probability
distribution, it must be normalized (gelation is not assumed to occur
in this case). This means
\begin{equation}
\int_0^\infty x\Phi(x)dx=1.
\label{eq:3.1305}
\end{equation}
This condition then determines $b$ in \ref{eq:3.1310}) and hence
$\Phi(x)$ uniquely. Finally, the absolute scale of
$\Phi(x)$ is related to that of $\rate(x,y)$, which is itself
fixed by the choice of time scale. This can be stated as follows.
If $\Phi(x)$ is a solution with rate $\rate(x,y)$, then $b\Phi(x)$ is
a solution to the equation with rates $\rate(x,y)/b$.

Note that a slight difficulty arises when $\lambda=1$, since the
transformation defined by (\ref{eq:3.1310}) then leaves $\int x\Phi(x)dx$
invariant. In this case we must keep the constant $\constsep_1$ explicitly
in (\ref{eq:a.5}) and set the normalization by allowing it to vary.

The above relations were all derived under the hypothesis that the
first moment of the cluster size distribution tends to a scaling
form. This is not always the case. In fact, it is easy to verify
that, when $\tau>2$, it is impossible to have such convergence.
Therefore, in the gelling case, which always has $\tau>2$, we must
consider convergence of higher order moments. In this case one finds
\begin{eqnarray}
&&\int dx\left[
(3-\tau)f(x)+xf^\prime(x)
\right]x^2\Phi(x)
=
\int dx\,dy\,\rate(x,y)\times\nonumber\\
&&\qquad\qquad\times\Phi(x)\Phi(y)x\left[
(x+y)f(x+y)-xf(x)
\right].
\label{eq:3.14}
\end{eqnarray}
Here we have used the strong scaling assumption to show that
\begin{equation}
\frac{{\dot{\mom[2]}}(t)}{{\mom[2]}(t)}=(3-\tau)\frac{\dot s(t)}{s(t)}
\label{eq:3.15}
\end{equation}
The appearance of the exponent $\tau$ in this equation may at first
appear disconcerting, but there is no real problem:
(\ref{eq:3.14}) can be solved for any $\tau$. It then remains to see
for which value of $\tau$ the condition (\ref{eq:3.910}) obtains. We
now show how these equations can be used to determine $\tau$ from the
knowledge of the exponents $\lambda$ and $\mu$ characterizing the
reaction kernel $\rate(x,y)$.
\subsection{The Solutions of the Scaling Equation}
\label{subsec:sol-scal}
\subsubsection{The Non-gelling Case: Small-$x$ Behaviour}
\label{subsubsec:small}
A very useful mathematical concept in the determination of the
behaviour of $\Phi(x)$ near the origin is that of regular behaviour.
We say that a function $\Phi(x)$ has regular behaviour at the origin
if the limit
\begin{equation}
h(a)=\lim_{x\to0}\frac{\Phi(ax)}{\Phi(x)}
\label{eq:3.16}
\end{equation}
exists. If this is the case, it is easy to see  that a number $\tau$
must exist such that
\begin{equation}
h(a)=a^{-\tau}.
\label{eq:3.17}
\end{equation}
This is shown by proving that $h(a)$ must satisfy the functional
equation
\begin{equation}
h(ab)=h(a)h(b),
\label{eq:3.18}
\end{equation}
for which power laws are the only solutions\footnote{I disregard
non-measurable functions throughout this paper.}.

We now assume that $\Phi(x)$ has regular behaviour near the origin.
That is, we assume that it has a power law behaviour, possibly
dressed by arbitrary logarithmic corrections, but no behaviour of
the type $\cos(b\ln x)$. We shall find that this assumption is
self-consistent for $\mu\geq 0$, but that it is not so when $\mu<0$.
In the former case it allows us to derive more or less explicit
expressions for $z$ and $\tau$. In the latter, we shall see that $\Phi(x)$
has a stretched exponential behaviour near the origin, which can be
determined by other techniques. It should be realized, however, that
self-consistency does not prove that the function $\Phi(x)$ does
display regular behaviour at the origin. In fact, there is substantial
numerical evidence \cite{lee01} that when $\mu>0$ the function
$\Phi(x)$ has non-regular behaviour at the origin, with (possibly)
undamped logarithmic oscillations.

Since we are interested in small values of $x$, we naturally consider
the large-$\rho$ behaviour of the two sides of (\ref{eq:3.12}).
We show in appendix \ref{app:large-rho}, see (\ref{eq:b.6}) and
(\ref{eq:b.7}), that
\begin{equation}
I(\rho)=\const_1\rho^{-(\lambda+3)}\Phi(1/\rho)^2+
\const_2\rho^{-2}\Phi(1/\rho)k(1/\rho),
\label{eq:3.19}
\end{equation}
where $\const_1$ and $\const_2$ are definite integrals defined by
(\ref{eq:b.7}) with the following peculiarity: of the two integrals
involved, one will diverge and the other one converge for any choice
of $\Phi(x)$. Only the term with the convergent integral should be
retained in (\ref{eq:3.19}).

{}From this follows, by matching $I(\rho)$ with the r.h.s. of
(\ref{eq:3.12})
which is of order $\rho^{-2}\Phi(1/\rho)$,
the following well-known three cases (the numbers
have become standard terminology):
\begin{itemize}
\item Case I: $\mu>0$. For these values of the parameters one finds
\begin{equation}
\tau=1+\lambda\qquad w=1.
\label{eq:3.20}
\end{equation}
In this case, however, there exists some doubt on the very existence of
a well-defined solution to (\ref{eq:3.10}). The issue is  the
following: it is readily seen that the function $x^{-(\lambda+1)}$
solves (\ref{eq:3.10}) quite generally. However, it clearly never
fulfils the essential normalization condition (\ref{eq:3.1305}). This
means that the condition of strong decay as $x\to\infty$ is essential
in determining the solution. This has cast some doubt on the existence
of such a solution. On  the other hand, recent numerical work by Lee
(see \cite{lee01}) strongly suggests that such a solution does
exist\footnote{contradicted, however, by earlier
work, see in particular \cite{kri95}}, though
it may not behave regularly at the origin: specifically, Lee observes
oscillations in $\Phi(x)$, which might well be of the
type $\cos\ln x$. These do not conform to the regularity hypothesis.
On the other hand, if $\Phi(x)$ really does behave regularly at the
origin, we may state that
\begin{equation}
\Phi(x)=\const_1 x^{-\tau},
\label{eq:3.21}
\end{equation}
where $\const_1$ is given by (\ref{eq:b.7}), which is a definite
integral that can be explicitly evaluated for every single kernel.

Finally note that the value for $w$ is the lowest possible for the
monomer decay
exponent in a system with binary reactions, since $c_1(t)$ can always
decay at least via a reaction with itself.
\item Case II: $\mu=0$. In this case, the exponent $\tau$ is
non-universal, but satisfies the inequality
\begin{equation}
\tau<1+\lambda.
\label{eq:3.22}
\end{equation}
If $\Phi(x)$ behaves regularly at the origin, then a comparison of
the prefactors of the r.h.s. and l.h.s. of (\ref{eq:3.10}) yields the
well-known relation \cite{don88}
\begin{equation}
\tau=2-\pref\int_0^\infty x^\lambda\Phi(x) dx=2-\pref\scalmom[\lambda],
\label{eq:3.23}
\end{equation}
where $\pref$ is defined by (\ref{eq:3.115}) and $\scalmom$
is defined as in (\ref{eq:3.905}).
Note that this result is invariant under the symmetry
transformation (\ref{eq:3.1310}) as well as under a common scale
change of $\Phi$ and $\rate$ (due to the presence of $\pref$).

Of course, (\ref{eq:3.23}) does not straightforwardly determine $\tau$,
since it requires in principle the knowledge of the full scaling
function $\Phi(x)$, from which the exponent $\tau$ could have been
derived without recourse to (\ref{eq:3.23}). However, as has been
shown in \cite{don88,cue97}, (\ref{eq:3.23}) can be used to determine
bounds on $\tau$ for any specific kernel independently of $\Phi(x)$.
In particular, Cueille and Sire \cite{cue97}
have developed a large number of very
elegant methods which allow to obtain extremely sharp rigorous
upper and lower bounds on $\tau$, to the extent that one may view
these as the equivalent of an exact determination. In some specific
cases, $\tau$ can also be displayed as the solution to a
transcendental equation \cite{don85}.

\item Case III: $\mu<0$. In this case, the assumption of regular
behaviour of $\Phi(x)$ near the origin leads to inconsistencies. We
may therefore assume a decay of $\Phi(x)$ near the origin that is
faster than every power. One may then approximate $I(\rho)$ as
shown in appendix \ref{app:large-rho}, with
the result
\begin{equation}
I(\rho)=\pref\scalmom[\lambda-\mu]\int_0^\infty x^{1+\mu}\Phi(x)e^{-\rho
x} dx.
\label{eq:3.24}
\end{equation}
{}From this and (\ref{eq:3.12}) follows, by inverting Laplace
transforms, that within our approximations
\begin{equation}
\left[x^2\Phi(x)\right]^\prime=\pref\scalmom[\lambda-\mu]x^{1+\mu}\Phi(x).
\label{eq:3.25}
\end{equation}
{}From this one finds the approximate form for $\Phi(x)$ near
the origin in this case:
\begin{equation}
x^2\Phi(x)=const.\cdot\exp\left[
-\pref\scalmom[\lambda-\mu]\frac{x^{-|\mu|}}{|\mu|}
\right]
\label{eq:3.26}
\end{equation}
Here only the exponential behaviour with the constants indicated are
reliable. The power law $x^{-2}$ does not follow at the degree of
accuracy we have used here. For a detailed treatment of corrections
to scaling for both Case II and Case III, see \cite{don88}. It is an
entertaining excercise for the reader to verify that (\ref{eq:3.26})
is indeed invariant under the symmetry transformations
(\ref{eq:3.1310}) as well as under a simultaneous change of scale
in $\rate$ and $\Phi(x)$.
\end{itemize}
Yet another dificulty arises when $\lambda=1$ and $\mu>0$: in this
case, since $\tau$ is predicted by (\ref{eq:3.20}) to be two, the
normalization condition (\ref{eq:3.1305})
for $\Phi(x)$ is not satisfied any more. Under these circumstances
the easiest solution is presumably to view this as an instance where
it is more appropriate to look at the convergence of the second
moment to a scaling form, so we defer consideration of this issue to
the next subsection, in which we consider also the gelling case,
which cannot be treated in any other way. For a different approach,
however, see \cite{don88}. On the other hand,
if $\mu\leq0$, we find ourselves either in Case II or in Case III.
Due to inequality
(\ref{eq:3.22}) in Case II and (\ref{eq:3.26}) in Case III,
no convergence problems arise even when $\lambda=1$.
\subsubsection{The Non-gelling Case: Large-$x$ Behaviour}
\label{subsubsec:large}
The large-$x$ behaviour in the non-gelling case has been treated by van
Dongen and Ernst in \cite{don87c,don88}. We present here a slightly
modified version of their approach. Let us first show that whenever
$K(\mass, \massp)$ is less than $\mass+\massp$, the scaling function
$\Phi(x)$ decays exponentially for large values of $x$. This can be done
as follows: consider the moments $\scalmom$ of the scaling function. These
satisfy the relation (\ref{eq:3.1302}), which implies, using the bound on
the kernel $K(\mass, \massp)$
\begin{equation}
\scalmom[\indmom+1]\leq\frac{1}{2\indmom}\sum_{k=1}^\indmom\left(
\begin{array}{c}
\indmom\\
k
\end{array}
\right)
\scalmom[k]\scalmom[\indmom-k+1].
\label{eq:3.7.1}
\end{equation}
If we now denote by $\scalmom^{(0)}$ the moments which satisfy
(\ref{eq:3.7.1}) as an equality, it is easy to show inductively that
\begin{equation}
\scalmom\leq\scalmom^{(0)},
\label{eq:3.7.2}
\end{equation}
for all $\indmom$. On the other hand, it is seen that the $\scalmom^{(0)}$
can be evaluated analytically, since they are the moments of the scaling
function of the kernel $\mass+\massp$, which is known exactly. One
therefore obtains
\begin{equation}
\scalmom\leq\scalmom^{(0)}=const.\cdot \indmom^{-3/2}\indmom!R^\indmom
\label{eq:3.7.3}
\end{equation}
for all $\indmom$ and some fixed value of $R$, which strongly depends
on various normalizations. (\ref{eq:3.7.3}) now immediately leads to
the desired conclusion by using standard results on the moment problem.
Similar lower bounds can also be shown assuming correponding lower bounds
on $K(\mass, \massp)$.

Assuming some such lower bound, it
follows that the integral equation (\ref{eq:3.12}) also
holds for negative values of $\rho$, up to a critical value $-\rho_c$
beyond which all integrals diverge. Introducing
\begin{equation}
h(x)=\Phi(x)e^{\rho_cx}
\label{eq:3.7.4}
\end{equation}
one obtains the following equation from (\ref{eq:3.12}) for $h(x)$
\begin{eqnarray}
\int_0^\infty dx \,x^2h(x)e^{-(\rho+\rho_c)x}&=&-\rho
\int_0^\infty
dx\,dy\,xK(x,y)h(x)h(y)e^{-(\rho+\rho_c)(x+y)}\times\nonumber\\
&&\qquad\times\left[
1-e^{\rho y}
\right]
\label{eq:3.7.5}
\end{eqnarray}
If we now assume that $h(x)$ behaves regularly at infinity with exponent
$\explarge$, that is,
\begin{equation}
\lim_{x\to\infty}\frac{h(ax)}{h(x)}=a^{-\explarge}
\label{eq:3.7.6}
\end{equation}
then one obtains, using altogether the same techniques as for the small-$x$
behaviour, that if $\nu<1$
\begin{equation}
\explarge=\lambda.
\label{eq:3.7.7}
\end{equation}
If, on the other hand, $\nu=1$, then the integral on the l.h.s of
(\ref{eq:3.7.5}) diverges. A treatment entirely similar to the one
corresponding to the indeterminate Case II for the small $x$ behaviour
shows that the exponent $\explarge$ in this case also cannot be determined
uniquely using only the type of scaling considerations discussed above.
Details on this subject are to be found in \cite{don87c}.
\subsubsection{The Dominant Singularity Hypothesis}
\label{subsubsec:dom-sing}
The following scenario for the explanation of scaling may, at this stage,
quite possibly have occurred to the thoughtful reader: if we consider the
generating function (or the Laplace transform) of the cluster size
distribution function $G(\lap,t)$ defined by
\begin{equation}
G(\lap,t)=\int_0^\infty c(\mass, t)e^{\lap\mass}d\mass,
\label{eq:3.11.1}
\end{equation}
one may assume that the appearance of a unique divergent size $s(t)$ in
the cluster size distribution $c(\mass, t)$ might translate itself in the
appearance of a single, dominant singularity of $G(\lap,t)$ on the positive
real axis. Certainly, standard theorems on Laplace transforms state that
the nearest singularity does indeed lie on the positive real axis, so
that such a hypothesis may seem plausible. This assumption might be
further strengthened by the careful study of a large number of exactly
solved models, since indeed the mechanism explaining the growth of the
typical size and the appearance of scaling is invariably of this nature.

Let us now use our understanding of the nature of the scaling function
$\Phi(x)$ in order to assess this possibility. Put in equations, our
assumption states:
\begin{equation}
G(\lap,t)=const.\cdot\left[
\lap_c-\lap s(t)
\right]^{\explarge+1}+\mbox{higher order terms},
\label{eq:3.11.2}
\end{equation}
where $\lap_c$ is some appropriate constant.

I have denoted the relevant exponent by $\explarge$ because, as we shall
now see, it has always that value. Indeed, we know from (\ref{eq:3.801})
that
\begin{equation}
\int_0^\infty x\Phi(x)e^{-\lap x}=
\left.
\frac{\partial G}{\partial\lap}(\lap,t)
\right|_{\lap=-\rho/s(t)}
\label{eq:3.11.3}
\end{equation}
The large $x$ behaviour of $\Phi(x)$ is given, as is well-known from
general theorems on Laplace transforms, by that of the nearest singularity
of $G(\lap, t)$. This nearest singularity is on the real axis, and we have
assumed it to be a power-law singularity. It therefore certainly always
describes the value of the $\explarge$ exponent.

Does it also yield the value of the $\tau$ exponent? Clearly, this
depends somewhat on the meaning we give to the phrase ``higher order
terms'' in (\ref{eq:3.11.2}). However, generally speaking, if we have
assumed only one singularity dominating the whole picture, it is hard to
see how it could be otherwise. One therefore deduces from the hypothesis
just described that
\begin{equation}
\tau=\explarge.
\label{eq:3.11.4}
\end{equation}
But this, we know from subsection \ref{subsubsec:large} is simply wrong in
the large majority of cases. Thus, when $\mu>0$, $\tau=1+\lambda$, which
cannot be equal to $\explarge$, which is $\lambda$. The only cases in
which this possibility exists are the loci $\mu=0$, for which $\tau$ is
non-universal, and $\nu=1$ for which $\explarge$ is non-universal. It is
certainly striking that all the exactly solved models known to
this day find
themselves on these loci, and that they indeed satisfy (\ref{eq:3.11.4}).
This should therefore serve as a warning: some features of exactly solved
models are highly atypical.

In subsection \ref{subsec:prod} we shall see that the constant kernel with
production term provides a counterexample to the dominant scaling
hypothesis. In this case, as will be shown later, one has $\tau=3/2$ and
$\explarge=0$. The generating function has infinitely many simple
poles tending simultaneoulsy to the origin as $t\to\infty$. The nearest
such pole dominates the large mass behaviour and leads to $\explarge=0$,
whereas the combined effect of all the poles together upon values of $\lap$
just below zero leads to a $-3/2$ singularity.
\subsubsection{The Gelling Case: Scaling Region}
\label{subsubsec:gel-scale}
In the gelling case, we define scaling using the approach to scaling of
second moments. This is appropriate as long as $\tau<3$. In order to have
no difficulty on that score, we limit ourselves in the following to the
following range of parameters
\begin{equation}
1<\lambda\leq2\qquad\nu\leq1.
\label{eq:3.8.1}
\end{equation}
The condition $\lambda>1$ is necessary to ensure gelation, whereas the
condition $\nu\leq1$ is required to avoid pathologies such as
instantaneous gelation or non-existence of solutions. The condition which
we really only put in for convenience is $\lambda\leq2$.

Our definition of scaling is therefore the following, there exist a
typical size $s(t)$ diverging at the gel time $t_c$ and a scaling
function $\Phi(x)$ such that
\begin{equation}
\lim_{t\nearrow t_c}\left[
\frac{1}{\mom[2](t)}\int_0^\infty
\mass^2c(\mass,t)f\left(
\frac{\mass}{s(t)}
\right)d\mass
\right]=
\constsep\int_0^\infty x^2\Phi(x)f(x).
\label{eq:3.8.2}
\end{equation}
One now proceeds as in the non-gelling case to derive an integral equation
for $\Phi(x)$. The details are carried out in Appendix
\ref{app:gel-equation}. If we now define
\begin{equation}
s(t)=const.\cdot(t_c-t)^{-1/\sigma},
\label{eq:3.8.3}
\end{equation}
It then follows from the work done in Appendix
\ref{app:gel-equation} that
\begin{equation}
\sigma=1+\lambda-\tau.
\label{eq:3.8.4}
\end{equation}
In other words, it is no more possible to fix the exponent describing the
growth of the typical size in terms of the exponents describing the
kernel. Instead, we have  a scaling relation between $\sigma$ and $\tau$.
Finally, for the integral equation determining $\Phi(x)$, one finds
\begin{eqnarray}
&&\int_0^\infty dx\,dy\,xK(x,y)\Phi(x)\Phi(y)\left[
(x+y)f(x+y)-xf(x)
\right]=\nonumber\\
&&\qquad
\int_0^\infty  dx\,x^2\Phi(x)\left[
(3-\tau)f(x)+xf^\prime(x)
\right]
\label{eq:3.8.5}
\end{eqnarray}
The novel feature here again is, of course, the appearance of $\tau$ in
the equation. At the basic level, presumably, it does not change much: one
can, in principle, solve (\ref{eq:3.8.5}) for each value of $\tau$ and then
see for which values of $\tau$ (if any) the corresponding solution, which we
may call $\Phi_\tau(x)$, behaves as $x^{-\tau}$ near the origin.

The difficulty of a theoretical treatment, however, is thereby considerably
increased. Let us first ask whether a formal solution of pure power-law
form exists, akin to the $x^{-(1+\lambda)}$ solution in the non-gelling
case. Let $x^{-\tau}$ be such a solution. In this case, it follows using
partial integration
that the r.h.s. of (\ref{eq:3.8.5}) vanishes identically for
arbitrary functions. Hence, putting $f(x)=e^{-x}$,
\begin{equation}
\int_0^\infty dx\,dy\,K(x,y)x^{1-\tau}y^{-\tau}e^{-x}\left[
(x+y)e^{-y}-x
\right]=0.
\label{eq:3.8.6}
\end{equation}
It is a straightforward exercise in gamma functions to show that, for
$K(x,y)$ of the form $x^\mu y^\nu+x^\nu y^\mu$, the solution of
(\ref{eq:3.8.6}) is
\begin{equation}
\tau=\frac{\lambda+3}{2}
\label{eq:3.8.7}
\end{equation}
Since, however, every kernel of homogeneity degree $\lambda$ can be
expressed in terms of kernels of this form, it would appear to follow, at
least on a formal level, that this value of $\tau$ is the one that
corresponds to the universal power-law solution
for arbitrary $K(x,y)$. As we shall see in the
next subsection, this value is indeed highly significant: it is the only
large-size behaviour compatible with a finite flow rate of mass to infinity. As
such, it is the only possible large-mass exponent for times beyond the gel
point. However, we are interested in the scaling region immediately {\em
before} the gel point. There is, in principle, no reason to assume that
this particular value of $\tau$ plays any role in this region.

Let us now attempt a scaling theory of the type we have achieved for the
non-gelling case. Our results, unfortunately, will be much more fragmentary
than in the regular case. As before, we use exponentials $e^{-\rho x}$
as functions $f(x)$ and investigate the large $\rho$ behaviour. Let us
assume that $\Phi(x)$ is of the form
\begin{equation}
\Phi(x)=Ax^{-\tau}\left[
1+Bx^\Delta+o(x^{\Delta})
\right]
\label{eq:3.8.8}
\end{equation}
where we have taken next-to-leading behaviour of $\Phi(x)$ into account,
since this is what determines the behaviour of the r.h.s. of
(\ref{eq:3.8.5}). One then obtains, under the hypothesis that the integral
(\ref{eq:3.8.6}) is different from zero, after some tedious
but straightforward calculations
\begin{equation}
\Delta=1+\lambda-\tau=\sigma.
\label{eq:3.8.9}
\end{equation}
On the other hand $\tau$ cannot be determined by straightforward scaling
arguments. Thus we have a general relation between the first correction to
scaling and the growth of the typical size, but neither can be determined,
both depending on the exponent $\tau$ which must be determined by
other considerations. It is not clear to me whether similar techniques
to the ones that were successful in determining $\tau$ to very high
accuracy in Case II of the non-gelling case \cite{cue97,cue98}
may be generalized to this case.

There are therefore two cases to be distinguished:
\begin{enumerate}
\item The ordinary case: The integral (\ref{eq:3.8.6}) is different from
zero. In this case we know nothing about the actual value of the exponent
$\tau$, but we know how both $\sigma$ and the leading next order
correction depend on it. The positivity of $\Delta$, which follows from
its definition, leads to the inequality for $\tau$
\begin{equation}
2<\tau<1+\lambda
\label{eq:3.8.10}
\end{equation}

\item The extraordinary case: The integral (\ref{eq:3.8.6}) vanishes. In
this case we have the ``standard'' value (\ref{eq:3.8.7}) for $\tau$. The
computation of the correction to scaling exponents is somewhat more
difficult, whereas the exponent $\sigma$ is given by
\begin{equation}
\sigma=\frac{\lambda-1}{2}.
\label{eq:3.8.11}
\end{equation}

\end{enumerate}
The nomenclature used above is meant to suggest that the ordinary case is
more common than the extraordinary. While I have no analytic arguments to
show it, the following evidence exists:
first, Lee \cite{lee01} has studied a large number of gelling kernels of
the form $x^\mu y^\nu+x^\nu y^\mu$, and has determined $\tau$ numerically
to high accuracy. In no instance except the case $\mu=\nu=1$ was $\tau$
given by the ``standard'' value (\ref{eq:3.8.7}) but was always smaller.
Second, even in the bilinear kernel, it can be shown
(see Appendix \ref{app:bilinear2}) that the case of power-law
initial conditions leads to continuously variable values of $\tau$ as well
as an exponent $\Delta$ satisfying (\ref{eq:3.8.9}). Therefore, even the
bilinear kernel sometimes yields the ordinary case. On
the other hand, the extraordinary case really does occur for the
bilinear kernel with initial conditions decaying rapidly with mass, since then
$\Delta=1$, which contradicts (\ref{eq:3.8.9}) with $\lambda=2$ and
$\tau=5/2$.

In principle, it would now be necessary to derive the value of $\tau$ in
the ordinary case from the overall structure of the reaction constants
$K(\mass, \massp)$. As far as I can see, this problem is of considerable
difficulty, and I have made no headway at all. Possibly the methods
developed in \cite{cue97,cue98} could be generalized to yield sharp upper
and lower bounds on the $\tau$ exponent. This is, of course, particularly
important since
all other exponents are expressed in terms of $\tau$ and are not known
explicitly. Lacking such a theory, we are led back to the numerical study
either of the gelation transition in the original equation (\ref{eq:2.2})
or else of the scaling function $\Phi(x)$ in (\ref{eq:3.8.5}).

Finally, let us note that the ordinary case can be characterized in an
amusing way as follows: consider the average concentrations defined as
follows
\begin{equation}
\overline C(\epsilon)=\frac{1}{\epsilon s(t)}\int_\epsilon^{2\epsilon}
c(\mass, t)d\mass.
\label{eq:3.8.12}
\end{equation}
It is then a straightforward consequence of the definition of scaling and
of the exponent $\tau$, that
\begin{equation}
\overline C(\epsilon)=const.\cdot \left[
\epsilon s(t)
\right]^{-\tau}
\label{eq:3.8.13}
\end{equation}
Now, it is easy to verify that the value of the correction to scaling
exponent $\Delta$ given by (\ref{eq:3.8.9}) corresponds exactly to the
fact that this limiting behaviour is reached linearly in time. This only
holds in the ordinary case, so that it should be violated in the case of
the bilinear kernel with initial conditions that decay rapidly as a
function of mass, as indeed it is:
there the average concentrations are readily seen to tend
to their limit quadratically in time.
\subsubsection{The Gelling Case: After Gelation}
\label{subsubsec:post-gel}
After gelation sets in, the power-law behaviour of the cluster size
distribution remains for all times. Since, by definition, mass must
continually decrease after gelation, the derivative must be a finite
number, that is,
\begin{equation}
0<\lim_{M\to\infty}\frac{d}{dt}\int_0^M \mass c(\mass, t)d\mass<\infty.
\label{eq:3.9.1}
\end{equation}
But one finds
\begin{eqnarray}
\frac{d}{dt}\int_0^M \mass c(\mass, t)d\mass&=&
\int_0^Md\mass_1\int_0^\infty
d\mass_2\Theta(\mass_1+\mass_2-M)\times\nonumber\\
&&\qquad\times\mass_1K(\mass_1,\mass_2)c(\mass_1,t)c(\mass_2,t)
\label{eq:3.9.2}
\end{eqnarray}
If one now inserts the power-law profile
$\mass^{-\tau_s}$ for $c(\mass, t)$ in
(\ref{eq:3.9.2}), one finds that
\begin{equation}
\frac{d}{dt}\int_0^M \mass c(\mass, t)d\mass=const.\cdot M^{\lambda+3-2\tau_s}
\label{eq:3.9.3}
\end{equation}
from which follows that, in order for (\ref{eq:3.9.1}) to hold, $\tau_s$
must indeed have the ``standard'' value discussed in the previous
subsection.

Concerning the large-time behaviour of the concentrations, it can be shown
rather generally that they all go as $t^{-1}$. In the discrete case, this
may be seen as follows: consider first the monomer equation
\begin{equation}
\dot c_1=-c_1\sum_{k=1}^\infty K(1,k)c_k\simeq-c_1\mom[\nu](t).
\label{eq:3.9.4}
\end{equation}
Since $\tau-\nu>1$, the sum in $\mom[\nu](t)$ converges, so that this
quantity behaves similarly to $c_1(t)$. Equation (\ref{eq:3.9.4}) then reduces
to
\begin{equation}
\dot c_1=-const.\cdot c_1^2,
\label{eq:3.9.5}
\end{equation}
which has the stated behaviour. The general statement then follows by
induction, since it is straightforward to show that the production terms
are of order $t^{-2}$. Since, however, the removal terms must dominate,
they must be of order $t^{-2}$ as well, implying the stated result.

In fact, it is seen that the {\em ansatz}
\begin{equation}
c_j(t)=\frac{a_j}{t+t_0}
\label{eq:3.9.6}
\end{equation}
can be made to satisfy (\ref{eq:2.2}) if the $a_j$ satisfy the following
algebraic relations
\begin{equation}
a_j=\half\left[
\sum_{k=1}^\infty K(j,k)a_k-1
\right]^{-1}
\sum_{k=1}^{j-1}K(k,j-k)a_ka_{j-k}.
\label{eq:3.9.7}
\end{equation}
It turns out to be quite hard to know whether these equations indeed have a
positive solution. For the case in which $K(k,l)$ is of the form
$(kl)^{\lambda/2}$, considerable progress can be made by reducing
(\ref{eq:3.9.7}) to a recursion together with a self-consistency condition.
It was shown \cite{ley83} that in this case the $a_j$ exist and that
for large $j$ they display a power-law behaviour
$j^{-\tau_s}$if $\lambda>1$. If $0<\lambda<1$, these
solutions presumably describe the behaviour of $c_j(t)$ at fixed $j$ and
$t$ large. They behave for large $j$ as $j^{-(1+\lambda)}$, in
agreement with the results derived previously for Case I.
\subsubsection{The Non-Gelling Case for $\lambda=1$}
\label{subsubsec:lam-one}
As has been pointed out above, if $\lambda=1$ and $\nu<1$, the ordinary
scaling theory suggests $\tau=2$. This, as first pointed out by
van Dongen and Ernst \cite{don88}, is a contradiction, since
one then has a divergent total mass at the origin. The way out is to
consider this as a borderline case and treat it via the convergence in
second moment, in the same way as we have already done with gelation.

The details are carried out in Appendix \ref{app:gel-equation} and lead to
the following relations
\begin{eqnarray}
s(t)&=&const.\cdot\exp\left(
const.\cdot\sqrt t
\right)\nonumber\\
\mom[2](t)&=&\frac{s(t)}{\ln s(t)}
\label{eq:3.12.1}
\end{eqnarray}
Convergence to scaling then takes  place as follows
\begin{equation}
\lim_{t\to\infty}\left[
\sqrt te^{-\sqrt t}\int_0^\infty
\mass^2c(\mass, t)f\left(
\mass e^{-\sqrt t}
\right)
\right]d\mass=\int_0^\infty x^2\Phi(x)f(x)dx.
\label{eq:3.12.2}
\end{equation}
One may then absorb one factor of $\mass$ into the function $f(\mass
e^{-\sqrt t})$ and obtains
\begin{equation}
\lim_{t\to\infty}\left[
\sqrt t\int_0^\infty
\mass c(\mass, t)f\left(
\mass e^{-\sqrt t}
\right)
\right]d\mass=\int_0^\infty x\Phi(x)f(x)dx
\label{eq:3.12.3}
\end{equation}
The function $\Phi(x)$ must then satisfy the following integral equation
derived in Appendix \ref{app:gel-equation}
\begin{eqnarray}
&&\int_0^\infty dx\,dy\,xK(x,y)\Phi(x)\Phi(y)\left[
(x+y)f(x+y)-xf(x)
\right]=\nonumber\\
&&\qquad
\int_0^\infty  dx\,x^2\Phi(x)\left[
f(x)+xf^\prime(x)
\right]
\label{eq:3.12.4}
\end{eqnarray}
As in the gelling case, $\Phi(x)$ need not have finite mass. This means
that, as in the gelling case, the total mass contained in the scaling
region vanishes.
\subsection{Crossover}
\label{subsec:cross}
Up to now, we have mainly considered the case in which
the reaction rate $K(\mass,\massp)$ are homogeneous in the
masses. The justification of this hypothesis lies, as stated
in the Introduction, in the fact that we are principally
interested in the behaviour of the system at large times
and for large aggregate sizes. Arguably, in such circumstances,
any simple physical mechanism for aggregation will have
a simple behaviour with respect to scaling of the masses in this
asymptotic regime. On the other hand, if two mechanisms act
jointly, we may expect that the one that scales with the lesser
homogeneity degree will be altogether negligible, so that
we may, again, limit ourselves to the homogeneous case.
These considerations already suggest, however, a possibly
important exception: Assume a fast mechanism which only
becomes appreciable at quite large sizes, superimposed
on another mechanism, with a lesser degree of homogeneity,
but effective at all sizes. The following illustrates such
a possibility:
\begin{equation}
K(\mass, \massp)=K_1^{(\lambda_1)}(\mass, \massp) +\epsilon
K_2^{(\lambda_2)}(\mass, \massp)
\label{eq:3.5.1}
\end{equation}
Here the $\lambda_{1,2}$ are two different degrees of homogeneity
and $\epsilon$ is a very small number. Under these circumstances,
if $\lambda_2<\lambda_1$, the second term in the sum can obviously
be neglected, but in the opposite case, it will eventually dominate.
{}From the point of view of pure scaling thery, it is in principle
sufficient to say this, and one has, indeed, a valid prediction
for ''sufficiently long'' times. For practical purposes, however,
this is often an excessive limitation, and one would like a
scaling theory which gives correct results when times are
large with respect to the scale determined by $K_1^{-1}$ as opposed
to $\epsilon^{-1}$. To this end, we need a theory of crossover.

To avoid unnecessary difficulties, we first fix the scale of the
parameter $\epsilon$: indeed, it is obviously irrelevant whether
we use $\epsilon$ or $f(\epsilon)$ as a crossover parameter, where
$f$ represents an arbitrary monotonic
function vanishing at the origin.
In order to fix the parametrization, I therefore make the
following convention
\begin{equation}
K(a\mass, a\massp;\epsilon/a)=a^\lambda K(\mass, \massp;\epsilon)
\label{eq:3.5.2}
\end{equation}
At a naive level, this means that $\epsilon$ has a dimension
of inverse mass, that is, the natural scale on which $\epsilon$
varies is given by $s(t)^{-1}$. This requirement does not determine
$\lambda$ uniquely, though: indeed, replacing $K$ by
$K^\prime=\epsilon^\alpha K$ changes $\lambda$ to $\lambda+\alpha$ while
respecting (\ref{eq:3.5.2}). We therefore impose the additional requirement
that $K(1,1)$ always remain of order one.
Thus, for example, the kernel
given in (\ref{eq:3.5.1}) is rewritten as
\begin{equation}
K(\mass, \massp)=K_1^{(\lambda_1)}(\mass, \massp) +\epsilon^{\lambda_2
-\lambda_1}
K_2^{(\lambda_2)}(\mass, \massp)
\label{eq:3.5.25}
\end{equation}
Note that this only gives a meaningful result if $\lambda_1<\lambda_2$,
as was to be expected on intuitive grounds anyway.

We now proceed as always: let
us first define what we mean by a scaling limit. This is given by
\begin{equation}
\lim_{t\to\infty}\int_0^\infty \massp c(\massp,y/s(t);t)
f\left(\frac{\massp}{s(t)}\right)\to\constsep\int_0^\infty x\Phi(x,y)dx,
\label{eq:3.5.3}
\end{equation}
where $c(\mass,\epsilon;t)$ denotes the value of the concentration
of clusters of mass $·\mass$ at time $t$ evolving under a dynamics using
a kernel with a fixed value $\epsilon$ of the crossover parameter.
We now use the standard approach in order to derive the following
equation for the typical size and the scaling function.
For the typical size $s(t)$ one obtains
\begin{equation}
\dot s(t)=\constsep s(t)^\lambda,
\label{eq:3.5.4}
\end{equation}
where $\lambda$ is defined by (\ref{eq:3.5.2}). Note that, for the case
described by (\ref{eq:3.5.25}), this means that the degree of
homogeneity is $\lambda_1$, that is the {\em lesser} degree of
homnogeneity. From this follows that we will be looking at the
size distribution on a scale defined altogether by the slow
aggregation process defined by $K_1^{(\lambda_1)}$. This
may at first appear strange, but it should be noted that, in
the scaling limit (\ref{eq:3.5.3}), the crossover parameter $\epsilon$
goes to zero, so that the large-time behaviour
of any specific system is not the one observed in the crossover
limit.

For the scaling function, we obtain the following equation in
exactly the same way as we obtained (\ref{eq:3.10}):
\begin{eqnarray}
&&\int_0^\infty dx_1\,dx_2\,x_1K(x_1,x_2;y)\Phi(x_1,y)\Phi(x_2,y)
\times\\
&&\qquad\times\left[f(x_1+x_2)-f(x_1)\right]=
\int_0^\infty dx\,xf(x)
\left[
y\Phi_y(x,y)-x\Phi_x(x,y)-2\Phi(x,y)
\right],\nonumber
\label{eq:3.5.5}
\end{eqnarray}
which must hold for all continuous function $f(x)$ and for all $y$.
{}From this we again derive an equation involving exponentials similar
to (\ref{eq:3.12}):
\begin{eqnarray}
&&\int_0^\infty dx_1\,dx_2\,x_1e^{-\rho x_1}
K(x_1,x_2;y)\Phi(x_1,y)\Phi(x_2,y)\times\\
&&\qquad\times\left[1-e^{-\rho x_2}\right]=
\int_0^\infty dx\,x e^{-\rho x}
\left[
2\Phi(x,y)+x\Phi_x(x,y)-y\Phi_y(x,y)
\right],\nonumber
\label{eq:3.5.6}
\end{eqnarray}
If one now takes the limit of large $\rho$ and makes exactly
similar considerations as in the case without crossover, it
is straightforward to show that the exponent $\tau$ defined
by
\begin{equation}
\lim_{a\to0}\frac{\Phi(ax,y)}{\Phi(a,y)}=x^{-\tau},
\label{eq:3.5.7}
\end{equation}
where $y$ is taken to be fixed, is the same as for the kernel
$K(x_1,x_2;0)$, that is, it is identical to the value
which would be established if the fast mechanism dominating
at long times did not exist. In other words, the small
size end of the cluster size distribution is not affected
by the existence of the fast aggregation mechanism in the crossover
limit as defined by (\ref{eq:3.5.3}). Again, this is not at
variance with the prediction that the whole size
distribution will eventually be determined by the fast
process: as time goes on, at fixed $\epsilon$, the value of $y$
diverges and the range of $x$ for which the exponent $\tau$
of the slow process yields a good description, tends to zero.

The general properties of (\ref{eq:3.5.7}) are not easy to work
out, however. It is seen that there is a formal solution of the form
\begin{equation}
\Phi(x,y)=y^{\lambda+1}\chi(xy),
\label{eq:3.5.8}
\end{equation}
where $\chi(x)$ is some function satisfying an equation which
is readily obtained from (\ref{eq:3.5.7}). This equation,
however, turns out to have no meaningful solutions,
since setting $\rho$ to zero in it leads to a contradiction.
The solution given in (\ref{eq:3.5.8}) must therefore
be viewed as a purely formal one, similar to the solution of
the form $\Phi(x)=x^{-(\lambda+1)}$ in the usual theory.

Frequently, in practice, one  is interested in the time evolution
of the moments of the size distribution function, since these
are often the  only accessible quantities. In this case, if
the system is at a fixed value of $\epsilon$, the resulting
expression is given by
\begin{equation}
\mom(t;\epsilon)=\constsep\int_0^\infty x^\indmom
\Phi(x,\epsilon s(t))dx,
\label{eq:3.5.9}
\end{equation}
where $s(t)$ is given by the solution to (\ref{eq:3.5.4}).
Note, of course, that  under these circumstances it is
not possible to define $s(t)$ via moment ratios, as is
otherwise often done.
\subsection{Extensions of the Model}
\label{subsec:ext}
\subsubsection{Monomer Production and Diffusion}
In many situations, it happens that monomers are continuously supplied
to the system, so that total mass is not conserved any more. If this
lasts only for a short time, the evolution of the system after the
monomer injection can be viewed as ordinary aggregation
starting from a somewhat polydisperse initial condition. This case is
therefore not essentially new. If, on the other hand, the supply of
monomer takes place in such a way that the total mass supplied
diverges, new phenomena arise.

Let us specifically look at the following equations also considered
by various authors, see  in particular
\cite{lus01,lus01a} and references therein, as well as
\cite{dav99,ley80}:
\begin{eqnarray}
\partial_tc(\mass,t)&=&\int d\mass_1\,d\mass_2\,
K(\mass_1,\mass_2)\,c(\mass_1, t)
c(\mass_2, t)\times\nonumber\\
&&\times\left[
\delta(\mass_1+\mass_2-\mass)-\delta(\mass_1-\mass)-\delta(\mass_2-\mass)
\right]+\nonumber\\
&&+p(t_0+t)^\omega\delta(\mass-\mass_0).
\label{eq:3.3.1}
\end{eqnarray}
Here the time shift $t_0$ is introduced merely in order to avoid
spurious divergences if $\omega$ is negative. From (\ref{eq:3.3.1})
one immediately obtains for the total mass, assuming that gelation
does not take place, that is, that $\lambda\leq1$,
\begin{equation}
\int_0^\infty d\mass\,\mass c(\mass,
t)=\frac{p\mass_0(t_0+t)^{1+\omega}}{1+\omega}.
\label{eq:3.3.2}
\end{equation}
{}From this follows the following natural analogue to the scaling
definition (\ref{eq:3.7})
\begin{equation}
\frac{1+\omega}{p\mass_0(t_0+t)^{1+\omega}}
\int_0^\infty \mass c(\mass, t) f[\mass/s(t)]d\mass
\mathop{\longrightarrow}_{t\to\infty}
\constsep\int_0^\infty
x\Phi(x)f(x)dx
\label{eq:3.3.3}
\end{equation}
{}From this we may again, using methods entirely similar to those
employed in Appendix \ref{app:deriv-scaling}, obtain the following
result for the typical size $s(t)$:
\begin{equation}
s(t)\approx(\constsep t)^{(\omega+2)/(1-\lambda)}.
\label{eq:3.3.4}
\end{equation}
For the scaling function $\Phi(x)$ one obtains the following relation
similar to (\ref{eq:3.10}):
\begin{eqnarray}
&&\frac{1+\omega}{\constsep}\int_0^\infty dx\,x\Phi(x)f(x)+
\int_0^\infty x^2\Phi(x)f^\prime(x)dx=\int_0^\infty dx\,dy\,K(x,y)x\Phi(x)
\Phi(y)\times\nonumber\\
&&\qquad\times\left[
f(x+y)-f(x)
\right]+\constsep^{-2}f(0),
\label{eq:3.3.5}
\end{eqnarray}
If one now substitutes $f(x)$ by
$e^{-\rho x}$ one obtain in the limit of large $\rho$ the following
condition for $\Phi(x)$:
\begin{equation}
I(\rho)=\constsep^{-2},
\label{eq:3.3.6}
\end{equation}
where $I(\rho)$ is defined by (\ref{eq:3.12}). If we now use the
results of Appendix \ref{app:large-rho} on the large $\rho$
asymptotic behaviour of $I(\rho)$, we find
\begin{equation}
\tau=\left\{
\begin{array}{lr}
\displaystyle\frac{\lambda+3}{2}&\qquad(\nu\leq\mu+1)\\
\mu+2&\qquad(\nu\geq\mu+1)
\end{array}
\right.
\label{eq:3.3.7}
\end{equation}
Assuming strong scaling we can state how the small clusters behave.
If $\nu\leq\mu+1$ one finds
\begin{equation}
c_j(t)\approx j^{-(\lambda+3)/2}t^{\omega/2}
\label{eq:3.3.8}
\end{equation}
as stated in \cite{dav99}. In particular, if $\omega=0$, that is, if
monomers are injected into the system at a constant rate, one
approaches a stationary distribution with a power-law exponent
$(\lambda+3)/2$. That this value is ubiquitous when a stationary state
is reached is well kmown (see e.g. \cite{whi82}).
On the other hand, if $\nu\geq\mu+1$, one finds
\begin{equation}
c_j(t)\approx j^{-(\mu+2)}t^{(\mu+1-(\omega+1)\nu)/(1-\lambda)}
\label{eq:3.3.9}
\end{equation}
which leads to decaying solutions when $\omega=0$. Therefore, in this
case, constant monomer input does {\em not} lead to a stationary
cluster size distribution, although no gelation takes place. A case
where this in fact occurs is given, for instance, by the kernel
$\mass_1^{-2}+\mass_2^{-2}$. In this case all concentrations decay as
$t^{-1/3}$ irrespective of the power law $\omega$, that is,
independently of the rate at which monomer is fed into the system.
The systematic nature of the approach developed here is worth emphasizing:
while I believe most of the results stated in this section to be
well-known, I am not aware of a similarly straightforward unified
derivation.
\subsubsection{Spatially inhomogeneous systems}
Let us now look at spatially inhomogeneous systems. As it frequently
happens that monomer injection occurs at a well localized position,
such systems are of genuine interest. At first, one might argue that
they must lie outside the domain of validity of mean field theory.
This is not generally the case, however: In order for the
mean field approximation to hold, one requires that there be no correlations
at the molecular level. Therefore, we must certainly exclude cases in
which the density varies significantly on the scale of typical interaggregate
distances. It often happens, though, that one has homogeneity
over such scales and yet a slow spatial variation giving rise to
diffusive dynamics being superimposed upon the reaction process. This
is the situation we shall consider. For closely
related work, see e.g \cite{che89}, but the literature on this kind of
problems is considerable.

We are therefore led to the following kinetic equations
\begin{eqnarray}
\partial_tc(\mass, \vec r)&=&\half\int_0^\infty d\mass_1\,d\mass_2\,
K(\mass_1,\mass_2)c(\mass_1, \vec r)c(\mass_2, \vec r)\times\nonumber\\
&&\times\left[
\delta(\mass_1+\mass_2-\mass)-\delta(\mass_1-\mass)-\delta(\mass_2-\mass)
\right]+\nonumber\\
&&+D(\mass)\Delta_{\vec r}c(\mass, \vec r)+p\delta(\vec
r)\delta(\mass-\mass_0)(t_0+t)^\omega.
\label{eq:3.4.1}
\end{eqnarray}
In the scaling approach we now introduce two time dependent growing
quantities, namely $s(t)$ which is the typical size of aggregates,
and $L(t)$, which the typical distance over which aggregate
concentrations vary significantly. We then require as a consistency
condition on our approximations, that $L(t)$ be much larger than the
radius of a typical aggregate. As we have discussed in the Introduction,
the relation betwen mass and radius is by no means a
trivial issue and depends sensitively on the particular model.
However, in a large variety of cases, the asymptotic relationship
between the mass $s(t)$ and the radius $R(t)$ can be
described by
\begin{equation}
s(t)\approx R(t)^{D_f},
\label{eq:3.4.2}
\end{equation}
where $D_f$ is the fractal dimension of the aggregate and lies
between one and the space dimension $d$. The minimal requirement for
the applicability of mean-field therefore becomes
\begin{equation}
L(t)\gg s(t)^{1/D_f}
\label{eq:3.4.3}
\end{equation}
for al $t$. In particular, if the growth exponent of $L(t)$ is less
than $z/D_f$, then mean-field will at best be applicable during a
finite range of times, and non-mean-field behaviour will dominate the
system at large times. In the opposite case, van Dongen has determined the
limit of validity of mean-field theory by considering the size of the
fluctuations and the efficiency of their transport. If one assumes that
the diffusion constants $D(\mass)$ decay with an exponent
$\mass^{-\gamma}$, where $\gamma>0$, it is found that the critical
dimension $d_c$ above which mean-field holds, at least qualitatively, is
given by
\begin{equation}
d_c=\frac{1}{1-\lambda-\gamma}
\label{eq:3.13.1}
\end{equation}
if the r.h.s. is positive, and is infinite otherwise.
\subsubsection{Scaling Theory for Inhomogeneous Systems}
Let us develop a scaling theory for (\ref{eq:3.4.1}). We limit ourselves
to the case in which the injection occurs at the origin and we transform
to spherical coordinates in order to exploit the existing symmetry.
As definition of the scaling limit we choose
\begin{equation}
\int_0^\infty d\mass\,\mass\int_0^\infty dr\,r^{d-1}c(\mass, r;t)f\left(
\frac{\mass}{s(t)},\frac{r}{L(t)}
\right)
\to\int_0^\infty dx\,x\Phi(x,y)f(x,y)
\label{eq:3.13.2}
\end{equation}
Here $s(t)$ denotes, as usual, a typical size and $L(t)$ a typical length.

We use the same
procedure as always and find for the scaling function $\Phi(x,r)$
\begin{equation}
I_1(\rho)=I_2(\rho)+I_3(\rho)+1,
\label{eq:3.13.3}
\end{equation}
where the various $I_j(\rho)$ are defined as
\begin{eqnarray}
I_1(\rho,\sigma)&=&\int_0^\infty
dx_1\,dx_2\,dr\,r^{d-1}K(x_1.x_2)x_1\Phi(x_1.r)
\Phi(x_2.r)\times\nonumber\\
&&\times e^{\rho x_1-\sigma r}\left(
1-e^{-\rho x_2}
\right)\nonumber\\
I_2(\rho,\sigma)&=&\sigma^2\int_0^\infty
dx\,dr\,r^{d-1}x^{1-\gamma}\Phi(x,r)e^{-\sigma r}
\left(
1-\frac{d-1}{\sigma r}
\right)
\label{eq:3.13.4}\\
I_3(\rho,\sigma)&=&\int_0^\infty dx\,dr\,r^{d-1}x\Phi(x)\left[
1+\omega-z\rho-\nu\sigma
\right]e^{-\rho x-\sigma r}.\nonumber
\end{eqnarray}
Here $z$ and $\nu$ stand for the exponents with which the typical size and
length grow with time respectively:
\begin{equation}
s(t)=const.\cdot t^z\qquad L(t)=const.\cdot t^\nu
\label{eq:3.13.5}
\end{equation}
and their values are found as in the usual case during the derivation of
(\ref{eq:3.13.3}). These are
\begin{equation}
z=\frac{\omega+2}{1-\lambda}\qquad\nu=\half(1-\gamma z).
\label{eq:3.13.6}
\end{equation}
The last relation is physically easy to inetrpret if one remarks that
it is equivalent to
\begin{equation}
L(t)\simeq\sqrt{
D\left[
s(t)
\right]t
}.
\label{eq:3.13.7}
\end{equation}
The fundamental length scale is therefore the one on which a typical
cluster diffuses in time $t$.

What can one deduce from (\ref{eq:3.13.3})? These are still
quite formidable equations, and it is not clear how to get results
out of them. The following approach yields some results: consider an
arbitrary exponent $\alpha>0$ and look at the curve defined by
$\sigma=\rho^\alpha$, where both $\rho$ and $\sigma$ go to infinity.
If we further assume that
\begin{equation}
\lim_{\sigma/\rho^\alpha=s}\Phi\left(
\frac{x}{\rho},\frac{r}{\sigma}
\right)=\phi(s)\rho^{\tau(\alpha)},
\label{eq:3.13.8}
\end{equation}
we find straightforwardly the following orders of magnitude at large $\rho$
and $\sigma$ for the $I_j(\rho)$:
\begin{eqnarray}
I_1(\rho, \sigma)&=&\rho^{-\lambda-3+2\tau(\alpha)-d\alpha}\nonumber\\
I_"(\rho, \sigma)&=&\rho^{(2-d)\alpha-2+\gamma+\tau(\alpha)}
\label{eq:3.13.9}\\
I_3(\rho, \sigma)&=&\rho^{-2-d\alpha+\tau(\alpha)}\nonumber
\end{eqnarray}
Matching these and taking also the constant term into account one
eventually finds the following relation between $\tau(\alpha)$ and
$\alpha$: first, if $\alpha>-\gamma/2$, then we define
\begin{equation}
\alpha_c=\frac{1-2\gamma-\lambda}{4-d}
\label{eq:3.13.10}
\end{equation}
Note that we shall always limit ourselves to the case $d<4$, since
otherwise it follows from well-known results that the aggregation process
becomes irrelevant and the whole system reduces to a non-interacting
diffusion.

One then finds for $\tau(\alpha)$
\begin{equation}
\tau(\alpha)=\left\{
\begin{array}{ll}
\half(d\alpha+\lambda+3)&\qquad(\alpha<\alpha_c)\\
1+\lambda+2\alpha+\gamma&\qquad(\alpha>\alpha_c)
\end{array}
\right.
\label{eq:3.13.11}
\end{equation}
and similarly, if $\alpha<-\gamma/2$, we find a different value of
$\alpha_c$:
\begin{equation}
\alpha_c=\frac{1-\lambda}{d}
\label{eq:3.13.12}
\end{equation}
and the values of $\tau(\alpha)$ are given by
\begin{equation}
\tau(\alpha)=\left\{
\begin{array}{ll}
\lambda+1&\qquad(\alpha<\alpha_c)\\
\half(d\alpha+\lambda+3)&\qquad(\alpha>\alpha_c)
\end{array}
\right.
\label{eq:3.13.13}
\end{equation}

What do these results mean? A dependence $\tau(\alpha)$ suggests a singular
behaviour at the origin of the type
\begin{equation}
\phi(x,r)\simeq \exp\left[
-\ln x\cdot\tau\left(
\frac{\ln r}{\ln x}
\right)
\right].
\label{eq:3.13.14}
\end{equation}
For the sake of comparison with
known results, consider the case of the constamt kernel with constant
diffusion. We are therefore always in the first case, since $\alpha>0$. The
singularity structure predicted by the theroy is therefore
\begin{equation}
\Phi(x,r)\simeq x^{-3/2}r^{-d/2}+x^{-1}r^{-2}
\label{eq:3.13.15}
\end{equation}
and the critical line which separates the regions in which one or the
other summand dominates is given by the relation $r=x^{1/(4-d)}$. The exact
connection of these results with the related ones published in \cite{che89}
is not entirely clear to me, but it is certainly
very close. There it is claimed
that in the stationary case, the scaling variable is $\mass/r^{4-d}$ and
that the $\tau$ exponent is given by $(6-d)/(4-d)$, which is exactly the
value of $\tau(\alpha_c)$ in our formalism.
\subsubsection{Higher Order Reactions}
Let us now consider the case in which reactions of higher order than
two may also be present. The reaction term of order $\order$ is given
by
\begin{eqnarray}
&&\frac{1}{\order!}
\int d\mass_1\cdot\ldots\cdot d\mass_\order K_{\order}
(\mass_{1},\ldots,\mass_\order)
c(\mass_1,t)\cdot\ldots\cdot c(\mass_r,t)\times\nonumber\\
&&\qquad\times\left[
\delta(\mass_1+\ldots+\mass_\order-\mass)-\sum_{\order^\prime}
\delta(\mass_{\order^\prime}-\mass)
\right].
\label{eq:3.1.1}
\end{eqnarray}
Denoting (in this subsection only) the degree of homogeneity of the
kernel $K_{\order}$ by $\lambda_{\order}$, we obtain by a formal
reasoning entirely analogous to that used in Appendix
\ref{app:deriv-scaling} the following conditions on the typical
size $s(t)$:
\begin{equation}
\dot s(t)=\constsep^{\order-1}s(t)^{\lambda_\order-\order+2}
\label{eq:3.1.2}
\end{equation}
At this stage, however, I must emphasize that this derivation is not
rigorous for reactions of order $r$, when $r\geq3$,
for reasons explained in detail in
Appendix \ref{app:deriv-scaling-3}. Let us nevertheless assume it for the
time being.

If therefore two different orders $\order_1$ and $\order_{2}$ are
relevant in the same physical process, we might expect that
one will invariably dominate the other in the scaling limit,
unless
\begin{equation}
\lambda_{\order_1}-\lambda_{\order_2}=\order_1-\order_2.
\label{eq:3.1.3}
\end{equation}
As we shall see later, however, the situation is more complex. The
crucial issue is (\ref{eq:3.1.2}).
As stated above and set out in detail in Appendix
\ref{app:deriv-scaling-3}, this derivation  is subject to certain
objections which do not arise in the two-body case.
It is perfectly possible for a scaling limit to
exist, with a well-defined scaling function $\Phi(x)$ satisfying some
form of an equivalent scaling equation, and yet for (\ref{eq:3.1.2})
not to hold. Under such circumstances, which we will discover in
certain exactly solvable models, a more complicated situation
arises, which we shall not discuss in the general framework of scaling
theory because of its complexity. We shall fully
discuss the exactly solvable case, however, in subsection
\ref{subsec:3body}.

Let us now look more carefully at the pure case of three-body
reactions. Let us first describe the various
exponents describing a three-body kernel. We define $\lambda_3$,
$\mu_1$ and $\mu_2$ as follows
\begin{equation}
\begin{array}{ll}
K(x,y,z)=z^{\lambda_3}k_2(x/z,y/z)&\\
k_2(y,z)=z^{\lambda_2}k_1(y/z)&\qquad(y,z\ll1)\\
k_2(y,z)=\pref_1(y)z^{\mu}&\qquad(z\ll1).
\end{array}
\label{eq:3.1.4}
\end{equation}
Note that, since the arguments of $K(x,y,z)$ can always be written in
ascending order, it is enough to specify $k_2(x,y)$ on the triangle
$0\leq x\leq y\leq1$. Conversely, $k_2(x,y)$ can be chosen
arbitrarily on this domain, so that no limitations
on the values of the above exponents hold. From this
follows that (\ref{eq:3.1.4}) gives all the
relevant information concerning the possible limiting
behaviours of $K(x,y,z)$. In particular, one readily
obtains the following limiting behaviour for $\pref_1(x)$:
\begin{equation}
\pref_1(x)=const.\cdot x^{\lambda_2-\mu}\qquad(x\to0)
\label{eq:3.1.4.1}
\end{equation}
Note that (\ref{eq:3.1.4}) is not the most general behaviour
imaginable: it would be
possible, for example, to define $k_2(y,z)$ in such a way that its
decay exponent $\mu$ as $z\to0$ depended on the value of $y$.
We disregard such cases and leave their treatment to the interested
reader.

We now proceed quite similarly to the case of binary reactions and
obtain an equation for the scaling function $\Phi(x)$ of the
following form
\begin{equation}
I_3(\rho)=\rho\int_0^\infty x^2\Phi(x)e^{-\rho x}dx,
\label{eq:3.1.5}
\end{equation}
where $I_3(\rho)$ is given by
\begin{equation}
I_3(\rho)=\half\int dx\,dy\,dz\,K_3(x,y,x)\Phi(x)\Phi(y)\Phi(z)
xe^{-\rho x}\left[
1-e^{-\rho(y+z)}
\right]
\label{eq:3.1.6}
\end{equation}
A fundamental difference between this case and
the two-body reaction
case should again be observed: in the derivation of the scaling
equation for $\Phi(x)$ in the two-body case, we saw in appendix
\ref{app:deriv-scaling} that the behaviour of the small aggregates
does not couple to that of the aggregates in the scaling regime: in
other words, we did not need to invoke anything beyond scaling to
justify the scaling equation (\ref{eq:3.10}). On the other hand, in
the case of (\ref{eq:3.1.6}), a problem may arise from the three-body
reactions involving two large particles and one small one.
The technical issue involved is discussed in detail
in Appendix \ref{app:deriv-scaling-3} for the three body case.
As we shall see in exactly solvable models,
such cases actually occur and signal the
necessity of choosing a different value for the growth exponent $z$
than the one predicted by (\ref{eq:3.1.2}). This is a qualitatively new
phenomenon, which does not occur in the two-body case.

The large-$\rho$ behaviour can again be estimated as in appendix
\ref{app:large-rho}. This leads to
\begin{eqnarray}
I_3(\rho)&=&\const_1\rho^{-(\lambda_3+4)}\Phi(1/\rho)^3
+\const_2\rho^{-(\lambda_2+3)}\Phi(1/\rho)^2+\nonumber\\
&&\qquad+\const_3\rho^{-2-\mu}\Phi(1/\rho).
\label{eq:3.1.7}
\end{eqnarray}
Here $\const_i$ are three integrals given in appendix
\ref{app:large-rho3}. One now proceeds as in the binary reaction
case. The following cases arise, excepting always the possibility
that a divergence at the origin occurs leading to a different value of
the exponent $z$:
\begin{enumerate}
\item Case I: $\mu>0$ and $\lambda_3-2\lambda_2<0$: This essentially
corresponds to Case I in the binary situation. One finds
$\tau=1+\lambda_3/2$ and the corresponding value of $1/2$ for the
decay exponent $w$ for the small aggregates, which is the lowest
possible value for $w$ in the purely ternary case. For this to be
consistent we additionally need the condition $\lambda_3<2\mu$. If
this is violated, we have a divergence at the lower end of the
integral and a different value for the exponent $z$.

\item Case II: $\mu>0$ and $\lambda_3-2\lambda_2>0$: This case is new.
One finds $\tau=1+\lambda_2$, and correspondingly
\begin{equation}
w=\frac{1-\lambda_2}{2-\lambda_3}
\label{eq:3.1.8}
\end{equation}
In this case, the second term in (\ref{eq:b.4}) dominates. Again, for
consistency we need $\lambda_2<\mu$, otherwise we again have
divergence at the lower end of the integral and a modified value of
the exponent $z$.
\item Case III: $\mu=0$: This case corresponds to Case II for
binary reactions. In this case, $\tau$ may take any value within a
certain range: in order for there to be no divergence at the origin,
we require $\tau<1$, and on the other hand, in order for the third
term in (\ref{eq:b.4}) to dominate one needs
\begin{equation}
\tau<\min(1+\lambda_3/2,1+\lambda_2).
\label{eq:3.1.9}
\end{equation}

\item Case IV: $\mu<0$: In this case, we cannot have a finite value
of $\tau$. If we have normal behaviour, then the function $\Phi(x)$
vanishes at the origin faster than any power law.
\end{enumerate}
We shall find the above theory indeed to be violated in the following
exactly solved cases:
\begin{eqnarray}
K_1(\mass_1,\mass_2,\mass_3)&=&\mass_1+\mass_2+\mass_3
\label{eq:3.1.10}\\
K_2(\mass_1,\mass_2,\mass_3)&=&\mass_1\mass_2+\mass_2\mass_3+\mass_1\mass_3
\label{eq:3.1.11}
\end{eqnarray}
$K_1$ has $\lambda_3$ equal to 1 and $\lambda_2$
and $\mu$ equal to zero, but the $z$ exponent is equal to 2 and
$\tau=3/2$\footnote{As if we were in fact dealing with Case I. Is this more
than a coincidence? I make no claim to know}. For $K_2$
we have $\lambda_3$ equal to 2, $\lambda_2$ equal to one and $\mu$
equal to zero. In this case we find gelation with a divergence
exponent $z$ of $-2$ and $\tau=2.5$, that is, behaviour quite similar
to the product kernel in the binary case.  These results will be shown in
detail in subsection \ref{subsec:3body}.
\subsubsection{Multicomponent Aggregation}
An interesting generalization of the aggregation model consists in
allowing the mass parameter $\mass$ to become a vector \cite{lus76,kra96a}.
The meaning of
the components could be, for instance, the concentration of various
species present in the aggregates. Similarly, we can think of one
species as being charged and the other neutral, so that the reactivity
of an aggregate depends not only on its total mass, but on the total
number of monomers of the charged species. The reaction rates
$K(\vmass_1,\vmass_2)$ then depend on two additive parameters and are
therefore an instance of the possibility of multicomponent aggregation.

Apart from its relevance in realistic aplications
(for an interesting use of related concepts concerning the coagulation of
charged aggregates, see for example
\cite{ivl02}),
the real interest in studying this case comes from the possibility of
looking at a different kind of scaling behaviour: Indeed, so far, we
have looked only at the average size of the aggregates. Here, as I
shall show shortly, under quite  general circumstances the limiting
distribution consists of aggregates which have a fixed composition
determined by the masses of the various species initially present. A
natural question then follows: on what scale does the distribution of
compositions vary? It is less than the typical size, so we obtain a
scaling in which two different sizes play an important role. In fact,
we show that this distribution is quite generally a L\'evy stable
distribution\footnote{possibly an ordinary Gaussian.}. Further, it can
be shown that the index of the L\'evy distribution is given by the ratio of
the exponents characterizing the growth of composition fluctuations to the
exponent $z$ for the growth of the typical cluster size.
These results are
in fact to be expected: indeed, we may consider the following
stochastic model for aggregation. Take a finite but large set of
aggregates of various compositions. Then choose two at random
according to the rates $K(\vmass_1, \vmass_2)$ and join them. Let
this process be repeated indefinitely. It is then quite likely that
an appropriate generalization of
the law of large numbers will apply, so that all aggregates tend
to a fixed composition. If one then wishes to analyze further on what
scale the composition varies, one needs some version of the
central limit theorem. (Similar random coagulation processes have been
looked into in \cite{nor00}, but not, it seems to me, to the extent of
proving such claims.)
Such a theorem would state that the resulting distribution is
always L\'evy and that
the typical size of the fluctuations is connected to the index of the
L\'evy law involved \cite{bouchaud}. Note that the constant kernel
corrresponding to such a case was already studied in \cite{lus76,kra96a}.
Defining
\begin{equation}
|\vec\mass|=\sum_k \mass_k
\label{eq:3.2.05}
\end{equation}
the scaling assumption now reads
\begin{equation}
\lim_{t\to\infty}\int_0^\infty d\vmass\,|\vmass|c(\vmass,t)f\left(
\frac{\vmass}{s(t)}\right)=\int_0^\infty d\vec x\,|\vec x|\Phi(\vec x)
f(\vec x).
\label{eq:3.2.1}
\end{equation}
Using exactly the same techniques as in the one component case, one
derives the following equivalent of the scaling equation (\ref{eq:3.10}):
\begin{eqnarray}
\int d\vec x\,\vec x\cdot\vec\nabla f(\vec x)\, \vec x&=&\int d\vec x_1
d\vec x_2K(\vec x_1, \vec x_2)\Phi(\vec x_1)\Phi(\vec x_2)\times\nonumber\\
&&\qquad\times\left[
f(\vec x_1+\vec x_2)-f(\vec x_1)
\right]
\label{eq:3.2.2}
\end{eqnarray}
In order to simplify notation, I limit myself to the case of two
components, though the general case presents no difficulties of
principle. A moment's thought shows that the following is a solution
of (\ref{eq:3.2.2}):
\begin{equation}
\Phi(\vec x)=\Phi_1(x_1)\delta(x_1-x_2),
\label{eq:3.2.3}
\end{equation}
where $x_i$ denotes the $i$-th component of $\vec x$ and $\Phi_1(x)$
is the scaling function corresponding to the rates
$K(x_1, x_1;x_2, x_2)$. There exist
further solutions differing from (\ref{eq:3.2.3}) by trivial
normalizations, but since we can always fix units so as to have the
same mass of both components initially present, this is irrelevant. I
do not know whether further solutions exist, but the plausible
considerations made at the beginning of this subsection concerning the
law of large numbers appear to speak against such a possibility.

Let us now look at the variations in composition and the way in which
they scale. To fix ideas, we start from a system in which the mass
initially contained in both components is the same, and we assume that
the scaling function given in (\ref{eq:3.2.3}) is indeed the relevant
one. This permits us to neglect the dependence of the reaction rates on
composition, since a fixed composition is reached at the scale
defined buy the typical size $s(t)$. .
We then define a scaling function $\Psi(\suma, \difer)$ for the
sum and difference between $x_1$ and $x_2$. We define it
as follows:
\begin{eqnarray}
&&\lim_{t\to\infty}\int_0^\infty d\mass_1d\mass_2(\mass_1-\mass_2)
c(\vec\mass, t)f\left(
\frac{\mass_1+\mass_2}{s_1(t)},\frac{\mass_1-\mass_2}{s_2(t)}
\right)=\nonumber\\
&&\qquad\int_0^\infty d\suma\int_{-\infty}^\infty d\difer\,\difer\,\Psi(\suma,
\difer)f(\suma, \difer).
\label{eq:3.2.4}
\end{eqnarray}
Here $s_1(t)$ corresponds to the typical size of aggregates, whereas
$s_2(t)$ gives the scale for the variations in composition.
In Appendix \ref{app:multi}. we derive a scaling equation (\ref{eq:k.1})
for $\Psi(\suma, \difer)$ and show that
\begin{equation}
\Psi(\suma, \difer)=\suma^{-\alpha}\Phi_1(\suma)\chi_\alpha\left(
\frac{\difer}{\suma^\alpha}.
\right)
\label{eq:3.2.5}
\end{equation}
is a solution. Here $\Phi_1(x)$ is the one-component scaling function
of (\ref{eq:3.2.3}), $\chi_\alpha(x)$ is the symmetric L\'evy stable
distribution of index $1/\alpha$ given by the Fourier transform of
$\exp(-|q|^{1/\alpha})$ with respect to $q$ and $\alpha$ is given by
\begin{equation}
\alpha=\lim_{t\to\infty}\frac{\dot s_2(t)\,s_1(t)}{\dot
s_1(t)\,s_2(t)}.
\label{eq:3.2.6}
\end{equation}
In order to determine $\alpha$ unambiguously, we argue as follows:
All L\'evy distributions except for the Gaussian have power-law tails.
Further, if $\alpha\geq1$, it is known that the first moment of the
distribution diverges. This can certainly not happen if, as we have
always supposed, we start from initial conditions having finite mass.
It is also known that the $\chi_{\alpha}$ are only positive if
$\alpha\geq1/2$. We therefore have $1/2\leq\alpha<1$. The value $1/2$
corresponds to a gaussian distribution, whereas all other allowable
values correspond to a power-law decay of $x^{-1-1/\alpha}$.
Thus a value of $\alpha$ different from $1/2$ implies that the
differences in composition have power-law tails extending to infinity
in the large time limit. Since no mechanisms are known to produce these
unless they be present in the initial conditions, we may state
that if the initial conditions do not have power-law tails, the
final distribution must be Gaussian%
\footnote{This is maybe a little less obvious than it seems: as we shall
see later in detail, the large $x$ asymptotics in the scaling regime has
no simple relationship to the large $\mass$ asymptotics at fixed times.
It is, however, a very reasonable assumption.}.
If the initial distribution did
contain a power-law tail, then it is quite likely that the
corresponding L\'evy stable law will be the relevant one in the
scaling regime. We therefore see that initial conditions are relevant
in this case to determine the asymptotic behaviour of the typical
size of the variations in the aggregate composition. We shall later
find similar behaviour in ballistic aggregation, which is, in fact, a
closely related problem.
\subsection{Non-scaling Asymptotics}
\label{subsec:non-scal}
The above is a subject which, from the peculiarly biased viewpoint
I have been taking in this paper, will naturally receive rather short
shrift. Nevertheless, it is rather obvious that the following questions
are relevant:
\begin{enumerate}
\item How does $c(\mass,t)$ behave at large times for fixed $\mass$?

\item Conversely, how does $c(\mass, t)$ behave for large $\mass$
at fixed $t$? Obviously, the decay should, in some sense be
exponential, or at least fast enough, but more detailed
information concerning power-law corrections to the exponential decay
is also important.

\item What is the intermediate asymptotics of the $c(\mass,t)$?
By this I mean the following: frequently, there exists a time dependence
$f(t)$ such that
\begin{equation}
c(\mass, t)=g(\mass)f(t)[1+o(1)]\qquad(t\to\infty)
\label{eq:3.6.1}
\end{equation}
for {\em fixed\/} $\mass$. In this case, one may ask what is
the large $\mass$ behaviour of $g(\mass)$. In the absence of such a function
$f(t)$, the whole issue becomes rather ambiguous.

\item  Similarly, the problem of the large time behaviour of
the moments of the cluster-size distribution is of considerable
physical relevance, since these are usually its most easily
accessible characteristics. The problem splits into two
disjoint issues.

First, the large time behaviour of moments
$\mom(t)$ of order $\indmom$ such that
\begin{equation}
\int_0^\infty x^\indmom \Phi(x)=\infty,
\label{eq:3.6.2}
\end{equation}
falls outside the purview of standard scaling theory.
The problem, of course. lies in the fact
that these low-order moments are dominated
by the low-$\mass$ part of the cluster-size distribution, whereas
the higher-order moments are determined by the distribution
as a whole.

On the other hand, high-order moments may also cause difficulties, though
perhaps of  a more trivial kind: indeed, if the initial condition
contains a power-law contribution, then some sufficiently
high-order moment will diverge from the very beginning., which is
a clear sign that we may not blindly apply the scaling
approach in evaluating moments. This is seen, for example, in the
exactly solved model of the constant kernel, for which I show that
scaling is approached whenever the initial distribution has finite first
moment. Nevertheless, it is obvious that the second moment has
no well-defined large-time asymptotics if its initial value
is divergent. One therefore sees that stronger assumptions
than merely the existence of the scaling limit are required
in order to justify the usual scaling expressions even for high-order
moments.
\end{enumerate}
In the literature, such issues have often been (mistakenly) discussed as
so-called ``violations of scaling'' (see e.g. \cite{ley86,ley87} for
a characteristic example of such confusion. The issue of scaling in
parity dependent kernels will be taken up again later in this paper.)
In the following, we shall always refer to our standard definition of
scaling, and will only consider as non-scaling such systems as
do not approach a limiting form in the sense defined by (\ref{eq:3.7}).
In fact, as we shall see, all of the quantities described above
may exhibit unexpected behaviour while the scaling hypothesis
remains fulfilled. We shall, in our terminology, call such deviations
from expected behaviour, violations of strong scaling. Since I have never
been too explicit about what is meant by strong scaling, it is clear
enough that anything peculiar enough to attract some attention might be
brought under this heading. In the following, I shall largely follow
the approach sketched in \cite{ley86a}\footnote{The reader should be warned
that this paper thoroughly
confuses the two concepts of scaling regime and non-scaling asymptotics.}.

Let us first turn to the large time behaviour of $c(\mass, t)$ at fixed
$\mass$. In this case, production of clusters of mass $\mass$ can be
neglected, and I may approximate (\ref{eq:2.2}) by
\begin{equation}
\dot c(\mass, t)=-c(\mass, t)\int_0^\infty
K(\mass, \massp)c(\massp, t)d\massp.
\label{eq:3.6.3}
\end{equation}
Using the definitions of $\mu$ and $\nu$ given in (\ref{eq:3.115})
and (\ref{eq:3.120}) as well as the fact that the relevant masses $\massp$
in (\ref{eq:3.6.3}) are much larger than $\mass$, one eventually obtains
the following approximate differential equation for $c(\mass, t)$
\begin{equation}
\dot c(\mass, t)=const.\cdot \mass^\mu\mom[\nu](t).
\label{eq:3.6.4}
\end{equation}
In order to understand the behaviour of this system, we must first know the
behaviour of $\mom[\nu](t)$. Assume strong scaling, after which one may
check for self-consistency. In this case, one finds for kernels of type III
\begin{equation}
\mom[\nu](t)=const.\cdot t^{(\nu-1)/(1-\lambda)},
\label{eq:3.6.5}
\end{equation}
from which follows that $c(\mass, t)$ decays as a stretched exponential of
the form
\begin{equation}
c(\mass, t)=const.\cdot\exp\left[-const.\cdot\mass^\mu t^{-\mu/(1-\lambda)}
\right]
\label{eq:3.6.6}
\end{equation}
Note that this is in perfect agreement with the predictions of
strong scaling, so that no violations are to be expected in Case III. Of
course, this might not be viewed as quite conclusive, since we need a
hypothesis to determine the large-time behaviour of $\mom[\nu](t)$. As we
shall see, however, the main hypothesis is that the expression of the
moment as an integral of the scaling function should have no singularities
at the origin, which is trivially satisfied in Case III. Another source of
problems might arise from power.law initial conditions, but again, as long
as $\nu<1$ no divergence is possible.

In case I, the above approach fails, since it is readily seen that
$\mom[\nu](t)$ cannot be described by a convergent integral of the scaling
function. The following observation, on the other hand, yields interesting
results \cite{lus73}: if one introduces the following
new variables in the discrete
representation, which turns out here to be more convenient,
\begin{eqnarray}
\phi_j(\tres)&=&c_j(t)/c_1(t)\nonumber\\
d\tres&=&c_1(t)dt,
\label{eq:3.6.7}
\end{eqnarray}
the original equations (\ref{eq:2.2}) become
\begin{equation}
\frac{d\phi_j}{d\tres}=\half\sum_{k=1}^{j-1}K(k,j-k)\phi_k\phi_{j-k}-
\phi_j\sum_{k=1}^\infty\left[
K(k,l)-K(1,l)\right]\phi_l.
\label{eq:3.6.8}
\end{equation}
These equations have a non-trivial equilibrium solution $a_j$ which can
sometimes be
determined recursively. It can be worked out for kernels of the form
$K(j,k)=(jk)^{\lambda/2}$,(see \cite{ley84}), as well as more generally
\cite{don85a} that these
constants behave as $j^{-\tau}$ with $\tau=1+\lambda$, which is therefore
in good agreement with the qualitative behaviour predicted by
strong scaling. This also settles the issue of intermediate
asymptotics as defined above, showing that it indeed exists, and that
it coincides with the asymptotic behaviour of small clusters in the
scaling limit.

One further finds using (\ref{eq:2.2}) that $c_1(t)$ goes as
$t^{-1}$, from which follows that all $c_j(t)$ do.
This demonstration has been put on a firmer footing by a
rigorous inductive proof by van Dongen and Ernst \cite{don85a}.
On the other hand, there remain discrepancies between the behaviour of the
$a_j$ at large $j$ and that of the scaling function $\Phi(x)$ at small $x$:
in particular, at least under the assumption of regular behaviour for
$\Phi(x)$, the prefactors of both differ.

Finally, for case II, no such arguments are available. In particular,
neither an approach based on (\ref{eq:3.6.4}) nor one based on recursion
relations can be justified. Further, a large number of counterexamples
are known, which precisely belong to this case. In fact, both
non-gelling exactly
solved classical models, the constant and the sum kernels, belong to case
II, as also do some variations on the constant kernel for which the
strangest deviations from strong scaling have been reported.

It remains to discuss shortly the issue of the large $\mass$ behaviour at
fixed time \cite{ern84a}. In this case one may safely neglect the reaction
of particles of mass $\mass$ with larger aggregates, so that we have the
following approximate equations, for the discrete case:
\begin{equation}
\dot c_j=\half\sum_{k=1}^{j-1}K(k,j-k)c_kc_{j-k}.
\label{eq:3.6.9}
\end{equation}
Again, for monodisperse initial conditions, this can be solved in a
straightforward way by recursion, using the {\em ansatz}:
\begin{equation}
c_j(t)=a_jt^{j-1}.
\label{eq:3.6.10}
\end{equation}
One
finds (for details see \cite{ern84a}) that $a_j$ goes as $j^{-\lambda}$
for all kernels $K(k,l)$ with $\nu<1$. A comparison with the results obtained
for the large $x$ end of $\Phi(x)$  shows that a difference between the
two can, and in fact does occur for the case $\nu=1$ \cite{don87c}.
It therefore follows from these
results that the large mass end of the cluster
size distribution at fixed times {\em always} decays exponentially, a fact
which we shall find to be of some importance in assesssing the possibility
to describe certain models by mean-field equations.

Let us finally consider the scaling prediction for the moments. Overall, the
strong scaling prediction for the asymptotic behaviour of $\mom(t)$ is
\begin{equation}
\mom(t)=\scalmom s(t)^{\indmom-1}\qquad(t\to\infty)
\label{eq:3.6.11}
\end{equation}
The immediate case in which this cannot hold is, of course, for values
of $\indmom$ such that $\scalmom$ does not exist (diverges). Clearly, this
can only occur when $\Phi(x)$ has a power-law  singularity at the origin,
since the exponential decay of $\Phi(x)$ at infinity shown in subsection
\ref{subsubsec:large} precludes any divergence there. Thus there are no
problems in Case III, but whenever $\tau$ is finite, the asymptotic
behaviour (\ref{eq:3.6.11}) is invalid in the range
\begin{equation}
\indmom\leq-1+\tau.
\label{eq:3.6.12}
\end{equation}
However, as we shall see later in exactly solved examples, this behaviour
can also fail {\em outside} the range stated in (\ref{eq:3.6.12}). Indeed,
in various kernels of type II one finds that for large $t$ at fixed $\mass$
\begin{equation}
c(\mass,t)=t^{-\wpr(\mass)}[1+o(1)]\qquad(t\to\infty).
\label{eq:3.6.13}
\end{equation}
Here $\wpr(\mass)$ varies continuously with $\mass$ and bears no necessary
relation to either $z$ or $\tau$. In this case, it may well happen that
$(p-1)z$ be less than $\max_\mass\wpr(\mass)$, so that the decay of
$\mom(t)$ is dominated by that of a specific type of aggregate. We shall
in fact see such examples in the next section. This definitely cannot
happen, however, for moments which grow in time, that is whenever
$\indmom>1$.

An altogether different violation of the behaviour stated in
(\ref{eq:3.6.11}) occurs whenever the initial conditions have a power-law
tail. In this case, as one sees in many exactly solved cases, the large
mass behaviour always maintains this tail, which cannot be eliminated by
any aggregation process except gelation. In this case, clearly,
whenever $\indmom$ is large enough to cause the divergence of $\mom(0)$,
the moment $\mom(t)$ is identically infinite, so that (\ref{eq:3.6.11})
becomes meaningless. Nevertheless, scaling often holds also under these
conditions. In such cases, we must reinterpret (\ref{eq:3.6.11}) to mean
\begin{equation}
\int_0^{\Lambda s(t)}\mass^\indmom c(\mass,t)d\mass=\left[
s(t)
\right]^{\indmom-1}\int_0^\Lambda x^\indmom\Phi(x).
\label{eq:3.6.14}
\end{equation}
where $\Lambda$ is an arbitrary constant. The time necessary for
(\ref{eq:3.6.14}) to become valid goes to infinity as $\Lambda$ does.
\subsection{The Moment Equations}
\label{subsec:mom-eq}
The moments $\mom(t)$ of the cluster size distribution function satisfy
some general exact equations, which are in many cases of considerable use,
though we shall not make much explicit use of them in this paper. Indeed,
from (\ref{eq:2.2}) follows immediately
\begin{eqnarray}
\frac{d\mom}{dt}&=&\int_0^\infty d\mass_1\,d\mass_2\,
K(\mass_1,\mass_2)c(\mass_1,t)
c(\mass_2,t)\times\nonumber\\&&\qquad\times\left[
(\mass_1+\mass_2)^\indmom-\mass_1^\indmom
-\mass_2^\indmom
\right]
\label{eq:3.10.1}
\end{eqnarray}
For integer values of $\indmom$, these take a particularly striking
form:
\begin{equation}
\frac{d\mom}{dt}=\sum_{k=1}^{p-1}
\left(
\begin{array}{c}
\indmom\\
k
\end{array}
\right)\int_0^\infty d\mass_1\,d\mass_2\,
\mass_1^k\mass_2^{\indmom-k}K(\mass_1,\mass_2)c(\mass_1,t)
c(\mass_2,t).
\label{eq:3.10.2}
\end{equation}
The two principal uses of these relations are the following: first, in the
case of the bilinear kernel (\ref{eq:4.5.1}), they yield closed equations,
which can be solved recursively. Their solution for low values of
$\indmom$ is one of the quickest ways to obtain qualitative information
concerning the solution. In particular, it readily yields the exponent $z$
as well the gel time in gelling systems.

On the other hand, they can also be used for estimating the solution from
above or below. Thus White's proof of global existence for the solutions
of (\ref{eq:2.2}), see \cite{whi1}, rests essentially on such an estimate
which follows from the estimate
\begin{equation}
K(\mass, \massp)\leq C(\mass+\massp)
\label{eq:3.10.3}
\end{equation}
The possibility of such estimates is a result of the positivity of the
right-hand side of (\ref{eq:3.10.2}). One then obtains
a closed set of differential inequalities, which can then,
because of their recursive character, be solved to yield upper bounds on
the $c(\mass, t)$. Similarly, we used inequalities of the same nature in
subsection \ref{subsubsec:large} to estimate the high moments of the
scaling function $\Phi(x)$. Note carefully that the moment equations
(\ref{eq:3.10.2}) are not at all equivalent to the corresponding
equations (\ref{eq:3.1302}) for the moments of $\Phi(x)$: the latter only
describe the moments if the system approaches scaling, whereas the former
hold in every case. In fact, the knowledge of all moments is equivalent, in
principle, to knowing the full cluster size distribution, so that the
solution of (\ref{eq:3.10.2}) is in principle of the same degree of
difficulty as the full solution of the original Smoluchowski equations
(\ref{eq:2.2}). The solution of (\ref{eq:3.10.2}) therefore also contains
indications of all corrections to scaling and non-scaling asymptotics,
which, as we have seen, often differ considerably from the scaling
behaviour. These two equations cannot, therefore, be identified.
Nevertheless, it is certainly true that the leading behaviour of $\mom(t)$
for $\indmom$ sufficiently large
only depends on the scaling function $\Phi(x)$. It might
therefore be possible to prove the scaling hypothesis by a careful study of
the leading terms for the large-time behaviour of (\ref{eq:3.10.2}).

\section{Exactly Solved Models}
\setcounter{equation}0
\label{sec:exact}
In the following, I shall discuss a large variety of exactly solved models.
One can justify interest in such cases in several ways: they are in
themselves a source of pleasure, they are an ideal test ground
for numerical work and they allow us better to understand the structure of
solutions in general. While I sympathize to some extent with all of these
motives, in this paper I shall be primarily concerned, as I have been
throughout, with the issue of the validity of scaling theory. Exact models
are, of course, invaluable in this respect, since we have ordinarily a
complete overview of the solution«s behaviour. It is therefore
straightforward to say in which way the system does ineed satisfy the
scaling hypothesis, and to what extent it may occasionally deviate from
what is expected. As we shall see, the scaling hypothesis as I have
defined it, see (\ref{eq:3.7}), holds essentially for all systems we shall
have occasion to look at\footnote{The solitary exception I have in mind is
the crossover between the sum kernel and the product kernel, for which I
have not been able to find a scaling description. However, this may still
be a soluble problem. In any case, this only concerns a crossover
phenomenon, for true scaling behaviour I am not aware of {\em any}
counterexamples}. On the other hand, as I shall show in an extensive set of
examples, the set of assumptions I have denoted by the general name
of ``strong scaling'', are in general not fulfilled. To be more accurate,
it is frequently possible to find more or less natural exactly solved
cases for which these assumptions fail in several ways. The detailed
illustration of such failure I consider to be instructive, since it shows
with great detail exactly how much we may, and more importantly what we
may not, deduce from the existence of a scaling limit.

The exactly solved kernels fall in two categories. First, one might
mention the so-called ``classical'' kernels
\begin{eqnarray}
K_1(\mass, \massp)&=&1
\label{eq:4.0.1}\\
K_2(\mass, \massp)&=&\mass+\massp
\label{eq:4.0.2}\\
K_3(\mass, \massp)&=&\mass\massp,
\label{eq:4.0.3}
\end{eqnarray}
which are known as the constant, sum and product kernels respectively.
Their solution can be generalized to the case of an arbitrary linear
combination of the three, see (\ref{eq:4.5.1}), which is known as the
general bilinear kernel. The three kernels described in
(\ref{eq:4.0.1}-\ref{eq:4.0.3}) are all quite different from each other,
as should be expected from the general scaling theory exposed in the
previous section: the first has $\lambda=0$, the second has $\lambda=1$
and the third has $\lambda=2$ and is accordingly a gelling kernel. In the
first two cases, on the other hand, we have regular growth, proportional
to time in the first case, exponentially fast in the second. All these
predictions are indeed borne out by the exact solutions.

As for the scaling theory, these kernels in general show a reasonable
behaviour, that is, they usually satisfy scaling in an extremely
strong form. However, even in these cases, it can be shown that many
issues depend rather sensitively on initial conditions. We shall see, in
particular, that initial conditions with power-law tails at large masses
can significantly modify the non-scaling behaviour, even when the
scaling function is unaffected.

I shall also discuss some natural extensions of these kernels to ternary
reactions. This will show the way in which the scaling approach may
actually break down altogether if collisions involving particles outside
the scaling regime dominate the physics.

Further, since not only the three kernels above can be solved, but also the
general bilinear kernel (\ref{eq:4.5.1}), we can clearly study various types
of crossover. In fact, as I shall show, two of the three possibilities can
be analyzed rather extensively and one arrives eventually at a
crossover function as defined in subsection \ref{subsec:cross}.

Finally, there are non-classical kernels, such as the $q$-sum or the
parity-dependent kernels. These are characterized by the fact that they
are not exactly homogeneous. As we shall see, they display a considerable
variety of unexpected behaviours in their non-scaling asymptotics, but all
satisfy scaling in the sense specified in this paper. Furthermore, as we
shall see, even these satisfy the Dominant Singularity Hypothesis, that
is, the value for the small-size exponent $\tau$ is the same as that of the
large size exponent $\explarge$. Nevertheless, the array of non-scaling
anomalies shown in these cases is instructive, as it shows us that certain
features one takes for granted are, in fact, peculiar features of the
simpler exactly solved kernel and do not generalize to more general cases.
\subsection{The Constant Kernel}
\label{subsec:const}
The constant kernel defined by
\begin{equation}
K(\mass_1,\mass_2)=1
\label{eq:4.2.1}
\end{equation}
was first solved by von Smoluchowski \cite{smo16} and its solution
has been often rederived. The approach we follow is an
elementary, yet instructive, use of the generating function approach:
define
\begin{equation}
F(\lap,t)=\int_0^\infty c(\massp,t)e^{\lap\massp}d\massp.
\label{eq:4.2.2}
\end{equation}
Here the integral must be replaced by a sum whenever $c(\mass,t)$ is
a sum of delta functions. The fundamental equation (\ref{eq:2.2}) now
takes the form
\begin{equation}
F_t=\frac{1}{2}F(\lap,t)^2-F(\lap,t)F(0,t).
\label{eq:4.2.3}
\end{equation}
After making the replacement
\begin{equation}
G(\lap,t)=F(\lap,t)-F(0,t),
\label{eq:4.2.4}
\end{equation}
(\ref{eq:4.2.2}) becomes
\begin{eqnarray}
&&G_t=\frac{1}{2}G^2\nonumber\\
&&G(\lap,t=0)=g(\lap):=\int_0^\infty
c(\massp,0)\left(e^{\lap\massp}-1\right)d\massp.
\label{eq:4.2.5}
\end{eqnarray}
{}From this one readily obtains the following explicit form for
$G(\lap,t)$ in terms of $g(\lap)$, which in turn is explicitly
determined by the initial concentrations. The connection between the
actual concentrations at time $t$ and the initial values is, as in
many other cases, quite opaque. We shall see, however, that for many
relevant issues the specific nature of this connection is not needed.
This expression for $G(\lap,t)$ is
\begin{equation}
G(\lap,t)=\frac{2g(\lap)}{2-tg(\lap)}.
\label{eq:4.2.6}
\end{equation}
If, using the notations introduced in (\ref{eq:2.205}) and
(\ref{eq:2.210}), I wish to compute $c_j(t)$ for monodisperse initial
conditions, it is a straightforward excercise in power series
manipulation to show that
\begin{equation}
c_j(t)=\frac{4}{(t+2)^2}\left(\frac{t}{t+2}
\right)^{j-1}.
\label{eq:4.2.7}
\end{equation}
This is readily seen to approach a scaling form: if we namely set
$s(t)=t$, we obtain using the definition (\ref{eq:3.7}) that
(\ref{eq:4.2.7}) yields as scaling function
\begin{equation}
\Phi(x)=4e^{-2x}.
\label{eq:4.2.8}
\end{equation}
It therefore follows that $\tau=0$, $z=1$ and $w=2$. This manifestly
satisfies the scaling relation (\ref{eq:scaling-rel}) derived in
\cite{vic84}. To tidy up various details, we point out that (\ref{eq:4.2.8})
is indeed a solution of (\ref{eq:3.12}): all functions of the form
$ae^{-x/2}$ satisfy it for $K(x,y)=1$ and only (\ref{eq:4.2.8})
satisfies the normalization condition (\ref{eq:3.1305}). Furthermore,
setting $s(t)$ equal to $t$, as we did, corresponds because of
(\ref{eq:a.5}) to setting $\constsep$ equal to one. Everything
therefore confirms the scaling calculations down to the smallest
detail. Note that, in the following, we shall usually not verify the
normalization issues in great detail. Rather, we shall usually implicitly
assume $\constsep=1$, which involves a certain choice of time scale such
that $s(t)$ satisfies excatly $\dot s=s^\lambda$, with no further
prefactors.

Now we still need to show that this scaling form
holds for arbitrary initial conditions. In itself, this is
non-trivial: estimating the asymptotic behaviour of the
formal expressions for $c(\mass, t)$ for
arbitrary initial conditions is quite cumbersome, even in this most
elementary case.
We therefore rely on an observation made at the end of subsection
\ref{subsec:scal1}: namely, in order to show scaling, it suffices to
show convergence as defined by (\ref{eq:3.8}). In our case, it amounts
to showing that
\begin{equation}
\lim_{t\to\infty}\left.G_\lap(\lap,t)\right|_{z=\lapp/t}
\label{eq:4.2.9}
\end{equation}
exists. But this is quite straightforward using (\ref{eq:4.2.6}), if
we may assume that $g(\lap)$ has finite first derivative  at $\lap=0$,
or, what amounts to the same, that the $c(\mass, 0)$ have finite first
moment. Under this hypothesis, one finds
that the limit defined in (\ref{eq:4.2.9}) is given by
\begin{equation}
\frac{4}{(2-\lapp)^2}
\label{eq:4.2.10}
\end{equation}
which is, according to (\ref{eq:3.801}),
the Laplace transform of $x\Phi(x)$ evaluated at $-\lapp$.

This system provides various interesting
examples for several of the phenomena discussed in
Section \ref{sec:scaling}: Let us, for example, consider initial
conditions for which $c_j(0)$ decays as $\const j^{-\alpha}$, with
$2<\alpha<3$. In this case, the total mass is finite, but none of the
higher integer moments converge. $G_\lap(\lap,t)$ has therefore
two singularities in $\lap$ at large times $t$: one at, and
the other near, the origin. This last is the one giving rise to the
scaling contribution, since we have just shown that convergence to a
scaling form does take place under these circumstances. On the other
hand, the second and higher moments do not exist, due to the presence
of the former singularity. It is therefore rather
misleading to say that $c(\mass, t)$ is ``well approximated'' by
$t^{-2}e^{-j/t}$. A more correct statement is that at large times, one
has
\begin{equation}
c_j(t)\simeq t^{-\alpha}\Phi_1(j/t)+4t^{-2}e^{-j/(2t)},
\label{eq:4.2.11}
\end{equation}
where $\Phi_1(x)$ is a function which can, in principle be computed
explicitly. In fact, one has, using a similar approach
as in the derivation of (\ref{eq:3.801}):
\begin{equation}
\laplace[x\phi_1(x)]=\frac{4C\Gamma(1-\alpha)s^{\alpha-2}\left[
2-2\alpha+(3-\alpha)s
\right]}{
(2+s)^3},
\label{eq:4.2.1105}
\end{equation}
where $\laplace$ denotes the Laplace transform.
It behaves as $x^{2-\alpha}$ as $x\to0$, but becomes
asymptotically equal to $x^{-\alpha}$ as $x\to\infty$.
Since $\alpha>2$, it follows that the second term of(\ref{eq:4.2.11})
always dominates when $j/t$ is kept fixed and $t\to\infty$.
On the other hand, the first term
eventually dominates as $j\to\infty$ for fixed $t$. Choosing an
appropriate definition of $\taup$ and $\wpr$, however, one finds that
$\taup=\tau$ and $\wpr=w$, so that strong scaling still holds in this
particular case.

Quite generally speaking, one should not confuse the fact that the
distribution approaches a scaling limit, which is a global property of
the full distribution at large times, with good approximations for
individual values of $c_j(t)$. The exact solution for monodisperse
initial conditions could easily lead to wrong impressions on that
subject: indeed, for (\ref{eq:4.2.7}), one has
\begin{equation}
c_j(t)=t^{-2}\exp(j/t)\left[
1+O(jt^{-2})+O(t^{-1})
\right]
\label{eq:4.2.12}
\end{equation}
for all $j$ and $t$. This does not generalize to arbitrary initial
conditions, however. In particular, for fixed values of $j$
the constant
\begin{equation}
\cinf_j=\lim_{t\to\infty}\left[
t^2c_j(t)
\right]
\label{eq:4.2.13}
\end{equation}
is in general not equal to $4$, and depends on the initial
data, even when these decay exponentially, as is
readily verified in specific instances \cite{don88}. If
the initial condition does decay exponentially, however, then
$\cinf_{j}$ also approaches $4$ exponentially fast \cite{ley99}, as
a straightforward estimate of the contour integrals yielding $c_j(t)$
from (\ref{eq:4.2.6}) will show.
\subsection{The Sum Kernel}
\label{subsec:sum}
The sum kernel is given by
\begin{equation}
K(\mass, \massp)=\mass+\massp.
\label{eq:4.3.1}
\end{equation}
It has been solved exactly first by \cite{mel53,sco68}, see also
\cite{drake} for further references on the subject. I will
show two approaches, which can both be used to obtain the full
solution: the first involves a transformation, introduced by Lushnikov
\cite{lus73}, which can always be performed when the kernel is
of the form
\begin{equation}
K(\mass,\massp)=f(\mass)+f(\massp),
\label{eq:4.3.2}
\end{equation}
where $f$ is an arbitrary function. On the other hand, the usual
approach using generating functions works quite well also. The reason
for showing both approaches is, as stated in the Introduction, to
give a broad overview of the different ways to attack a given
problem.

Let us therefore first consider the discrete case with monodisperse
initial conditions. We may then define, for arbitrary kernels
of the form (\ref{eq:4.3.2}),
\begin{equation}
\phi_j(\tres)=\frac{c_j(t)}{\sum_{k=1}^\infty c_k(t)}\qquad
d\tres=dt\sum_{k=1}^\infty c_k(t),
\label{eq:4.3.3}
\end{equation}
where the $c_j(t)$ are defined from the $c(\mass, t)$ via
(\ref{eq:2.210}). One then sees that (\ref{eq:2.2}) becomes
\begin{equation}
\frac{d\phi_j}{d\tres}=\sum_{k=1}^{j-1}f(k)\phi_k\phi_{j-k}-f(j)\phi_j.
\label{eq:4.3.4}
\end{equation}
This set of equations is recursive and can sometimes be solved
explicitly. In the case at hand, when $f(j)=j$, the additional
substitution
\begin{equation}
\psi_j=\phi_je^{j\tres}
\label{eq:4.3.5}
\end{equation}
does the job, leading to the system of equations
\begin{eqnarray}
\frac{d\psi_j}{d\tres}&=&\sum_{k=1}^{j-1}k\psi_k\psi_{j-k}
\label{eq:4.3.6a}\\
\psi_j(0)&=&\delta_{j,1}
\label{eq:4.3.6b}
\end{eqnarray}
These are solved via the {\it ansatz}:
\begin{equation}
\psi_j(\tres)=a_j\tres^{j-1}.
\label{eq:4.3.7}
\end{equation}
The coefficients $a_j$ are now determined by the following recurrence
\begin{eqnarray}
(j-1)a_j&=&\sum_{k=1}^{j-1}ka_ka_{j-k}\nonumber\\
a_1&=&1.
\label{eq:4.3.8}
\end{eqnarray}
This recursion is readily solved, for example using generating
functions, see Appendix \ref{app:fin-sum}, and yields finally
\begin{equation}
a_j=\frac{j^{j-2}}{(j-1)!}.
\label{eq:4.3.9}
\end{equation}
The sum over all concentrations obeys a closed equation (if we assume
that mass is conserved for all times, which will turn out to be
justified here). This leads to
\begin{equation}
\sum_{k=1}^\infty c_k(t)=\sinit e^{-t}\qquad\sinit=\sum_{k=1}^\infty
c_k(0).
\label{eq:4.3.905}
\end{equation}
Note that now $\sinit=1$.
Combining all this eventually yields for the concentrations $c_j(t)$
\begin{equation}
c_j(t)=\frac{j^{j-2}}{(j-1)!}\left(1-e^{-t}\right)^{j-1}\exp\left[
-j(1-e^{-t})
\right]
e^{-t}
\label{eq:4.3.10}
\end{equation}

Note here an amusing connection: if one considers (\ref{eq:4.3.4})
for $f(j)=j$ and make the substitution $x_j=\phi_j/j$, one obtains
\begin{equation}
x_j=\half\sum_{k=1}^{j-1}k(j-k)x_kx_{j-k}-jx_j
\label{eq:4.3.10.1}
\end{equation}
the $x_j$ therefore satisfy the Smoluchowski equations (\ref{eq:2.2}) for
the kernel $K(k,l)=kl$, at least as long as the total mass $\mom[1](t)$ is
conserved, that is, before the gelation time. This is particularly useful
to evaluate the exponent $\tau$, which does not depend on the time scale.
A generalization of this remark was shown in \cite{zif83a}.

One finds that the appropriate scaling variable is $je^{-2t}$ and the
scaling function is given by
\begin{equation}
\Phi(x)=\frac{x^{-3/2}e^{-x/2}}{\sqrt{2\pi}}.
\label{eq:4.3.11}
\end{equation}
To show that scaling occurs for all initial conditions, however, it is
easier to use the generating function approach without further
transformations. The equations for $c(\mass, t)$ can be cast in the form
\begin{eqnarray}
\partial_{t}c(\mass, t)&=&\int_0^\mass d\massp \massp c(\massp, t)
c(\mass-\massp,t)d\massp-\nonumber\\
&&\qquad-\sinit e^{-t}\mass c(\mass,t)-c(\mass,t),
\label{eq:4.3.12}
\end{eqnarray}
where both mass conservation and (\ref{eq:4.3.905}) have been used to
simplify the equations. If we define $F(\lap,t)$ as
\begin{equation}
F(\lap,t)=\int_0^\infty c(\mass, t)e^{\lap\mass}d\mass,
\label{eq:4.3.13}
\end{equation}
one obtains using (\ref{eq:4.3.12}):
\begin{eqnarray}
&&\partial_tF=(F-\sinit e^{-t})\partial_\lap F-F
\label{eq:4.3.14a}\\
&&F(\lap,t=0)=f(\lap).
\label{eq:4.3.14b}
\end{eqnarray}
This can be solved by the method of characteristics, with the
following result, shown in appendix \ref{app:sum}
\begin{eqnarray}
&&F(\lap,t)=f(\lap_0)e^{-t}
\label{eq:4.3.15a}\\
&&\lap=\lap_0-\left[
f(\lap_0)-f(0)
\right]
(1-e^{-t}).
\label{eq:4.3.15b}
\end{eqnarray}
{}From this it follows after some algebra, that, if we assume that the
initial cluster size distribution $c(\mass, t)$ has finite second
moment, then
\begin{equation}
\lim_{t\to\infty}
\left.\left[
\partial_\lap F(\lap,t)
\right]\right|_{\lap=se^{-2t}}=\frac{1}{\sqrt{1-2f^{\prime\prime}(0)s}}.
\label{eq:4.3.16}
\end{equation}
which gives for the scaling function $\Phi(x)$:
\begin{equation}
\Phi(x)=\frac{x^{-3/2}e^{-x/[2\mom[2](0)]
}}{\sqrt{2\pi\mom[2](0)}},
\label{eq:4.3.17}
\end{equation}
where we have used the fact that the second derivative of $f(\lap)$
is the second moment of the initial distribution. It is certainly both
remarkable and somewhat anomalous that the scaling function depends
(albeit in a somewhat trivial fashion) on the initial condition.

This dependence on the initial condition, as well as the appearance
of the second derivative of the generating function in the expression
for the solution make it of particular importance to study the effect
of highly polydisperse initial conditions, that is, such initial
conditions as have $c(\mass,0)$ behaving as $C\mass^{-\alpha}$ as
$\mass\to\infty$, where $2<\alpha<3$, as we saw in the case of the
constant kernel. To avoid tedious repetitions, we use the remark made in
(\ref{eq:4.3.10.1}) in order to map this problem onto the corresponding
problem for the product kernel, which will be treated in Appendix
\ref{app:bilinear2}. There we shall find that
the typical size does not grow as $e^{2t}$ any more,
but rather as $\exp[(\alpha-1)t/(\alpha-2)]$. This is the first sign
of non-universal behaviour in this system. One also finds
\begin{equation}
\tau=\frac{\alpha}{\alpha-1}.
\label{eq:4.3.21}
\end{equation}
Note how, for $\alpha=3$, the traditional value $3/2$ is recovered. For
$\alpha>3$ the above computations cease to be valid. On the other hand,
for $\alpha\to2$, one sees that $\tau$ approaches the limiting (and not
allowable) value of two, for which the total mass diverges.
\subsection{Parity-dependent Kernels}
\label{subsec:par}
We now consider a variation of the constant kernel which is specific to the
discrete case, namely a kernel in which the reaction rates depend on
the parity of the sizes of the reactants but are otherwise constant. We
shall therefore use the discrete notation throughout this subsection.
This kernel is defined as follows:
\begin{equation}
K(k,l)=\left\{
\begin{array}{ll}
K&\qquad\mbox{if $k$ and $l$ are odd}\\
L&\qquad\mbox{if $k$ and $l$ are even}\\
M&\qquad\mbox{otherwise}
\end{array}
\right.
\label{eq:4.10.1}
\end{equation}
This model was introduced in \cite{ley86,ley87}. The physical justification
given there was that such parity effects might arise in the
diffusion-limited aggregation of alternating linear copolymers. While
this seems
at best questionable, there is no doubt the model provides an important
testing ground for the universality assumption implicit in the use of the
scaling approach. Indeed, such an alternation, within constant bounds, of
reactivity according to parity is a perfect example of the sort of
detail that ought to be irrelevant if we want to be able to
apply scaling theory in concrete situations. It does not matter that this
particular instance of deviation from constancy may not
appear reasonable
physically: there  is every reason to believe that similar deviations may
produce at least similar effects, so that the exact solutions we can find
in this case are important because they teach us, to what extent
the scaling theory may be expected to hold, and in what aspects it is
likely to fail, once we leave the familiar ground of exactly solved
systems. The discussion of the solution falls therefore squarely within the
principal purpose of this review.

It turns out to be impossible to give a full and explicit solution for
the kernel (\ref{eq:4.10.1}). What can be done, however, is the following:
on the one hand, one can prove in full generality that the kernel
(\ref{eq:4.10.1}) satisfies scaling as defined in (\ref{eq:3.7}), with
a scaling function that is, however, different from the one given in
(\ref{eq:4.2.8}) for the constant kernel. It is also possible to give some
of the non-scaling asymptotics exactly for the general kernel
(\ref{eq:4.10.1}). On the other hand, it is also possible fully to solve
some particular instances. I shall restrict myself to the case in
which $L=4M$,
for which simple and completely explicit formulae are obtainable, as shown
in \cite{cal00}. For
these, then, everything is known and can be discussed in full detail.

Let us first consider the scaling limit as  defined in (\ref{eq:3.7}).
As stated above, the scaling limit differs from the one obtained for
the constant kernel. This invalidates to some extent the usual claims
of universality of the scaling theory. However, it does not
contradict what
was stated in Appendix \ref{app:deriv-scaling}: indeed, there, we had
shown that if a kernel behaves asymptotically as a homogeneous function,
then the scaling function should be given by a solution of equation
(\ref{eq:3.10}). But the kernel (\ref{eq:4.10.1}), which oscillates for
ever around the constant kernel, does not satisfy this hypothesis. Still,
this change in the scaling function is quite
unexpected. The scaling function $\Phi(x)$ for the kernel (\ref{eq:4.10.1})
is evaluated in detail in Appendix \ref{app:parity-scal}. As we shall see,
however, the set of odd and even
concentrations obey completely different asymptotics. How, then, can we
have scaling for the full distribution, without having to divide it into
odd and even clusters?
The answer lies, again, in the very particular nature
of the way in which the scaling limit is defined, see (\ref{eq:3.7}).
Indeed, what is considered there is the average of a function which varies
on a {\em macroscopic} scale, namely the typical size $s(t)$. Such an
average is clearly unaffected by any even-odd oscillations, since they
are effectively averaged over. One might at first then argue
that this is an artefact of an inappropriate definition. However, if one
thinks about it from the purely physical point of view, the very reverse
is the case: any reasonable physical measurement, say light scattering,
will measure such an average and will be quite insensitive to the
oscillations. In fact, the only measurement likely to feel such effects is
a complete measurement of the cluster size distribution function at the
level of accuracy of the monomer size unit. This is only very rarely
practicable\footnote{A determination of this nature has, however, been
achieved in \cite{bro90}.},
so that one usually has to be satisfied with the
coarser measures which do not perceive the oscillations. However, the
mathematical solution certainly has these oscillations, so that, as we
shall see, no statement can be made stating that the individual
concentrations are well approximated by the scaling form.

In Appendix \ref{app:parity-scal} it is shown that both the odd and even
cluster-size distributions approach scaling in the sense of
(\ref{eq:3.7}) separately. This is given precise meaning in
(\ref{eq:s.11}). The scaling functions $\Phi_d(x)$ and $\Phi_p(x)$ for the
odd and even distributions respectively cannot be evaluated analytically,
but the $\tau$ and $\explarge$ exponents can. Here, as throughout
in this subsection, the letter $d$ refers to the
odd ({\em dispari}) clusters, whereas the letter $p$ refers to the even
({\em pari}) clusters\footnote{This somewhat singular notation arose first
in  \cite{cal00}. It is unambiguous and easy to remember after some time,
so I decided to stick with it here.}. One first defines
\begin{equation}
\tpinf=\frac{K-M}{L}\left[
1+\sqrt{
1+\frac{KL}{(K-M)^2}
}
\right].
\label{eq:4.10.105}
\end{equation}
The expressions for $\tau_{d,p}$ and $\explarge_{p}$ are then found to be:
\begin{eqnarray}
\tau_d&=&1-\frac{M}{K}\tpinf
\label{eq:4.10.110}\\
\tau_p&=&\min\left[
2-\frac{M}{K}+\frac{L}{K}\tpinf,2\tau_d-1\right]
\label{eq:4.10.120}\\
\explarge_p&=&0,\label{eq:4.10.140}
\end{eqnarray}
whereas for $\explarge_d$ we only know that it is positive.
Note finally that from the definition of scaling given in (\ref{eq:s.7})
immediately follows that the full cluster size distribution tends to a
scaling limit with the scaling function $\Phi(x)$ given by
\begin{equation}
\Phi(x)=\half\left[
\Phi_d(x)+\Phi_p(x)
\right]
\label{eq:4.10.150}
\end{equation}
which is quite an extreme illustration of the fact that the scaling
function need not yield a good approximation to any of the concentrations
$c_j(t)$ at all.

Let us now turn to the non-scaling asymptotics: as usual, define the exponents
\begin{eqnarray}
c_{2j+1}(t)&=&const.\cdot t^{-\wpr_d}
\label{eq:4.10.2}\\
c_{2j}(t)&=&const.\cdot t^{-\wpr_p}
\label{eq:4.10.3}
\end{eqnarray}
It is then straightforward to
check, as is also shortly discussed in Appendix \ref{app:parity-scal}, that
\begin{eqnarray}
\wpr_d&=&1+\frac{M}{K}\left(
1+\tilde p_\infty
\right)\nonumber\\
\wpr_p&=&\min\left(
1+\frac{2M}{K}\tilde p_\infty,2w_d-1
\right)
\label{eq:4.10.4}
\end{eqnarray}
Since these coincide with the values of $w_d$ and $w_p$ that are obtained
from the values of (\ref{eq:4.10.110}) and (\ref{eq:4.10.120}) for
$\tau_d$ and $\tau_p$ via the scaling relation (\ref{eq:scaling-rel}), we
may state that strong scaling holds.

The unexpected feature in this case
is rather the non-universal nature of the result. That is, in general, one
does not expect a simple odd-even oscillation in the reaction rates to
produce such massive changes in the scaling function. However, one must
remember that the constant kernel is in case II, which is quite
non-universal in itself. In fact, as is shortly pointed out in Appendix
\ref{app:parity-scal}, these peculiarities may be typical of case II.
\subsubsection{Two Fully Solvable Subcases}
We may cite two cases that have been solved more or less fully, which
therefore allow for a test of the above computations. The first is given by
\begin{equation}
L=4M
\label{eq:4.11.1}
\end{equation}
In this case a completely explicit solution has been found in \cite{cal00}
for the initial condition
\begin{equation}
c_j(0)=\alpha\delta_{j,1}+\half(1-\alpha)\delta_{j,2}.
\label{eq:4.11.2}
\end{equation}
It is then given explicitly by the formulae
\begin{eqnarray}
c_{2j+1}(t)&=&2^{-2j}\left(
\begin{array}{c}
2j\\
j
\end{array}
\right)\frac{\alpha\left[
s(t)
\right]^{-1/2}}{1+\alpha Kt}\left\{
1-[s(t)]^{-1}
\right\}^j
\label{eq:4.11.3}\\
c_{2j}(t)&=&\frac{2\dot s(t)}{Ls(t)d(t)}
\left\{
1-[s(t)]^{-1}
\right\}^j
\label{eq:4.11.4}\\
s(t)&=&\frac{L}{\alpha(L+4K)}(1+\alpha Kt)+\frac{(\alpha-1)L+\alpha}{
\alpha(L+4K)}(1+\alpha Kt)^{-L/(4K)}.
\label{eq:4.11.5}
\end{eqnarray}
The quantity $s(t)$ defined in (\ref{eq:4.11.5}) is therefore the most
natural definition of typical size in this system, since the forms which
define $c_{2j+1}(t)$ and $c_{2j}(t)$ are already very close to scaling form
in their exact expression. One immediately verifies from
(\ref{eq:4.11.3},\ref{eq:4.11.4}) that $\tau_d=1/2$ and $\tau_p=0$, all of
which indeed confirms exactly the results obtained in the previous
subsection. Similarly, one verifies the value of zero for $\explarge_p$.
The exponent
$\explarge_d$ on the other hand, which could not be evaluated in general,
has in this case the value $1/2$. It
therefore coincides with $\tau_d$ in this particularly simple case, once
again satisfying the Dominant Singularity Hypothesis. It is not clear in
general whether this will always be the case for these kernels, however.

Another case which can to some extent be solved exactly is the one in which
\begin{equation}
M=\half(K+L)
\label{eq:4.11.6}
\end{equation}
In this case the kernel is of the form of a sum kernel and can be solved
using the techniques mentioned in subsection \ref{subsec:sum}. The
solution obtained in \cite{ley87} is extremely opaque, however, and does
not yield such explicit formulae for the concentrations as are given for
the case $L=4M$. However, after some complicated calculations one obtains
values of $\tau_d$ and $\tau_p$ again in full agreement with those found
in the previous subsection.
\subsubsection{The addition-aggregation kernel}
In a recent paper \cite{mob03} a minimal model showing anomalies very
similar to those found in the parity-dependent kernel was studied. One
defines constant rates, with the only additional feature that monomers
are singled out and react differently from other aggregates. more precisely
\begin{eqnarray}
&&K(1,1)=\epsilon\nonumber\\
&&K(1,k)=K(k,1)=1\\
&&K(k,l)=\gamma,\nonumber
\label{eq:4.14.1}
\end{eqnarray}
where $k,l\geq2$. For this model one obtains a fairly bewildering array of
large time behaviours for monomers of fixed size, yet scaling in the sense
in which we have defined it continues to hold, as the authors show in an
Appendix. Furthermore, the scaling function is the same as for the
constant kernel, thereby vindicating the usual claims of universality.
\subsection{The $q$-sum kernel}
\label{subsec:q-sum}
In a recent paper \cite{cal99}, a model quite different from the classical
ones was solved, namely
\begin{equation}
K(k,l)=2-q^k-q^l=2-e^{-bk}-e^{-bl}.
\label{eq:4.12.1}
\end{equation}
where $q=e^{-b}$ is an arbitrary number between zero and one. We shall
use $q$ and $b$ interchangeably in the following. Again the model is
restricted to discrete values of the masses only. This model is interesting
for at least two different reasons:
\begin{enumerate}
\item It is a non-trivial variant on the constant kernel, in which the
reactivity of the small aggregates is diminished with respect to that of
larger ones. The difference is  only significant
for rather small aggregates, at least
in the general case in which $q$ is fixed and not too near to one, so that
one expects no major deviations from the ordinary scaling picture. As we
shall see, this is indeed the case if we interpret scaling in the sense of
this paper, that is, as the convergence in first moment as defined in
(\ref{eq:3.10}). On the other hand, for non-scaling asymptotics, quite
unexpected results are encountered.

\item On the other hand, it is also possible to consider a simultaneous
asymptotic behaviour in which $q\to1$ at the same time as $t\to\infty$. In
this case we obtain a highly peculiar crossover situation in which the sum
kernel drives the aggregation of small clusters, whereas the large
clusters obey a constant kernel kinetics. We shall not
have much to say about it, as the evaluation of the crossover function is a
daunting task which is left to the inspired reader. It can be shown,
however, that the formalism developed in \ref{subsec:cross} indeed yields
the correct scaling variables.
\end{enumerate}
In order to obtain the solution, we note that the kernel (\ref{eq:4.12.1})
is of the form $f(k)+f(l)$, with $f(k)$ given by $1-q^k$.
One therefore proceeds along the same lines as
described in subection \ref{subsec:sum} for the sum kernel, namely we use
the transformation given by (\ref{eq:4.3.3}). This leads to the equations
\begin{equation}
\frac{d\phi_j}{d\tres}=\sum_{j=1}^{j-1}(1-q^k)\phi_k\phi_{j-k}
-(1-q^j)\phi_j.
\label{eq:4.12.2}
\end{equation}
These equations can in principle be solved recursively. The solution
becomes very soon quite intricate, however, and it is not possible to
obtain much information in this way.

The solution makes heavy use of the $q$ factorial and $q$ exponential,
defined as follows
\begin{eqnarray}
(a;q)_n&=&\prod_{l=0}^{n-1}(1-aq^l)
\label{eq:4.12.3}\\
e_q(x)&=&\prod_{l=0}^\infty (1-xq^l)^{-1}\nonumber\\
&=&\sum_{k=0}^\infty \frac{x^r}{(q;q)_r},
\label{eq:4.12.4}
\end{eqnarray}
where the last identity is familiar from the theory of the $q$ exponential
\cite{gas90}.

The solution, which is derived in detail in \cite{cal99} is given as
follows: one defines
\begin{equation}
H(\lap, \tres)=\sum_{j=1}^\infty \phi_j(\tres)\left(
e^{bj\lap}-1
\right)
\label{eq:4.12.5}
\end{equation}
This function satisfies a non-linear differential--difference equation, the
solution of which is shortly sketched in Appendix \ref{app:q-sum}. The
final result is given by
\begin{eqnarray}
H(\lap, \tres)&=&-\frac{\partial}{\partial\tres}\ln\left[
1+S(\lap, \tres)
\right]
\label{eq:4.12.6}\\
S(\lap,
\tres)&=&\frac{1}{e_q(q^{-\lap})}\sum_{r=0}^\infty\frac{q^{-r\lap}}{(q;q)_r}
\left[
\exp(\tres q^r)-1
\right]
\label{eq:4.12.7}
\end{eqnarray}
{}From these results and the definition of $\tres$, see (\ref{eq:4.3.3}) one
readily obtains
\begin{equation}
t=\frac{1}{e_q(q)}\sum_{r=0}^\infty\frac{\exp(\tres q^r)-1}{(q;q)_r}.
\label{eq:4.12.8}
\end{equation}
{}From these results essentially everything can be computed, though this can
sometimes be difficult.

Let us first concentrate on the case in which $q$ is fixed. In this case,
the only limits of interest are the large-time limit at fixed $j$ and the
scaling limit. It is quite straightforward to obtain from
(\ref{eq:4.12.2}) the asymptotic order of magnitude of the $c_j(t)$. With
some additional work one obtains the more precise result
\begin{equation}
c_j(t)=\frac{t^{-(2-q^j)}}{e_q(q)^{1-q^j}(q;q)_{j-1}}\left[
1+o(1)
\right]\qquad(t\to\infty)
\label{eq:4.12.9}
\end{equation}
It is therefore clear that the large time asymptotics is markedly
different from that of the constant kernel. In particular, the exponent
$\wpr$ is not even defined, since the large-time decay law depends on the
size $j$ of the cluster. The only feature that remains is the fact that the
exponent tends to the value $2$ for large values of $j$. On the other
hand, an elementary analysis shows that this behaviour is only attained
when
\begin{equation}
t\gg\exp(q^{-j})
\label{eq:4.12.10}
\end{equation}
which increases extremely fast with $j$. We have then obviously no
contradiction with the scaling statement, which asserts that the decay goes
as $t^{-2}$ as long as $j$ is of order $t$. Along similar lines, we see
that the moments $\mom(t)$ for $\indmom<q-1$ are dominated by $c_1(t)$ and
hence decay as $t^{-2+q}$ instead of going as $t^{\indmom-1}$ as predicted by
scaling theory. This is the case even though, as long as $\indmom>-1$, the
scaling expression for the moments {\em converges} at the origin.

The scaling limit can also be shown to exist. This is proved by showing
that the function $H(\lap, \tres)$ has only one singularity which
approaches zero as $t\to\infty$, which is a simple pole. Due to this the
scaling characteristics of the constant kernel are obtained. The detailed
proof of this fact is quite technical and is given in \cite{ley99}.

Let us now turn to the issues involved in the crossover regime, when
$b\to0$. This has been treated fairly extensively in \cite{ley99}, to
which the interested reader is referred for more details. As things stand,
this appears to be an instance in which the formalism developed in
subsection \ref{subsec:cross} could be applied
if we substitute $K(k,l)$ by $K(k,l)/b$, so as to satisfy the requirement
that $K(1,1)$ be always of order one: indeed one has $\lambda=1$
with $b$ playing the role of the small parameter $\epsilon$ in
(\ref{eq:3.5.25}). One therefore expects a typical size of order $e^t$ and
scaling parameters $je^{-t}$ and $bj$. I have not been able to carry out
this evaluation successfully, however.

The results which have been obtained are the following:
\begin{enumerate}
\item When $bt$ is of order one, the sum kernel is a good approximation
both for clusters of fixed size $j$ and for clusters in the scaling
region. On the other hand, considerable discrepancies always exist in the
large-size limit, no matter how short times are. This is related to the
fact that the nearest singularity of the $q$-sum kernel is a pole,
so that the $\explarge$ exponent is zero. Similarly, we may say that
moments of fixed order $\indmom$ are, in this range, well approximated by
those of the sum kernel. However, large order moments will show deviations
earlier than low order moments.

\item For sufficiently large times, all features of the constant kernel
are recovered. This is an immediate consequence of the statements made
about the existence of the scaling limit. However, the time needed to reach
in all respects this scaling limit is surprisingly
large: the necessary condition is
\begin{equation}
t\gg\exp(1/b).
\label{eq:4.12.11}
\end{equation}

\item A more accurate description of the regime lying between these two
extremes would clearly be highly desirable, but I am not aware of a way to
obtain such results. It can be shown that the crossover limit in the sense
of subsection \ref{subsec:cross} exists with the scaling
variables $je^{-2t}$ and $bj$. This is shown in Appendix
\ref{app:q-cross}.
\end{enumerate}
\subsection{The Bilinear Kernel}
\label{subsec:bil}
The most general kernel which has been solved in classical works on the
subject is the so-called bilinear kernel:
\begin{equation}
K(\mass,\massp)=A+B(\mass+\massp)+C\mass\massp.
\label{eq:4.5.1}
\end{equation}
The history of the various solutions to this kernel is too long to be told
in detail. The crucial issue concerns the case in which $C>0$. As we have
already stated in subsection \ref{subsubsec:gel-scale}, such a case, involving
as
it does a homogeneity degree $\lambda>1$, necessarily leads to violation
of mass conservation in finite time. While it was originally believed that
no solutions could be found beyond that time (see for example
\cite{McLeoda,McLeodb}), it was eventually realized \cite{zif80,ley81} that
such
solutions existed and were physically meaningful. The
decrease in total mass was interpreted as the flow of mass to an infinite
cluster, in a way quite similar to that which is observed
in systems such as
percolation: as the size of a finite system is taken to infinity, the
total mass comprised in those clusters which are of a size that diverges as
the system size may or may not be zero. In the latter case, this mass does
not appear in the final bookkeeping once the system has been taken as
infinite. There is hence no contradiction involved. Furthermore, from the
mathematical point of view, it can in fact be shown that if the second moment
of the cluster size distribution diverges, mass conservation can be
violated. We shall further show explicitly how this happens in the
bilinear model
(\ref{eq:4.5.1}).

Let us first limit ourselves to the simpler model
\begin{equation}
K(\mass,\massp)=\mass\massp
\label{eq:4.5.2}
\end{equation}
We shall prove in Appendix \ref{app:bilinear1} that for
purposes of scaling (\ref{eq:4.5.1}) and (\ref{eq:4.5.2}) are equivalent.
The possibility of solving the general kernel (\ref{eq:4.5.1}) exactly
does, however, open new vistas concerning crossover, as we show in the
next subsection.

In order to gain some insight into the behaviour of (\ref{eq:4.5.2}), we
introduce the following generating function
\begin{equation}
G(\lap,t)=\int_0^\infty \massp c(\massp,t)e^{\lap\massp}.
\label{eq:4.5.3}
\end{equation}
This satisfies the following PDE
\begin{equation}
G_t-\left[
G-\mom[1](t)
\right]G_\lap=0.
\label{eq:4.5.4}
\end{equation}
Here we have {\em not} assumed that $\mom[1](t)$ is constant, since
this hypothesis cannot be sustained over all
times, though a solution to the Smoluchowski equations (\ref{eq:2.2}) with
the kernel (\ref{eq:4.5.2}) does indeed exist for all times.
Let us first consider the case in which $\mom[1](t)$ is equal to one. In
this case, a straightforward application of the characteristic method
yields
\begin{eqnarray}
G\left[\lapbar(t)
\right]&=&g(\lap_0)\nonumber\\
\lapbar(t)&=&\lap_0-t(g(\lap_0)-1)
\label{eq:4.5.5}
\end{eqnarray}
where $g(\lap)$ is the generating function of the initial conditions.
For each value of $t$, the function $\lapbar(t)$ takes
its maximum value at a value of $\lap_0$ such that $g^\prime(\lap_0)$
equal $1/t$. Above this value $G(\lap,t)$ is not defined.
But $G(\lap,t)$ must clearly be defined for all negative
values of $\lap$. The maximum value must therefore be at least equal to
zero. But it is easy to verify, using the convexity properties of $g(\lap)$
following from the positivity of the initial conditions, that if
$t>1/g^{\prime}(0)$, the maximum value of $\lapbar(t)$ is strictly
negative, so that the function $G(\lap,t)$ is not defined for all
negative $\lap$ for such a value of $t$. This would imply
a cluster size distribution $c(\mass, t)$ that is exponentially increasing
with $\mass$, an utter absurdity. For lesser values of $t$, however,
everything is well-defined and the solution has all the
expected properties. As
an example we may cite the instance of monodisperse initial conditions, for
which the exact computation is an easy excercise in contour integrals
performed in Appendix \ref{app:bilinear}. The result is
\begin{equation}
c_j(t)=\frac{j^{j-3}}{(j-1)!}t^{j-1}e^{-jt}.
\label{eq:4.5.6}
\end{equation}
This is indeed a well-defined solution, but it is obvious that for
$t>1$ the corresponding generating function stops
being a solution to (\ref{eq:4.5.4}) with $\mom[1](t)$ equal to one, since
all functions $c_j(t)$ decrease. In fact, these are the so-called Flory
solutions of the Smoluchowski equations corresponding to (\ref{eq:4.5.2}).
They are obtained by starting from the assumption that $\mom[1](t)$ is
always equal to one, and therefore eliminating it from the Smoluchowski
equations. In other words, instead of solving the original problem
(\ref{eq:2.2}) for the kernel (\ref{eq:4.5.2}), one solves instead
\begin{equation}
\dot c(\mass, t)=\half\int_0^\infty \massp(\mass-\massp)
c(\massp, t)c(\mass-\massp, t)d\massp-\mass c(\mass, t).
\label{eq:4.5.7}
\end{equation}
Here $\mom[1](t)$ has been set equal to one by fiat. It is then
straightforward to check that (\ref{eq:4.5.6}) is the solution of
(\ref{eq:4.5.7}) for all times. However, one then finds for this solution
\begin{equation}
\mom[1](t)=\left\{
\begin{array}{ll}
1&\qquad(t\leq1)\\
\exp[-\lap(t)]&\qquad(t>1),
\end{array}
\right.
\label{eq:4.5.8}
\end{equation}
where $\lap(t)$ is given as the solution to the equation
\begin{equation}
\frac{\lap(t)}{1-e^{-\lap(t)}}=t.
\label{eq:4.5.9}
\end{equation}
Clearly, (\ref{eq:4.5.8}) casts some doubt on the validity of
(\ref{eq:4.5.7}) beyond $t=1$. It has, however, been
defended on the following grounds: beyond the gel point, it is no longer
possible to identify the finite systems and its infinite limit as we have
always implicitly done so far. Rather, in the finite system, for times
larger than the gelation time, we have two cluster populations: on the  one
hand, we have the ones
that are small with respect to system size, the concentrations of which
tend to a well defined limit as it is made to go to infinity.
These are known as the {\em sol} particles. On the other hand, the large
clusters of the finite system have sizes that go to infinity with the system
size, and contain a finite portion of the mass. These so-called {\em
gel} particles do interact, in the finite system, with the sol. Since, in
the finite case, mass is certainly conserved, we might expect some
variation of (\ref{eq:4.5.7}) to be derivable for the finite system. If one
then goes to the limit in this fashion, the solution (\ref{eq:4.5.8}) is
obtained for the sol for all times, and (\ref{eq:4.5.9}) simply tells us
how the mass contained in the finite clusters decreases with time. As we
shall see later, however, there exist exactly solved finite systems which
tend either to the Flory solution (\ref{eq:4.5.7}) or to the so-called
Stockmayer solution, which is the post-gel solution of (\ref{eq:4.5.4}).
It is therefore not straightforward to decide which of the two solutions
is the ``correct'' one. The issue is, to some extent, simplified by the
fact that, in any realistic situation, the mean-field assumptions made in the
derivation of the Smoluchowski equations (\ref{eq:2.2}) certainly break
down near the gel point, so that the whole issue becomes irrelevant. A
detailed discussion of the difference between the Flory and the
Stockmayer solutions is found in \cite{zif80}.

Let us now derive this latter solution. The central issue is a geometric
one, discussed in greater detail in Appendix \ref{app:bilinear}: it is
found that the function $G(\lap,t)$ covers the $(\lap,t)$-plane twice and
has a singularity over some curve on the $(\lap,t)$ plane, on which the
two branches of the surface join smoothly. This structure is a universal
feature of all solutions of (\ref{eq:4.5.4}), independently of the shape
of $\mom[1](t)$. However, since it is manifest that only the sheet of
$G(\lap,t)$ that connects to $\lap=-\infty$ without crossing the
singularity, has meaning as a generating function of a solution of
(\ref{eq:4.5.4}), it is clear that we must choose $\mom[1](t)$ in such a
way that the singular curve always lie above $\lap=0$. On the other hand,
after $t=1$, it follows from the previous dicussion that we cannot make
$\mom[1](t)$ to remain constant. We must therefore allow a non-trivial
time-dependence of $\mom[1](t)$. But it is clear that such a
dependence can only obtain if the second moment of the cluster size
distribution diverges, that is, if the singular curve {\em coincides}
with the $\lap=0$ axis. This computation is carried out in detail in
Appendix \ref{app:bilinear} for monodisperse initial conditions, yielding
\begin{equation}
c_j(t)=\left\{
\begin{array}{ll}
\displaystyle\frac{j^{j-3}}{(j-1)!}t^{j-1}e^{-jt}&\qquad(t\leq1)\\
\\
\displaystyle\frac{j^{j-3}e^{-j}}{(j-1)!}t^{-1}&\qquad(t\geq1)
\end{array}
\right.
\label{eq:4.5.10}
\end{equation}
Note that both the values and the first derivatives of the $c_j(t)$ are
continuous, so that they really are (strong)
solutions of the set of first order ordinary differential equations
(\ref{eq:2.2}).

Finally let us make some remarks on the applicability of the
scaling approach. As
always, one must first define the sense in which we want scaling to hold.
Since most of the mass is contained in clusters of size of order one, even
as the gel time is approached, the usual definition of scaling given by
(\ref{eq:3.7}) does not hold.
As we have already stated before, the definition of second moment
scaling (see (\ref{eq:3.810}) for the precise definition) is the most
appropriate one. The scaling function as $t$ approaches $t_c$ is computed
in Appendix \ref{app:bilinear}, with the result
\begin{eqnarray}
\Phi(x)&=&\frac{1}{\sqrt{2\pi\alpha}}x^{-5/2}e^{-x/(2\alpha)}\nonumber\\
\alpha&=&\frac{\gpp}{\gp^3}.
\label{eq:4.5.11}
\end{eqnarray}
Two remarks are in order at this stage; First, the ordinary scaling we
have just derived rests upon the finiteness of the {\em third\/} moment of
the initial conditions. If this diverges and yet the second moment is
finite, we are led to an interesting situation: the scaling
behaviour is then given by different exponents, but the gel time remains
finite. After this gel time, however, the asymptotic behaviour must
necessarily have the exponent $-5/2$, since that is the only one that
allows a finite rate of mass transport. In Appendix \ref{app:bilinear2} we
therefore study in detail what happens under these circumstances, as well
as what happens in the case of a divergent second moment, in which case
instantaneous gelation occurs.
\subsection{Crossovers in the Bilinear Kernel}
\label{subsec:cross-bil}
The bilinear kernel (\ref{eq:4.5.1}) has three variants which display
crossover behaviour. Using the type of parametrization defined in
(\ref{eq:3.5.2}), these are
\begin{eqnarray}
K(\mass, \massp)&=&1+\epsilon(\mass+\massp)
\label{eq:4.6.1}\\
K(\mass, \massp)&=&1+\epsilon^2\mass\massp
\label{eq:4.6.2}\\
K(\mass, \massp)&=&\mass+\massp+\epsilon\mass\massp
\label{eq:4.6.3}
\end{eqnarray}
Let me first recall the definition stated in (\ref{eq:3.5.3}) of the
crossover limit and the corresponding scaling function
\begin{equation}
\lim_{t\to\infty}\int_0^\infty \massp c(\massp,y/s(t);t)
f\left(\frac{\massp}{s(t)}\right)\to\constsep\int_0^\infty x\Phi(x,y)dx.
\label{eq:4.6.4}
\end{equation}
Here, as in subsection \ref{subsec:cross}, the function $c(\mass,
\epsilon;t)$ denotes the solution of the Smoluchowski equations
(\ref{eq:2.2}) with a kernel characterized by a given value of $\epsilon$.

For the first case, the scaling function is evaluated exactly in Appendix
\ref{app:crossover2}. It is given by the following expression
\begin{equation}
\Phi(x,y)=\frac{
4y^2(2xy)^{2xy}\exp\left[
-2xy/(1-e^{-y})
\right]
}
{
(1-e^{-y})^{2+2xy}\Gamma(2+2xy)
}
\label{eq:4.6.5}
\end{equation}
This expression reduces to the scaling function (\ref{eq:4.2.8}) for the
constant kernel when $y\to0$, as of course it should. Here $s(t)$ is taken
to be $t$. The limit of large $y$ is somewhat more subtle: taking the
limit $x\to\infty$ and $y\to\infty$ with $xye^{-2y}=z$ constant, one
obtains the correct limit apart from a divergent prefactor due to issues
of normalization. This illustrates nicely the difficulties involved in
studying the large $y$ behaviour of such functions.
However, the use of (\ref{eq:4.6.5}) is not, of course,
in the large $y$ regime, where everything is best described by the scaling
function (\ref{eq:4.3.17}) of the sum kernel, but rather at intermediate
times, for which the aggregates have already become large but not large
with respect to $\epsilon^{-1}$.

\begin{figure}
\begin{center}
\includegraphics*[scale=0.4,angle=270]{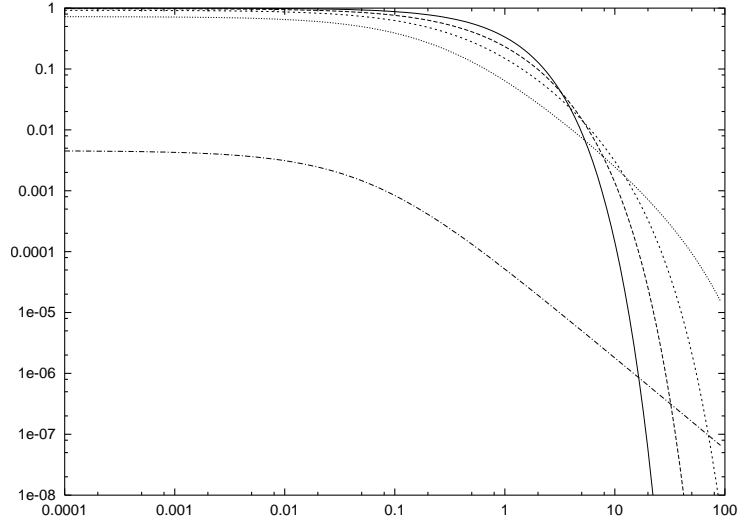}
\end{center}
\caption{
The crossover scaling function is plotted as a function of $x$ for
various values of $y$, namely $0.1$, $0.5$, $1$, $2$ and $10$.Note how
a long $x^{-3/2}$ regime is eventually established. It requires, however,
the very largest value of $y$.
}
\label{fig:1}
\end{figure}

On the other hand, for any finite value of $y$,
$\Phi(x,y)$ tends to a constant as $x\to0$, as stated in subsection
\ref{subsec:cross}. Figure \ref{fig:1} shows the scaling function
(\ref{eq:4.6.5}) as a function of $x$ for various values of $y$. Note how
the initial values at low $x$ are always constant, corresponding to a
$\tau=0$ regime, whereas a $x^{-3/2}$ behaviour arises over a very
extended range for large values of $y$.

It is also possible to go part of the way toward evaluating the crossover
functions for one of the two other cases arising in the full bilinear kernel,
namely (\ref{eq:4.6.2}).
For the crossover from the constant to the product kernel given by
(\ref{eq:4.6.2}), one finds the following result, the derivation of which
is sketched in Appendix \ref{app:crossover3}: we first define
\begin{eqnarray}
q(s,y)&=&\frac{2}{2+s_0^2-(s_0^2\cos y+2s_0\sin y)}
\label{eq:4.6.6}\\
s&=&s_0-t+\int_0^ydy^\prime\,q(s,y^\prime).
\label{eq:4.6.7}
\end{eqnarray}
Here (\ref{eq:4.6.7}) must first be inverted and the result substituted
into (\ref{eq:4.6.6}) to yield the desired result. One then has for the
crossover scaling function $\Phi(x,y)$
\begin{equation}
q(-s,y)=\int_0^\infty x\Phi(x,y)e^{-sx}dx
\label{eq:4.6.8}
\end{equation}
The analytical (or even numerical) evaluation of $\Phi(x,y)$ from
these equations appears to be a non-trivial task, which is left to
future work. It is easy to verify, however, that this function has a
singularity at $y$ equal to $\pi/2$, which corresponds to the gelling
singularity. Indeed, as will be remembered from the definition of the
crossover limit, the parameter $y$ in this case is nothing else than a
rescaled time $\epsilon t$. But it is easy to verify, using for example
the moment equations, that the gel time for the reaction rates
(\ref{eq:4.6.2}) is of order $\pi/(2\epsilon)$ as $\epsilon\to0$.

On the other hand, computing the scaling function in the case
(\ref{eq:4.6.3}) of a crossover betweeen the sum and the product kernel
presents considerable difficulties. A naive approach does not appear
to lead to true scaling behaviour in a straightforward way. While this is
in itself a matter of some interest, it should be left as a subject for
future work, as the issue is not settled.
\subsection{Finite Systems}
\label{subsec:fin}
It has been shown by Lushnikov and Piskunov \cite{lus78} that the
following finite versions of the classical
reaction kernels $K_1$, $K_2$ and$K_3$ as defined in (\ref{eq:4.0.1},
\ref{eq:4.0.2}, \ref{eq:4.0.3})
can be solved exactly at least in the discrete
case:
\begin{equation}
\Kfin_\alpha(k,l)=\left\{
\begin{array}{cc}
K_\alpha(k,l)&\qquad(\min(k,l)\leq N)\\
0&\qquad(\min(k,l)>N)
\end{array}
\right.
\label{eq:4.9.2}
\end{equation}
These have all been solved exactly in \cite{lus78}. We shall first discuss
in some detail the case of $\Kfin_1(k,l)$, showing how the exact
solution can be obtained and discussing its behaviour also from the point
of view of the crossover theory devfeloped in the previous section. We
shortly state the exact solution for $\Kfin_2(k,l)$ and go on to analyze
the limiting behaviour of $\Kfin_3(k,l)$. This last is of particular
interest, as it tends to the kernel $K_3(k,l)$ as $N\to\infty$, which
displays gelation. On the other hand, it is obvious that $\Kfin_3(k,l)$
cannot display such a phenomenon. As we shall see, one can show that
$\Kfin_3(k,l)$ leads to the Stockmayer kinetics as defined in subsection
(\ref{subsec:bil}), that is, the exact solution to the equations defined
by the kernel $\Kfin_3(k,l)$ tend to the solution of the full Smoluchowski
equations, not to those for which one has performed the substitution of the
first moment by the constant value of the total mass (Flory solution). On
the other hand, as can also easily be shown (see \cite{bak94}) the
following finite version of $K_3(k,l)$ will lead to the Flory solution
\begin{equation}
\Kfinbar_3(k,l)=\left\{
\begin{array}{cc}
K_3(k,l)&\qquad\max(k,l)\leq N\\
0&\qquad\max(k,l)>N
\end{array}
\right.
\label{eq:4.9.3}
\end{equation}
We therefore have the very remarkable phenomenon, that the limiting
behaviour of the finite systems depends crucially on the nature of the
cutoff. The reason for this has been pointed out already by Ziff {\it et
al.} \cite{zif83,ern84}, where it is shown that the Stockmayer
solution corresponds to a situation in which large aggregates do not react
with small ones, whereas the Flory solution correpsonds to the case  in
which they do. Here, by ``large'', I mean comparable to the size of the
cutoff. In this case, it is understandable that $\Kfin_3$ should tend to
the Stockmayer solution, since reactions between large particles are
strongly restricted, whereas in $\Kfinbar_3$ this is not the case.

At this stage it is important to clear up a possible misunderstanding: with
the above kernels one aims
to study the gelation transition by viewing it as the
limit of a set of non-gelling kernels, which represent systems with only a
finite number of different {\em reacting species}. This should be sharply
contrasted with the regularization induced by considering only a finite
number of {\em reactive aggregates} in the first place. The effects of
such an approximation are much more complex, since one must then take into
account the qualitatively new phenomenon of {\em fluctuations} in the
aggregate numbers. This has been done by van Dongen and Ernst in
\cite{don87a}, which presents the whole issue with great clarity. This,
however, belongs to those interesting subjects, the treatment of which
falls outside of this review.
\subsubsection{The Finite Constant Kernel}
We shall here treat the case of the kernel $\Kfin_1(k,l)$ as defined by
(\ref{eq:4.9.2}). The exact solution can only be found in a
straightforward way for the case of monodisperse initial conditions, which
is why we have limited ourselves to the discrete case. The solution in this
case is given by
\begin{eqnarray}
c_j(t)&=&\tres^{j-1}\exp\left[
-2\sum_{k=0}^N\frac{\tres^k}{k}
\right],
\label{eq:4.9.3a}\\
t&=&2\int_0^\tres dy\exp\left[
2\sum_{k=0}^N\frac{y^k}{k}
\right].
\label{eq:4.9.3b}
\end{eqnarray}
For completeness' sake the derivation of this solution is given in
Appendix \ref{app:const-fin}. It should, by the way, be emphasized that the
solution here given is, in a sense, incomplete: Indeed, the functions
$c_j(t)$ for $N<j\leq2N$ are different from zero and are not given by the
expressions (\ref{eq:4.9.3a}). These are inactive clusters formed by the
aggregation of clusters of size less than $N$. Eventually, of course, all
the mass will come to be concentrated in those clusters, whereas the mass
contained in the active clusters goes to zero.

Qualitatively, one sees that the solution behaves as follows: as long as
$\tres{}$ remains significantly less than one, one finds that the solution
given
by (\ref{eq:4.9.3a}) is very close to the exact solution of the constant
kernel given in (\ref{eq:4.2.7}). On the other hand, as soon as $\tres{}$
becomes
significantly larger than one, the concentration profile becomes
exponentially increasing and a large amount of the total mass finds
itself in the inactive clusters. It also follows from (\ref{eq:4.9.3b}).
that the time at which this change in behaviour takes place is of the
order of the cutoff $N$.

It is, of course, quite obvious that such a kernel in the limit of large $N$
offers a good example of a crossover situation, as discussed in a similar
case in \cite{vic85}. From the exact solution we may obtain the crossover
behaviour as follows: first define a scaling variable $\xi$ given by
\begin{equation}
\xi=N(1-\tres{}).
\label{eq:4.9.4}
\end{equation}
We now take note of the following identity valid in the scaling regime in
which $\xi$ is of order one, also shown in Appendix
\ref{app:const-fin}:
\begin{equation}
\sum_{k=1}^N\frac{\tres{}^k}{k}=\left\{
\begin{array}{cc}
-E_1(\xi)-\ln\xi&\qquad(\xi>0)\\
\Ei(|\xi|)-\ln|\xi|&\qquad(\xi<0)
\end{array}
\right.
\label{eq:4.9.5}
\end{equation}
where the exponential integrals $E_1(x)$ and $Ei(x)$ are defined as in
\cite{hmf}.
If one now substitutes these expressions in
(\ref{eq:4.9.3a}) and (\ref{eq:4.9.3b}) one eventually obtains
\begin{eqnarray}
c_j(t)&=&\left\{
\begin{array}{cc}
t^{-2}\xi^2\left(\frac{t}{N}\right)^2
\exp(-x\xi t/N+2E_1(\xi))&\qquad(\xi>0)\\
t^{-2}\xi^2\left(\frac{t}{N}\right)^2
\exp(-x\xi t/N-2\Ei(|\xi|))&\qquad(\xi<0)
\end{array}
\right.
\label{eq:4.9.6a}\\
t/N&=&\left\{
\begin{array}{cc}
2\int_{\xi}^\infty\exp\left[
-2E_1(\eta)
\right]
\frac{d\eta}{\eta^2}&\qquad(\xi>0)\\
2\int_{-\infty}^{|\xi|}\exp\left[
+2\Ei(\eta)
\right]
\frac{d\eta}{\eta^2}&\qquad(\xi<0),
\end{array}
\right.
\label{eq:4.9.6b}
\end{eqnarray}
where $x$ is the scaling variable $j/t$. We see therefore that $t^2c_j(t)$
depend only on the scaling variables $x$ and $t/N$, or equivalently $\xi$.
The scaling variables required are
indeed those predicted by the crossover theory developed in subsection
\ref{subsec:cross}: the exponent $\lambda$ is equal to zero and
$\epsilon$ is equal to $1/N$, as is readily seen if we write $\Kfin_1$
as follows
\begin{equation}
\Kfin_1(k,l)=\chi_{[0,1]}\left(\frac{k}{N}\right)
\chi_{[0,1]}\left(\frac{l}{N}\right),
\label{eq:4.9.7}
\end{equation}
where $\chi_{[0,1]}(x)$ is the characteristic function of the unit
interval.

As to the qualitative behaviour of the crossover function described by
(\ref{eq:4.9.6a}) and (\ref{eq:4.9.6b}), it must be admitted that is is
not wholly transparent. However, it does describe the transition between an
exponentially decreasing cluster size distribution and an exponentially
growing one as $\xi$ changes sign, which it clearly must do, in view of
(\ref{eq:4.9.6b}): as soon as $t/N$ is larger than the (convergent)
integral from $\xi$ to $\infty$ given on the right-hand side, $\xi$ must
become negative, so that the concentration profile becomes exponentially
increasing. Note further that the scaling picture shown here implies that for
times of order $N$, that is in the scaling regime, the total mass in the
active range is still of order one, and can be given in terms of the
scaling variables by the following expression
\begin{equation}
M_1(t)=\left\{
\begin{array}{cc}
\exp[2E_1(\xi)]&\qquad(\xi>0)\\
\exp[-2\Ei(|\xi|)]&\qquad(\xi<0)
\end{array}
\right.
\label{eq:4.9.8}
\end{equation}
\subsubsection{The Finite Sum Kernel}
For the kernel $\Kfin_2(k,l)$, the results are entirely similar. I
therefore limit myself to stating the exact solution, which is given by
\begin{eqnarray}
&&\tres=\frac{1}{\Lambda}\exp\left[
-\int_\Lambda^\infty dw\frac{
\sum_{k=1}^N a_kw^{-(k+1)}
}{
1-\sum_{k=1}^Na_kw^{-k}
}
\right]\nonumber\\
&&t=\int_0^\tres\frac{d\tres^\prime}{
1-\sum_{k=1}^Na_k\Lambda(\tres^\prime)^{-k}
}
\label{eq:4.9.9}\\
&&c_j(t)=\frac{a_j}{
\tres\Lambda^j
}\left(
1-\sum_{k=1}^Na_k\Lambda(\tres)^{-k}
\right)\nonumber\\
&&a_k=\frac{k^{k-2}}{(k-1)!}\nonumber
\end{eqnarray}
The derivation of this solution proceeds on exactly the same lines as
that of the constant kernel and is given in Appendix \ref{app:fin-sum}.
The detailed analysis of the crossover is left
as an excercise to the enthusiastic reader.
Concerning the asymptotic behaviour, however, one can make the following
observations: as follows from the computations in Appendix
\ref{app:fin-sum}, for $N=\infty$, $\sum_{k=1}^\infty a_kw^{-k}$ has a
singularity at $w=1$, where it has a value of one. There is hence for
every finite $N$ a value $w_c(N)<1$ such that
\begin{equation}
\sum_{k=1}^Na_kw_c(N)^{-k}=1,
\label{eq:4.9.91}
\end{equation}
which is the limiting value to which $\Lambda$ tends as $\tres\to\infty$.
Since the singularity in the integral representing $\tres$ is a simple pole
at $\Lambda=w_c(N)$, $\tres$ diverges logarithmically with $t$, in
contradistinction to the $N=\infty$ case, for which it saturates at one.
As for the constant kernel, we must distinguish between times for which the
simple pole at $w_c(N)$ dominates and earlier times., for which we are far
enough from the singularity for the finite sum to be well approximated by
the infinite sum.
\subsubsection{The Finite Product Kernel}
The exact solution for $\Kfin_3(k,l)$ is surprisingly simple to obtain
using the same approach as in the previous two cases. The final result is
given by:
\begin{eqnarray}
&&c_j(t)=\frac{a_j\Lambda^j}{t}\nonumber\\
&&t=\Lambda\exp\left[
\int_0^\Lambda\frac{\sum_{k=1}^Na_kz^{k-1}}{1-\sum_{k=1}^Na_kz^k}dz
\right]\label{eq:4.9.10}\\
&&a_j=\frac{j^{j-2}e^{-j}}{(j-1)!}\nonumber
\end{eqnarray}
Using these formulae, one sees that $\Lambda$ grows monotonically with time
starting from zero. If $N$ is infinite then, upon reaching the critical
value $e^{-1}$, $\Lambda$ will yield a finite value of the time, but will
not be able to grow beyond this value, since the integrand then has an
integrable singularity. This critical time is indeed the gel time and is,
in this case, equal to one. Beyond this time, the above equations do not
quite state unambiguously what happens, but a closer look at the derivation
shows that $\Lambda$ remains at its critical value of $e^{-1}$.
We have therefore rederived the Stockmayer solution in this way.

If, on the other hand, we take $N$ to be finite, the behaviour becomes
qualitatively quite different: indeed, the nearest singularity of the
integrand on the positive real axis is now a simple pole, since the
denominator is a polynomial. Since this polynomial approximates the
corresponding analytic function quite well, we see that this singularity
lies beyond the value $e^{-1}$ by an amount of order $1/N$. THerefore, as
time goes on, $\Lambda$ will approach ever nearer to this pole without ever
reaching it. It then follows that the finite aggregates (those having
$j\ll N$) are given for all times very nearly by the Stockmayer solution.
This therefore shows our claim, that the kernel $\Kfin_(k,l)$ yields in
the limit $N\to\infty$ the Stockmayer kinetics for the finite aggregates.
Note, on the other hand, that for times larger than $t_c$, the aggregates
of size of order $n$ will be significantly different from the
value predicted by the infinite model, since $\Lambda$ is distant by a
quantity of order $1/N$ from its theoretical value, thus implying a change
for those aggregates by a factor of order one.

Let us now look at the kernel $\Kfinbar_3(k,l)$. Since it is
bounded from above by $const.\cdot N(i+j)$ it
cannot display gelation, so  that, if one considers {\em all\/}
aggregates, mass must be conserved. From this follows that the kinetic
equations (\ref{eq:2.2}) for $\Kfinbar_3(k,l)$ for the aggregates of size
$1\leq j\leq N$ read
\begin{equation}
\dot c_j=\half\sum_{k=1}^{j-1}k(j-k)c_kc_{j-k}-jc_j,
\label{eq:4.9.11}
\end{equation}
which are manifestly the same as the Flory equations (\ref{eq:4.5.7}) as
pointed out by Bak and Heilman in \cite{bak94}.
They therefore have the same solution for the active clusters, so that the
solutions of (\ref{eq:4.9.11}) indeed straightforwardly tend to the Flory
solution.

We are hence in the highly peculiar situation that the limiting
dynamics depends on the sequence by which the infinite system is approached.
It is intuitively clear that this can only happen in a gelling system.
The issue at hand is that of the interaction between finite aggregates
(having a size $j\ll N$) and ``gel aggregates'', of size of order $N$. In
fact, it follows from a careful study of the global existence results
proved in \cite{ley81} that such a situation can never arise in the
situation studied there, namely whenever the kernel is bounded from above
as follows:
\begin{equation}
K(k,l)\leq r_kr_l\qquad(r_k=o(k)).
\label{eq:4.9.12}
\end{equation}
and, in fact, more generally whenever $\nu<1$ with $\nu$ defined as in
(\ref{eq:3.120}). The peculiar phenomena described above can therefore only
arise for a kernel with $\nu=1$ and for which gelation occurs. (It has
been shown by van Dongen \cite{don87d} that kernels with $\nu>1$ have
instantaneous gelation, and in fact their solutions may in fact perhaps not
exist at all). The intuitive reason for this is as follows: when $\nu<1$,
the reaction rates between a $j$-mer and an $N$-mer, with $N\gg j$ is of
order $N^\nu$. If we now average this reaction rate over all gel aggregates
and use the fact that the gel mass is finite, we see that the average
reaction rate tends to zero. It is therefore essential that $\nu=1$
for these pathologies to occur. This may perhaps diminish the sense of
urgency in finding a ``solution'' to these problems. The above results
do show, however, that no resolution can be expected from the consideration
of finite systems when $\nu=1$.
\subsection{Pure Three-body Reactions}
\label{subsec:3body}
It is fairly straightforward to generalize the methods described in
the preceding subsections to higher order reactions. Let us first
describe what has already been done in the literature. In
\cite{jia89c,jia89d} Jiang {\it et al}. described the behaviour of
the following $n$-particle kernel:
\begin{equation}
K(\mass_1,\ldots\mass_n)=\prod_{l=1}^n(A\mass_l+B)
\label{eq:4.1.2}
\end{equation}
These show gelation with exponents $z=-2$ and $\tau=2.5$.
Furthermore, the case of monodisperse initial conditions can be
solved exactly in this case as well. These results are obtained
using an extension of the techniques described in
\cite{ley81}. This agrees quite well with the overall picture given
in Section \ref{sec:scaling} (see in particular subsection
\ref{subsec:ext}),
that as far as scaling properties are concerned, many-body
reaactions do not involve major changes from binary, apart from
replacing $\lambda$ by $\lambda-(n-2)$. It can also straightforwardly
be verified that the constant kernel reaches scaling from arbitrary
initial conditions with $\tau=0$ and $z=1/(n-1)$. In these examples
we therefore do not see any variation from the expected picture.

We now turn to the two other examples given in (\ref{eq:3.1.10}) and
(\ref{eq:3.1.11}). These, as we stated before, show some significant
peculiarities.
$K_1$, in which the rates are given by $\mass_1+\mass_2+\mass_3$,
can be straightforwardly solved by an extension of the
techniques described for the sum kernel, see \cite{kra94}.
As usual, let us define the generating function:
\begin{equation}
F(\lap,t)=\int_0^\infty c(\mass,t)e^{\mass\lap},
\label{eq:4.1.201}
\end{equation}
one finds the following PDE for $F(\lap,t)$
\begin{eqnarray}
\partial_tF&=&\half[F^2-S(t)^2]F_\lap-S(t)F
\label{eq:4.1.202a}\\
S(t)&=&F(\lap=0;t)
\label{eq:4.1.202b}\\
f(\lap)&=&F(\lap;t=0)
\label{eq:4.1.202c}
\end{eqnarray}
These equations can again be solved via the characteristic function
technique, as discussed in appendix \ref{app:3body1},
with the following result
\begin{eqnarray}
F(\lap,t)&=&\frac{f(\lap_0)}{1+f(0)t}
\label{eq:4.1.203a}\\
\lap&=&\frac{\lap_0}{1+tf(0)}-\half f^{\prime\prime}(0)\lap_0^2.
\label{eq:4.1.203b}
\end{eqnarray}
{}From this, the solution for monodisperse initial
conditions can be obtained explicitly and is given by
\begin{eqnarray}
c_{2k+1}(t)&=&\frac{(2k+1)^{k-1}}{2^kk!}\frac{1}{1+t}
\left(
\frac{t}{1+t}
\right)^k\exp\left(
-\frac{2k+1}{2}\frac{t}{1+t}
\right)\nonumber\\
c_{2k}(t)&=&0.
\label{eq:4.1.3}
\end{eqnarray}
One finds that the typical size $s(t)$ grows as $t^2$.
This leads immediately to the following scaling function:
\begin{equation}
\Phi(x)=\frac{x^{-3/2}e^{-x/4}}{2\sqrt\pi}.
\label{eq:4.1.4}
\end{equation}
The remarkable result, however, is not in the scaling function, but
rather in the rate of growth of the typical size. Ordinarily, as we have
seen in many occasions, this quantity is not difficult to find out and
its value rests on very simple-minded, and hence very robust, scaling
arguments. Here, however, this is not the case: a perfectly
straightforward scaling theory predicts that $z=1$ for this kernel,
but one finds through an exact solution that $z=2$. It is also easy to
see, using the same kind of techniques that were developed for other
systems, that this scaling function is in fact valid for arbitrary
initial conditions. We see, therefore, that this is an example of
the phenomenon discussed earlier, namely in subsection \ref{subsec:ext}:
one indeed readily verifies that
the condition $\tau-\mu<1$, which is necessary for the validity of
the scaling equation (\ref{eq:3.1.5}) for three-body reactions is
here violated. Since the derivation of (\ref{eq:3.1.5}) fails because
of a divergence of the integrals involved, we see that the expression
for the typical size will also be given incorrectly: in fact, the
real typical size is larger than the one predicted by
(\ref{eq:3.1.2}). The following rough argument may help to understand
what is happening: if we cut the integrals appearing in (\ref{eq:3.1.5})
at a lower limit of $1/s(t)$, these become time dependent quantities,
diverging as $s(t)^{1/2}$, thus leading to a modified version of the
equation (\ref{eq:3.1.2}) for $s(t)$, namely
\begin{equation}
\dot s(t)\propto s(t)^{\lambda-1}s(t)^{1/2}\propto s(t)^{1/2},
\label{eq:4.1.5}
\end{equation}
which yields the correct large-time behaviour for $s(t)$.

It is also straightforward to verify that, for strongly polydisperse
initial conditions, that is, if $c(\mass,0)$ goes as $\mass^{-\alpha}$
with $2<\alpha<3$, then both $z$ and $\tau$ vary continuously with
$\alpha$ (see Appendix \ref{app:3body1}). This is reminiscent of similar
properties for the
(two-body) sum kernel. Nevertheless, the peculiar behaviour of $z$ is
much more striking in this three-body case, since we are dealing with
continuously variable exponents, not a continuously variable
exponential growth rate. Of course, since the
various exponents are undetermined by our general considerations,
there is no {\em a priori}
reason why they should not vary continuously with the
nature of the initial condition.

Let us now consider instead the kernel $K_2$ given by
\begin{equation}
K_2(k,l,m)=kl+km+lm.
\label{eq:4.1.6}
\end{equation}
In order to avoid tedious repetitions, let us use a simpler though
less complete method of solution: we write down the equations for the
various moments $\mom$ as follows
\begin{equation}
\dot{\mom}=\half\sum_{
\stackrel{p_1,p_2,p_3}
{p_1+p_2+p_3=p}
}\!\!\!\!\!\!{}^\prime\frac{p!}{
p_1!p_2!p_3!}\mom[p_1+1]\mom[p_2+1]\mom[p_3],
\label{eq:4.1.7}
\end{equation}
where the primes indicates that the sum runs only over those indices for
which no two of the $p_i$ simultaneously vanish. From this follows in
particular
\begin{eqnarray}
\dot{\mom[0]}&=&-\half\mom[0]\nonumber\\
\dot{\mom[2]}&=&\mom[2]^2\mom[0]+2\mom[2].
\label{eq:4.1.8}
\end{eqnarray}
These equations can be solved exactly. However, it is almost
immediately apparent, without any further work, that $\mom[2]$
diverges at finite time. This is confirmed by the exact solution,
which reads, for the case of monodisperse initial conditions:
\begin{equation}
\mom[2](t)=\frac{e^{2t}}{2-e^{t}}.
\label{eq:4.1.9}
\end{equation}
We therefore find that this kernel gels, although the straightforward
formula for $z$ leads us to expect an exponential
growth of the typical size. From this follows, incidentally, that a
theorem such as that proved in \cite{whi1} cannot be generalized in
the expected way to the pure three-body case.

Summarizing, we have given two examples for which the naive scaling
arguments give a value for the growth exponent $z$ different from
$1/(2-\lambda_3)$, which is the result of the straightforward scaling
approach. We may now ask: What happens in the more realistic case, in
which the three-body term perturbs a given two-body aggregation
process. In the next subsection, we shall consider the cases in which
the kernel $K_1$ perturbs the constant kernel, and the kernel $K_2$
perturbs the sum kernel. Thus in both cases we consider the case in
which the two-body reaction rate have the same growth exponent $z$ as
$1/(2-\lambda_3)$ for the three-body kernel. In other words, we ask
whether the naive value for the growth exponent may not play some
role when the three-body dynamics is taken together with a
corresponding two-body aggregation. As we shall see, the answer is (in
part) affirmative.
\subsection{Three-body Reactions Perturbing Two-body Aggregation}
\label{subsec:3body2}
Let us now consider the case in which both two and three-body
reactions occur. We therefore consider the system of equations
\begin{eqnarray}
\dot c(\mass, t)&=&\half\int_0^\infty d\mass_1d\mass_2\,K^{(2)}(\mass_1,
\mass_2)c(\mass_1,t)c(\mass_2,t)\times
\nonumber\\
&&\left[\delta(\mass_1+\mass_2-\mass)-\delta(\mass_1-\mass)-
\delta(\mass_2-\mass)\right]+
\\
&&+\frac{\alpha}{6}\int_0^\infty d\mass_1d\mass_2d\mass_3\,K^{(3)}(\mass_1,
\mass_2,\mass_3)c(\mass_1,t)c(\mass_2,t))c(\mass_3,t)\times
\nonumber\\
&&\left[\delta(\mass_1+\mass_2+\mass_3-\mass)-\delta(\mass_1-\mass)-
\delta(\mass_2-\mass)-\delta(\mass_3-\mass)\right]\nonumber
\label{eq:4.4.1}
\end{eqnarray}
Here we shall discuss the two following cases
\begin{equation}
\begin{array}{ll}
K^{(2)}(\mass_1,\mass_2)=1&K^{(3)}(\mass_1,\mass_2, \mass_3)
=\mass_1+\mass_2+\mass_3\\
K^{(2)}(\mass_1,\mass_2)=\mass_1+\mass_2
&K^{(3)}(\mass_1,\mass_2, \mass_3)
=\mass_1\mass_2+\mass_1\mass_2+\mass_2\mass_3
\end{array}
\label{eq:4.4.2}
\end{equation}
In Appendix \ref{app:3body2} we show in some detail using the method
of characteristics that the first case does indeed satisfy scaling
with the following exponents
\begin{equation}
z=\left\{
\begin{array}{lc}
1&\qquad(\alpha<\half)\\
\displaystyle\frac{4\alpha}{1+2\alpha}&\qquad(\alpha\geq\half)
\end{array}
\right.
\label{eq:4.4.3}
\end{equation}
This can also be seen more easily using the method of moments.
Indeed, one finds
\begin{eqnarray}
\dot{\mom[0]}&=&-(\half+\alpha)\mom[0]^2\nonumber\\
\dot{\mom[2]}&=&1+\alpha+2\alpha\mom[0]\mom[2].
\label{eq:4.4.4}
\end{eqnarray}
{}From this the values of $z$ stated in (\ref{eq:4.4.3}) follow
straightforwardly if we {\em assume} scaling instead of proving it.
This shows therefore that when the sum kernel perturbs a two-body
kernel, it behaves very much as if it had the value of $z$ predicted
by a naive scaling theory: that is, it does not affect the growth
exponent of a two-body kernel which itself has $z$ equal to one
unless the strength of the perturbation reaches a certain threshold
value. Above this threshold value, on the other hand, it does affect
the growth exponent, but only to the extent of making it continuously
variable, rather than imposing its own value. This last only happens
in the limit in which the three-body perturbation becomes infinitely
strong. Yet another peculiar feature of this model is the following: it is
an instance in which adding to a given aggregation mechanism, namely the
three-body kernel, a very strong second mechanism, namely the constant
kernel, actually leads to a slowing down of the first mechanism. This
should be compared with a conjecture framed in \cite{ley83} in which it was
hypothesized that if
\begin{equation}
K_1(\mass, \massp)>K_2(\mass, \massp)
\label{eq:4.4.45}
\end{equation}
for all $\mass$, $\massp$, then the typical size of the system
corresponding to $K_1$ would grow faster than the one corresponding to
$K_2$. Counterexamples were given in \cite{buf91}, but they involved
channeling particles towards non-reactive states. The above example,
although not strictly to the point since it involves ternary reactions,
certainly shows the need for great care in making such claims.

{}From the results of Appendix \ref{app:3body2} we also obtain the
scaling function $\Phi(x)$, albeit in a very implicit form. However,
this allows to estimate its small $x$ behaviour and hence the
exponent $\tau$. One obtains the result
\begin{equation}
\tau=\left\{
\begin{array}{lc}
1&\qquad(\alpha<1/2)\\
\frac{3}{2}&\qquad(\alpha>1/2)
\end{array}
\right.
\label{eq:4.4.5}
\end{equation}
Note that from this and the scaling
relation (\ref{eq:scaling-rel}) follows that, for $\alpha<1/2$,
$w$ is equal to
one. However, as is easy to see, for monomers (and more generally
speaking, whenever $j$ is held fixed and $t$ goes to infinity)
the decay exponent $\wpr$ is given by
\begin{equation}
\wpr=\frac{2+2\alpha}{1+2\alpha}.
\label{eq:4.4.6}
\end{equation}
We therefore see that in this model strong scaling is violated. It is
arguable whether this is  a more ``natural'' model than the $q$ model
as an example of this phenomenon. In the $q$ model, we require a
non-homogeneous kernel, whereas here we have a homogeneous two-body
kernel perturbed by a three body kernel of a degree of homogeneity
which, according to a {\em naive\/} scaling argument, should yield
the same growth rate. The type of violation involved is also, in a
sense, less drastic, since the $\wpr$ exponent is at least defined
independently of $j$.

On the other hand, writing down the moment equations for the sum kernel
perturbed by the three-body sum-of-products kernels yields a gelling
system: this is easily seen by examining the equations for $\mom[2](t)$
and $\mom[0](t)$, which only involve one another and can be solved
elementarily. Again, it is easy to verify that the solution for the second
moment diverges no matter how small $\alpha$ is. We see that the behaviour
is sometimes the one that would have been expected on naive grounds, but
that one cannot count on it.
\subsection{Constant Kernel with Production Term}
\label{subsec:prod}
The constant kernel with a time-independent production term
can also be solved exactly. The results are quite instructive, so
we present them here in some detail. The equations are
\begin{eqnarray}
\partial_tc(\mass, t)&=&\half\int_0^\infty
c(\mass_1)c(\mass_2)[\delta(\mass-\mass_1-\mass_2)-\nonumber\\
&&\delta(\mass-\mass_1)-\delta(\mass-\mass_2)]d\mass_1d\mass_2
+P\delta(\mass-\mass_0).
\label{eq:4.7.1}
\end{eqnarray}
For simplicity, we assume that no aggregates are present at
time zero. From this follows that only aggregates with
mass an integer multiple of $\mass_0$ are produced, which
allows to use the notation for discrete initial conditions.

Under these circumstances a straightforward application
of the generating function technique yields the following
solution: define
\begin{equation}
G(\lap,t)=\sum_{k=1}^\infty c_k(t)\left(
e^{k\lap}-1
\right).
\label{eq:4.7.10}
\end{equation}
One then obtains after a straightforward calculation
\begin{equation}
G=-\sqrt{2P(1-e^\lap)}\tanh\left(
\half t\sqrt{2P(1-e^\lap)}
\right).
\label{eq:4.7.11}
\end{equation}
Note how this function is meromorphic with a countable set of simple
poles on the positive real axis which crowd toward the origin as
$t\to\infty$. Since all singularities are simple poles, one finds that
$\explarge=0$, whereas, as we have seen before, $\tau=3/2$, thereby
violating the Dominant Singularity Hypothesis.

We now use the identity
\begin{equation}
\tanh x=x\sum_{l=-\infty}^\infty\frac{1}{x^2+(2l+1)^2\pi^2/4},
\label{eq:4.7.12}
\end{equation}
in order to find the solution
\begin{equation}
c_k(t)=\frac{\pi^2}{Pt^3}\sum_{l=-\infty}^\infty
(2l+1)^2\left[
1+\frac{
(2l+1)^2\pi^2
}
{
2Pt^2
}
\right]^{-(k+1)},
\label{eq:4.7.2}
\end{equation}
which yields the various results  stated in the previous
section in a rather straightforward way: indeed, if we let
$t\to\infty$ at {\em fixed\/} values of $k$, one obtains
a constant value, since (\ref{eq:4.7.2}) is then a Riemann sum
which can be replaced by an integral, namely
\begin{equation}
\lim_{t\to\infty} c_k(t)=\frac{\sqrt{2P}}{\pi}
\int_{-\infty}^\infty\frac{x^2}{(1+x^2)^{k+1}}dx=\sqrt{\frac{2P}{\pi}}
\frac{\Gamma(k-\half)}{\Gamma(k+1)}.
\label{eq:4.7.3}
\end{equation}
This confirms the general results that such models tend to  a
stationary value, as  well as the specific value for the
limiting concentrations. Indeed these are well-known to be given
by the second expression in (\ref{eq:4.7.3}).

However, from (\ref{eq:4.7.2}) one also finds the general scaling form
for large sizes and large times. This goes beyond the well-known results
for the stationary distribution, giving a fairly complete overview
of the dynamical behaviour. The result is readily obtained by noting that,
when $k$ is of the order of $t^2$, only the first few terms of the
infinite sum in (\ref{eq:4.7.2}) contribute, and that these
can be approximated accordingly. However, a simpler approach is the following:
remembering the expression (\ref{eq:3.801}) for the scaling function
$\Phi(x)$, we  obtain the following expression for the Laplace transform of
$\Phi(x)$:
\begin{equation}
\int_0^\infty dx\,\Phi(x)\left(
e^{-\rho x}-1
\right)=-\sqrt{2s}\tanh\sqrt\frac{s}{2}
\label{eq:4.7.4}
\end{equation}
from which it is straightforward to obtain the scaling function.
\subsection{The constant kernel with a localized source term}
Let us now consider the case in which one has the constant kernel with a
localized source term and a diffusion term, that is
\begin{eqnarray}
\partial_tc(\mass,\pos; t)&=&\half\int_0^\infty K(\mass_1, \mass_2)
c(\mass_1)c(\mass_2)[\delta(\mass-\mass_1-\mass_2)-\nonumber\\
&-&\delta(\mass-\mass_1)-\delta(\mass-\mass_2)]d\mass_1d\mass_2+\nonumber\\
&&+P\delta(\mass-\mass_0)\delta(\pos)+D\Delta c.
\label{eq:4.8.1}
\end{eqnarray}
We essentially follow the treatment given in \cite{che89}.
In this case, it is an obvious observation that the moments obey a closed set
of equations, namely
\begin{equation}
\dot{\mom}(\pos,t)=\half\sum_{q=1}^{\indmom-1}\left(
\begin{array}{l}
\indmom\\
q
\end{array}
\right)
\mom[q]\mom[\indmom-q]+P\mass_0^{\indmom}\delta(\pos)+D\Delta\mom.
\label{eq:4.8.2}
\end{equation}
These are diffusion equations with a recursively given inhomogeneity, so
they can clearly be solved iteratively. Again, however, the explicit
expressions rapidly become so unwieldy that they do not yield any
understanding of the system. It is nevertheless possible to gain some
insight by looking for scaling solutions of (\ref{eq:4.8.2}) as
follows: define
\begin{equation}
\mom(t)=t^{\expmom_\indmom}\Psi_\indmom\left(
\frac{|\pos|}{\sqrt t}
\right).
\label{eq:4.8.3}
\end{equation}
Matching powers on both sides of (\ref{eq:4.8.2}) and noting that
\begin{equation}
\int \mom[1](\pos,t)d\pos=P\mass_0 t,
\label{eq:4.8.4}
\end{equation}
which sets the value of $\expmom_1$ to $1-d/2$, one finally obtains
\begin{equation}
\expmom_\indmom=(2-d/2)\indmom-1
\label{eq:4.8.5}
\end{equation}
The fact that the $\expmom_\indmom$ increase as $\indmom$
is crucial for the approximations to be consistent: indeed, the
production term in (\ref{eq:4.8.2}) contributes to the total
mass $\mom[1](\pos,t)$, and is neglected for higher values of
$\indmom$. This is only valid for $d\leq4$, so that the upper crittical
dimension of this system is 4. Indeed, for $d>4$, it can be verified by a
similar power counting argument that the aggregation terms can be discarded
and the whole problem reduces to diffusion with a constant source. On the
other hand, as we shall see in Section \ref{sec:non-mft}, for dimensions one
and two the hypotheses of mean-field theory are bound to fail at some
stage. This does not mean that the theory developed here is never
applicable: if the reaction probability upon contact is very small, there
will be a large intermediate regime for which the theory just developed
will apply. However, eventually, it will fail, and the issues discussed
in Section \ref{sec:non-mft} become of central importance.

A considerable simplification occurs in (\ref{eq:4.8.1}) if the
time-dependence of the l.h.s. is neglected or, in other words, if we
consider the stationary approximation. This is in fact what was done in
\cite{che89}: it is shown there that the
$c(\mass, r)$ have a scaling form defined as follows:
\begin{equation}
c(\mass, r)=\mass^{-\tau}\Phi(\mass/r^\ztil)
\label{eq:4.8.6}
\end{equation}
where the exponents $\tau$ and $\ztil$ are given by
\begin{eqnarray}
\ztil&=&4-d\nonumber\\
\tau&=&\frac{d-6}{d-4}
\label{eq:4.8.7}
\end{eqnarray}
The reader is referred to the original paper for further details. In
particular, an exact solution can be obtained for the one-dimensional
mean-field case, as well as for the diffusion-limited case, for which
techniques analogous to those to be discussed in the next section are
necessary.


\section{Beyond Mean-field}
\setcounter{equation}0
\label{sec:non-mft}
As has been stated many times, the mean-field equations (\ref{eq:2.2}) are
only a valid description if no spatial correlations between the aggregates
arise. The reason for this requirement is clear enough. Implicitly, we have
always had the following picture in mind: aggregates, in order to grow,
must first be brought into contact by some transport mechanism, and then
react. Both the efficiency of the former and that of the latter may depend
in an arbitrary manner on the masses $\mass$ and $\massp$ of the two
reacting aggregates. The reaction rate matrix $K(\mass, \massp)$ is
therefore a product of a transport-related collision cross-section
$k(\mass, \massp)$ and a reaction efficiency upon contact given by some
matrix $\sigma(\mass, \massp)$. In this picture, however, we take it for
granted that those pairs of aggregates which the transport mechanism is
likely to bring together represent an unbiased sample of all aggregates,
in other words, that no correlations between aggregates exist on the scale
determining their transport.

This is, as we shall see in the following, a non-trivial assumption. More
exactly, we shall see that in the majority of physically relevant cases, it
is likely that this hypothesis cannot be sustained for large times. On the
other hand, in very many systems, there is a large regime of intermediate
times for which the mean-field theory described in the first part of this
paper is valid. This is due to the fact that, for many systems of
practical relevance, the limiting factor in the reaction is not the
transport mechanism itself, but rather the efficiency of reaction upon
contact. In that case, it is straightforward to see that, over the times
necessary to produce one single aggregation event, the transport mechanism
will have randomized the environment of the reacting pair so efficiently
that the mean-field assumption is unproblematic. If, on the other hand,
the transport mechanism is the rate-limiting factor, then matters are much
more difficult: as we shall see, the transport and the reaction mechanisms
may then collaborate to create correlations between particles, leading to a
situation in which mean-field theory fails. In the following, we shall
therefore not be concerned with solutions of (\ref{eq:2.2}) any more.
Rather, we shall consider specific models for irreversible aggregation, in
which a given transport mechanism is specified, as well as a reaction
mechanism. We then proceed to analyze the cluster size ditributions
generated by these models, comparing them with possible solutions of
(\ref{eq:2.2}) and using the scaling techniques developed in the rest of
this paper to characterize the behaviour of these distributions.
\subsection{Diffusion-limited Cluster-Cluster Aggregation}
\label{subsec:dlca}
The simplest model for irreversible aggregation, inspired from the physics
of colloids, for which the transport mechanism is Brownian diffusion, is
known as the Particle Coalescence Model (PCM) and goes as follows:
consider point particles diffusing on a lattice and carrying
an integer which we call their mass. When two particles,
of mass $\mass$ and $\massp$ respectively, alight on the same
site, they combine at a given rate $k$ to form a particle of mass
$\mass+\massp$. The  rate $k$ is often taken to be infinite, that is, the
formation of the new particle is instantaneous. The new particle is again
pointlike. On the other hand, the rate at which it diffuses may
depend on the mass $\mass$ as an arbitrary function $D(\mass)$, which
goes asymptotically as $\mass^{-\gamma}$.

Note in passing that this model already contains a considerable amount of
structure. Its main unrealistic fetaure is that the particles remain
pointlike. The  great advantage of this assumption is, that it frees us
from considering issues of morphology. Indeed, if we assume that clusters
connect rigidly when they approach within one lattice spacing of each
other, the model becomes much more complex: in particular, it then
generates fractal structures, and the radius of the cluster will depend in
a highly non-trivial way on the mass. Such a model, simulated in three
dimensions, already deserves to be called a realistic model of colloid
formation in the Brownian regime. However, since nearly nothing
can be said about it at the analytical level, I shall not treat it any
further, but refer the  interested reader to the extensive literature
(see e.g. \cite{jul87,mea98} and references therein).

The PCM, introduced by \cite{kan84} can be treated rigorously as
long as the diffusion constant is mass-independent. In this case,
see \cite{lig85} it has
been shown that for space dimension $d\geq3$ the model behaves qualitatively
like the mean-field model with a constant kernel. While this may seem
encouraging, we shall in fact see that this feature is intimately linked
to the pointlike nature of the particles. On the other hand, in two space
dimensions, logarithmic corrections appear,
whereas in one dimension an explicit exact solution is available
\cite{spo88a,spo88b}, which displays a number of highly anomalous
features.

For arbitrary diffusion constants, on the other hand, no such rigorous
treatment is available. Let us shortly argue what mean-field theory
we should expect to hold: as is well known, the current which flows
into an immobile absorbing sphere of radius $R$ surrounded by a sea of
particles diffusing with a diffusion constant $D$ is $DR^{d-2}$ for $d>2$.
{}From this we may plausibly argue that the rate at which one diffusing
particle encounters another is proportional to
$D_1+D_2$ as well as to $(R_1+R_2)^{d-2}$. Here
$R_i$ are the radii of the two particles and $D_i$ their diffusion
constants. Thus, if $R_i$ is always the same, we may disregard it and we
are eventually left with
\begin{equation}
K(\mass,\massp)=D(\mass)+D(\massp).
\label{eq:5.1.1}
\end{equation}
If the space dimension $d\leq2$, matters are more complex, since any two
particles are sure to react eventually. As we shall see, however, for
$d\leq2$, mean field theory is not expected to work at all, so the precise
form of the reaction coefficients is a rather academic issue.

At this point, it may be worthwhile to point out that it is occasionally
possible to mimic the results produced by correlations through an
appropriate choice of reaction term, different from quadratic. Here I
shall systematically disregard such possibilities: when a reaction
involves two bodies, I shall take the attitude that only a
quadratic expression can be viewed as an adequate reaction term
describing it.
\subsubsection{Simple Scaling Arguments}
In the following, I shall show how very straightforward arguments of the
kind that have been used throughout this paper, can be used to predict the
behaviour of the PCM. In the following subsection, I shall briefly present
the exact one-dimensional solution due to Spouge
\cite{spo88a,spo88b}, as well as shortly
mention the corresponding rigorous results concerning the two-dimensional
model, thus partly confirming the results to be derived now.

First let us derive the growth of the average cluster size. Let us assume
that a ``typical'' particle starts out by being a monomer and grows in
such a way as to be typical, that is, of size $s(t)$ all the time. If we
consider the probability $P(x,t)$ of finding the particle at $x$ at time
$t$ given that it was at the origin at time $t=0$, one has
\begin{eqnarray}
&&\frac{\partial P}{\partial t}=D\left[
s(t)
\right]\Delta P=s(t)^{-\gamma}\Delta P\nonumber\\
&&P(x,0)=\delta(x).
\label{eq:5.1.2}
\end{eqnarray}
{}From this follows that the probability of finding the particle at a
distance of order $L$ from the origin is of order one if
\begin{equation}
L=O\left(
\sqrt{D\left[
s(t)
\right]t}
\right)=O(t^{(1-z\gamma)/2}).
\label{eq:5.1.3}
\end{equation}
Therefore we expect the particle to have covered a space of order $L$ in
time $t$. Since originally the total mass in this interval was of order
$L$, and it did not change except through diffusion at the boundaries,
which can be neglected, we expect the typical size to grow as $L$, or in
other words
\begin{equation}
z=\frac{1}{2+\gamma}
\label{eq:5.1.4}
\end{equation}

Now let us turn to the values of $\tau$ and $w$. At this (low) level of
sophistication, we shall assume that strong scaling holds throughout and
shall have no qualms in identifying the
decay exponent for monomers with
$w$. We may picture the monomer decay problem as follows: each monomer is
surrounded by two particles of typical size. These grow normally by
coalescing with other particles outside the interval which connects them
and which contains the monomer. These particles' growth in no way affects
the monomer's survival. The problem therefore reduces to a three particle
problem, where the two outer particles have a time-dependent diffusion
constant, whereas the middle particle diffuses normally. The question is
then to determine the decay law for the survival probability of the middle
particle. This can then be identified with the monomer concentration decay
at large times. This problem is somewhat technical and was studied in
\cite{hel02}, with the following result: if $\gamma<0$, that is if the
aggregate diffusion constant grows with size, there is a fair probability
that the two large particles will diffuse quite far away, letting the
middle particle survive for quite long times. The result then is given by
\begin{equation}
c_1(t)=const.\cdot t^{-2/(2+\gamma)}\qquad(t\to\infty)
\label{eq:5.1.5}
\end{equation}
For $\gamma=0$, on the other hand, a straightforward solution is known
\cite{kan84,fis84} which yields
\begin{equation}
c_1(t)=const.\cdot t^{-3/2}\qquad(t\to\infty).
\label{eq:5.1.6}
\end{equation}
If we now apply (\ref{eq:scaling-rel}) to the values of $w$ obtained
from (\ref{eq:5.1.5}) and (\ref{eq:5.1.6}), one obtains
\begin{equation}
\tau=\left\{
\begin{array}{cc}
0&\qquad(\gamma<0)\\
-1&\qquad(\gamma=0)
\end{array}
\right.
\label{eq:5.1.7}
\end{equation}
If we now compare this with the strict mean-field prediction for which we
use the kernel (\ref{eq:5.1.1}), which is really the only one with a more
or less fundamental significance for the PCM, we obtain for the mean-field
growth exponent $z_{MF}$
\begin{equation}
z_{MF}=\frac{1}{1+\gamma}.
\label{eq:5.1.8}
\end{equation}
With respect to $\tau_{MF}$ our knowledge is less complete, since kernel
(\ref{eq:5.1.1}) is of type II, for which no closed expression for $\tau$
is available. Nevertheless we know from (\ref{eq:3.22}) that it satisfies
the inequality
\begin{equation}
\tau_{MF}\leq1-\gamma.
\label{eq:5.1.9}
\end{equation}
Both (\ref{eq:5.1.8}) and (\ref{eq:5.1.9}) are, of course, incompatible
with our scaling theory.

At this stage, one might nonetheless wonder whether some appropriate
modification of the equations could possibly describe the system. From the
value of $z$ one deduces that the degree of homogeneity $\lambda$ should be
equal to $-\gamma/2$. The fact that $w\neq1$ excludes a type I kernel, so we
are left with the possibility of type II. From the inequality
(\ref{eq:3.22}) we see that $\gamma>\tau$, which is indeed always
satisfied.
But in fact, there is a far more serious objection to such attempts: let
us compute the probability of finding a particularly large cluster.
This is clearly related to the possibility that any given particle might
go particularly far in time $t$.  This probability decays as a Gaussian,
and not as an exponential. But we have seen (see
subsection \ref{subsubsec:large}),
that under very general hypotheses, the large size behaviour of a cluster
size distribution is {\em always} given by an exponential. We see
therefore that even in this apparently exceptionally simple case, no simple
adaptation of mean-field will give an adequate description of the system's
behaviour at large times.

Let us now turn back to the case in which $\gamma>0$. Then the two
particles surrounding the monomer diffuse apart at an ever decreasing rate,
leading thereby to very rapid trapping of the central monomer. More
specifically, one has
\begin{equation}
c_1(t)=const.\cdot\exp\left[
-const.\cdot t^\beta
\right]\qquad(t\to\infty)
\label{eq:5.1.10}
\end{equation}
The determination of $\beta$ involves severe analytical difficulties.
Hell\'en {\em et al.} \cite{hel02}
suggest the following approximate value obtained from
a Lifshitz tail argument
\begin{equation}
\beta=\frac{1+\gamma}{4+2\gamma}
\label{eq:5.1.11}
\end{equation}
Again this is in qualitative agreement with the Smoluchowski picture, but
in clear quantitative disagreement: indeed, for $\gamma>0$, kernel
(\ref{eq:5.1.1}) is of type III, so that we expect small monomers to decay
as stated in (\ref{eq:5.1.10}), with $\beta$ given by
\begin{equation}
\beta=\gamma,
\label{eq:5.1.12}
\end{equation}
in clear contrast to (\ref{eq:5.1.11}). The authors in \cite{hel02} also
state unambiguously that even fairly coarse numerical measurements
invalidate the mean-field prediction (\ref{eq:5.1.12}) quite clearly.
\subsubsection{An Exact Solution}
In the one-dimensional case, if the diffusion constants of the aggregates
are size-independent, that is, if $\gamma=0$ in the notation of the
previous subsection and if, additionally, reaction between aggregates is
instantaneous upon contact, an exact solution was found by Spouge in
\cite{spo88a,spo88b}. Here I wish to give an outline of the way in
which this solution can be obtained, as well as to point out that it does,
in fact, confirm the results suggested by the previous, admittedly coarse,
scaling analysis. We shall, however, follow a slightly different approach
in the spirit of that pioneered by Doering and ben-Avraham
\cite{doe88a,doe88b}.

It is clear, by the definition of the model described above, that we are
dealing with a Markov process. This implies that one can write down a master
equation for the joint probability distribution function at time $t$
for finding the first particle at $x_1$, the second at $x_2$, and so
on to the $N$-th, where the $x_k$ represent the lattice position at time
$t$ of the particle $k$. This equation, however, is far too complex to be
of any use, and also contains more information than is really needed.
The standard approach consists, as is well-known, in taking correlation
functions involving one or more particles. In this case, this leads to an
intractable set of equations, in which each correlation function couples
to one of higher order. If, on the other hand, we consider a different
kind of reduced description, which takes the particular one-dimensional
nature of the system fully into account, one can indeed find a closed set
of equations. This goes as follows: define $E(k,x)$ as the probability of
finding a total mass $k$ contained in an interval of length $x$. The
crucial simplifying feature about this particular choice of correlation
function is, that it remains unchanged by all reactions that take place
either inside or outside the relevant interval. In other words, it is only
affected by events in which particles move inside or outside the given
interval via diffusion. And this, understandably, can be solved.

Let us translate the above verbal arguments into equations. First define
$F(k,x)$ to be the probability of having a total mass $k$ inside a given
interval of length $x$, and additionally that the leftmost site of this
interval be occupied, irrespectively of the state of the rightmost site.
In this case, one has
\begin{equation}
F(k,x)=E(k,x)-E(k,x-a),
\label{eq:5.2.1}
\end{equation}
where $a$ is the lattice spacing, which also indicates the length which
particles jump in one diffusion step.

For simplicity we now restrict ourselves
to a continuum limit, in which the particles are far apart on the scale
$a$. In this case, we may discard the possibility of both ends of the
interval being occupied, so we have
for the time variation of $E(k,x)$
\begin{eqnarray}
\partial_tE(k,x)&=&F(k,x+a)-F(k,x)\nonumber\\
&=&D\Delta E(k,x),
\label{eq:5.2.2}
\end{eqnarray}
where $D$ denotes the macroscopic diffusion constant arising from the
continuum limit.  In the derivation of
(\ref{eq:5.2.2}) we have made essential use of the fact that sites cannot
be doubly occupied, which follows from the instantaneous reaction rate.
We now need initial and boundary conditions for
(\ref{eq:5.2.2}). One finds, after some careful considerations involving
the original discrete model, that
\begin{eqnarray}
&&E(k,0)=0\nonumber\\
&&\left.\partial_xE(k,x)\right|_{x=0}=c_k(t)
\label{eq:5.2.3}
\end{eqnarray}
for all times $t$. Furthermore, if the initial particle density is $\dens$,
one has
\begin{equation}
E(k,x;t=0)=\frac{(\dens x)^k}{k!}e^{-\dens x},
\label{eq:5.2.4}
\end{equation}
at least, if the monomers are initially distributed at random\footnote{All
the above restrictions can, and have, been lifted. Again, for greater
details see \cite{spo88a,spo88b}.}. It is now straightforward to solve
for (\ref{eq:5.2.2}) with initial conditions (\ref{eq:5.2.4}) and the
absorbing boundary condition at the origin (\ref{eq:5.2.3}). Using the
method of images yields
\begin{equation}
E(k,x)=\frac{\dens^k}{2k!\sqrt{\pi Dt}}\int_0^\infty
y^ke^{-\dens y}\left[
e^{-(x-y)^2/(4Dt)}-e^{-(x+y)^2/(4Dt)}
\right]
\label{eq:5.2.5}
\end{equation}
and hence
\begin{equation}
c_k(t)=\frac{\dens^{-2}}{2k!\sqrt{\pi}(Dt)^{-3/2}}
\int_0^\infty x^{k+1}e^{-x}\exp\left(
-\frac{x^2}{4D\dens^2t}
\right)dx
\label{eq:5.2.6}
\end{equation}
Analytically it is not quite straightforward to evaluate this expression, or
even to show that it scales. In terms of generating functions, however,
this is easy and is shown in Appendix \ref{app:scaling-pcm}. The
resulting scaling function $\Phi(x)$ is given by
\begin{equation}
\Phi(x)=\frac{4xe^{-x^2}}{\sqrt\pi}
\label{eq:5.2.7}
\end{equation}
and the typical size indeed grows as $\sqrt{Dt}$. We have therefore
confirmed that the decay at large $x$ of the scaling function is faster
than exponential, thereby precluding any identification with a mean-field
model of the type we have been discussing in the body of this paper. We
also show rigorously that the $\tau$ and $w$ values are in fact given by
the values predicted in the previous subsection, namely $-1$ and $3/2$.
Finally, we may note that (\ref{eq:5.2.6}) shows that strong scaling
holds in this system: indeed, if $t\to\infty$ in the r.h.s. of
(\ref{eq:5.2.6}) while $k$ remains fixed, the integral becomes
time-independent and one finds
\begin{equation}
\lim_{t\to\infty}\left[
(Dt)^{3/2}c_k(t)
\right]=\frac{\dens^{-2}(k+1)}{2\sqrt{\pi}}
\label{eq:5.2.8}
\end{equation}

What have we learned from this example? Primarily we see that in one
dimension the mean-field approach fails, essentially because of the buildup
of correlations between particles due to the
interplay between reaction and diffusion. In particular, one
finds that the probability of finding two unreacted aggregates nearby is
considerably smaller than expected on the assumption that the reactants are
uncorrelated: thus, if we ask for the probability of finding exactly
one monomer in each half of an interval of length $x$, that is
$E(2,x)-2E(2,x/2)$, one finds it to decrease as $x^3$ as $x\to0$, whereas the
square of the probability of finding a monomer in such an interval goes
as $x^2$. There is, therefore, a strong repulsion between nearby
particles, due to the efficiency with which diffusion  in one dimension
brings nearby particles to react.

It should be pointed out that we have assumed instantaneous reaction rate
whenever particles meet. Without this assumption, the model cannot be
solved exactly as far as is known. However, it is easy to realize that in
the limit where the reaction rate $k$ becomes very small, the transport
mechanism becomes irrelevant and mean-field theory holds. However, if
particles have an initial separation of order $L$, they will on average
meet $L$ times in a time of order $L^2$, that is, in a time sufficient for
them to come close at all. This follows from well-knnown properties
of one-dimensional diffusion. Therefore, if $kL$ is of order one,
the problem reduces to the one in which reaction is essentially certain.
Since the typical interparticle distance $L(t)$ is growing with time, we
should say that when $L(t_c)k$ is of order one, the results obtained
above should start to hold, whereas before the description using the
constant kernel should be valid. Numerical work does indeed confirm this
expectation. The time $t_c$ can be evaluated under the assumption that for
$t=t_c$. $L(t)$ can be evaluated both using the formulae of the constant
kernel as those of the diffusive model. One then has
\begin{eqnarray}
L(t_c)&=&\left(
\sum_{k=1}^\infty c_k(t)
\right)^{-1}=const.\cdot kt_c\nonumber\\
&=&const.\cdot(Dt_c)^{1/2}
\label{eq:5.2.9}
\end{eqnarray}
which implies that $t_c$ goes as $D/k^2$. One sees therefore that it may be
possible to apply the mean-field theory quite meaningfully, even in a
system for which we know that it cannot be applied in the true asymptotic
regime.

A similar reasoning leads one to conjecture that the mean-field description
for the PCM with $\gamma=0$ becomes accurate in dimensions $3$ and higher.
This is due to the fact that random walks are transient in this case,
that is, they visit each point only a finite number of times. One does not
expect, therefore, that the correlations created by the diffusion and the
reaction process should build up indefinitely in time. Indeed, these
expectations have been confirmed: specifically, it has been shown
(see e.g. \cite{lig85}) that the cluster size distribution can in this case
be scaled onto an exponential distribution, just as for the constant kernel,
which is the one we expect to model this particular system.

Finally, I believe this system also shows very nicely the extent to which
the concepts originally developed in the context of the
mean-field equations can be carried over to more general situations: thus
the basic scaling relationship (\ref{eq:scaling-rel}) connecting
$\tau$, $w$ and $z$  follows from mass conservation alone and holds in all
cases. Similarly, the distinction between power-law polydisperse and bell
shaped scaling functions also generalizes, as we have seen in the previous
subsection.
\subsection{Ballistic Aggregation}
\label{subsec:ball}
Another simple model of aggregation is the following: consider particles
of a given radius $a$ originally spread at random in ${\mathcal{R}}^d$ with
random velocities. These then move freely, and stick irreversibly upon any
two coming closer to each other than the capture radius $a$. In order to
specify the model completely, we require two additional pieces of
information: first, how does the radius grow upon irreversible sticking?
Here many solutions are in principle possible. We will focus entirely on
the two following: either the radius does not grow, or else it grows
deterministically in such a way as to maintain the total aggregate volume
constant. The second is of course more realistic, but considerably more
difficult to discuss, as well as more controversial. Second, we need to know
how the velocity of the compound particle is obtained. Since we are
physically assuming that this represents sticking of freely moving
particles, we shall always assume that momentum conservation determines the
final velocity, which it does in a unique way. Since the two particles
combine to one inelastically, we have, of course, no conservation of energy.

Let us first consider the one-dimensional case. The two growth rules for
the radius are then equivalent. This model was first considered by
Carnevale {\em et al.} \cite{car90} and later investigated further by
Jiang and Leyvraz \cite{jia93} using numerical simulations together
with qualitative scaling arguments. We present these first, and later
discuss shortly an elegant exact solution \cite{mar94,fra00}, which
confirms the results anticipated
in \cite{jia93} in a very satisfactory way.
\subsubsection{Simple Scaling Arguments}
\label{subsec:ball1}
As we have seen in the case of the diffusive one-dimensional model, it
is often a good strategy to search for the long time behaviour of the
typical size and the long-time decay of very small clusters, since this
contains essentially most of the relevant information. In \cite{car90} it
was shown that the typical size in one-dimension grows as $t^{2/3}$ using
the following argument: an aggregate of mass $\mass$ has arisen, in the
discrete picture, out of $\mass$ monomers. These all had momenta of order
one, which averaged, as vectors, to zero. If we therefore assume that they
are essentially independent, we are led to assume that the momentum
$p(\mass)$ of
the $\mass$-mer is of order $\mass^{1/2}$, which leads to a velocity
$v(\mass)$ of the order of $\mass^{-1/2}$. This leads to a typical
aggregate of size $s(t)$ crossing in a time $t$ a length of order
$v[s(t)]t$, which is of order $t/\sqrt{s(t)}$. Since the number of
particles initially contained in this interval will be of the same order
as $s(t)$, we wind up with
\begin{equation}
s(t)=const.\cdot(\dens t)^{2/3},
\label{eq:5.3.1}
\end{equation}
where $\dens$ again denotes the initial particle concentration.

This reasoning is quite robust, but does make an assumption concerning the
initial velocity distribution function (VDF), namely that it has no large
velocity power-law tails. Indeed, the argument which
invokes the central limit theorem fails, and $s(t)$ grows in a
different way when such tails are present.
Indeed, one finds \cite{jia93} that if the VDF has a
distribution with a behaviour of the type
\begin{equation}
p(v)=const.\cdot v^{-\alpha-1}\qquad(1<\alpha<2)
\label{eq:5.3.2}
\end{equation}
then one has using an obvious extension of the above reasoning, that the
exponent $z$ is given by
\begin{equation}
z=\frac{\alpha}{2\alpha-1}.
\label{eq:5.3.3}
\end{equation}
The assumption $\alpha>1$ is necessary, since otherwise the typical
velocities would grow on average, making the whole approach
questionable. No meaningful numerical results have been obtained for this
case, whereas (\ref{eq:5.3.3}) has been well confirmed numerically
for the range $1<\alpha<2$ \cite{jia93}.

To complete our scaling argument, we should find the asymptotic decay for
very small clusters. A completely general lower bound on $c_1(t)$ can be
derived, quite independently of the initial VDF, in the following way:
consider the aggregates at time $t$. It is possible, of course, to
identify the positions at time zero of the monomers conforming them. These
fall in consecutive intervals of typical length $s(t)$ separated by
intervals of length one. If both the following events
occur, the initial particle will survive until time $t$,
\begin{enumerate}
\item The particle should first have been placed on one of the empty
regions between two of the intervals constituting the two neighbouring
typical clusters. This has probability $1/s(t)$.

\item The particle should have been launched with a velocity small enough
that it will not reach either of the two neighbouring clusters in time $t$.
Since the two clusters are eventually separated by a distance $s(t)$,
this limiting velocity is of the order $s(t)/t$. If the VDF has a finite
probability density for velocity zero, this leads to a probability of the
order $s(t)/t$ as well.
\end{enumerate}
Combining the two probabilities one obtains
\begin{equation}
c_1(t)\geq  const.\cdot t^{-1}.
\label{eq:5.3.4}
\end{equation}
It further appears quite unlikely that monomers could survive under any
other circumstances than the ones stated above, so that (\ref{eq:5.3.4}) is
presumably the exact order of magnitude of the monomer decay. Note in
passing that we have assumed that the VDF is finite at zero velocities.
What happens, for example, if all initial velocities are taken with
values $\pm1$? At first sight, this changes a lot, since odd clusters
always have a lower (mass-dependent) bound on their velocities, whereas
even ones always have a finite probability of having velocity exactly zero.
Therefore, for a {\em fixed} even value of the mass one has
$c_{2j}(t)$ going
asymptotically as $1/s(t)$, which differs from (\ref{eq:5.3.4}). Similarly,
for fixed odd values of the mass, we get exponential decay of
$c_{2j+1}(t)$ for large times. However,
it is easy to see that in the scaling limit the continuum approximations
can in fact be made, so that the scaling exponent $w$ is $1$ in this
case also.

{}From this evaluation of $w$ follows via (\ref{eq:scaling-rel})  that
\begin{equation}
\tau=\left\{
\begin{array}{cc}
1/\alpha&\qquad(1<\alpha<2)\\
\half&\qquad(\alpha>2),
\end{array}
\right.
\label{eq:5.3.5}
\end{equation}
where the usual case is contained in the second. Numerically, these claims
have been verified\footnote{Persistent findings stating that $\tau=0$ in
the case of velocities with no power-law tail have repeatedly appeared
in the literature. Since they are now found to be contrary to exact results
\cite{mar94,fra00}, it may be enough to say here that they are erroneous.}
in \cite{jia93}.

Now let us compare with mean-field theory. To this end, we need to develop
a plausible expression for the reaction rates $K(\mass,\massp)$. A common
and frequently used one for related problems (see e,g, \cite{drake}),
is the following:
\begin{equation}
K(\mass, \massp)=|v(\mass)-v(\massp)|\left[
R(\mass)+R(\massp)
\right]^{d-1},
\label{eq:5.3.6}
\end{equation}
where $v(\mass)$ is the velocity of a typical aggregate of mass $\mass$
and $R(\mass)$ its radius. Using the standard approach outlined in this
paper, one finds for the growth exponent $z$ in the case where $v(\mass)$
goes as $\mass^{-1/2}$ and $R(\mass)$ constant, that $z$ is in fact $2/3$
as expected. On the other hand, in this case, the kernel (\ref{eq:5.2.6})
is of type III, so that the monomer concentration $c_1(t)$ should go as a
stretched exponential, specifically
\begin{equation}
c_1(t)=const.\cdot\exp\left(
-const.\cdot t^{1/3}
\right)
\label{eq:5.3.7}
\end{equation}
It is, of course. quite evident what is going wrong here: in
(\ref{eq:5.3.6}) we are assuming that monomers always react as if they were
moving at a velocity of order one, whereas it is clear that the velocity
of the surviving monomers is a decreasing function of time, due to fast
monomers being selectively eliminated
by the reaction. In fact, it is seen numerically that
the velocity profile $v(\mass)$ does not decay with mass, but is on the
contrary rather flat, decaying as $s(t)^{1/2}$ with {\em time}. There is
also (somewhat inconclusive) evidence \cite{jia93} that (\ref{eq:5.3.7})
actually holds for very small reaction rates, so one may in fact argue
that one is using the ``correct'' mean-field theory.

A more sophisticated approach involves considering the joint distribution
of mass and momentum. In this case we may view the vector $(\mass, p)$ as
a vector valued mass, in a way similar to the one in which we treated
multicomponent aggregation. One then finds for the reaction kernel
\begin{equation}
K(\mass, p;\massp, p^\prime)=\left|
\frac{\mass}{p}-\frac{\massp}{p^\prime}
\right|\left[
R(\mass)-R(\massp)
\right]^{d-1}.
\label{eq:5.3.8}
\end{equation}
Since the $p$ are on average zero, we are in the same situation as when
discussing the scaling of composition. However, it is easy to see that, at
least within the treatment of composition scaling that we have given, the
behaviour of small clusters is essentially the same. The gaussian behaviour
of the momenta is justified in detail by our approach. On the other hand,
it should be admitted that we have not shown that these are the only
possible solutions to the multicomponent scaling equations
(\ref{eq:3.2.4}).
\subsubsection{Exact Results}
Here I shortly review the exact solution derived in
\cite{mar94,fra99,fra00,fra01}. The precise form of the solution is quite
formidable and its derivation is an impressive feat indeed. I shall not,
however, go into the details, which are rather technical, and for
which the interested reader is referred to the original literature.

In order to state the result in a readily understandable form, let me
define some notation. I denote by $\trans(x_1.y_1;x_2,y_2)$ the following
probability: Let $y(x)$ be a random function of $x$ chosen according to the
Wiener measure, or, in other words, a Brownian motion
directed along the $x$ axis. $\trans$ is then
the probability that it satisfy
\begin{eqnarray}
&&y(x_1)=y_1\nonumber\\
&&y(x_2)=y_2\nonumber\\
&&y(x)<x^2\qquad(x_1<x<x_2)
\label{eq:5.4.1}
\end{eqnarray}
In other words, it is the probability that a directed random walk start
at $(x_1, y_1)$ and reach $(x_2,y_2)$ without having ever touched the
``absorbing parabola'' $y=x^2$.
This can also be descibed as the solution to the following partial
differential equation
\begin{equation}
\frac{\partial\trans(x_1,y_1;x,y)}{\partial x}=\frac{\partial^2
\trans(x_1,y_1;x,y)}{\partial y^2}
\label{eq:5.4.2}
\end{equation}
with the boundary conditions
\begin{eqnarray}
\trans(x_1,y_1;x_1,y)&=&\delta(y-y_1)\\
\trans(x_1,y_1;x,x^2)&=&0
\label{eq:5.4.3}
\end{eqnarray}
{}From this one now defines a few auxiliary functions: first I introduce
$\transit(x_1,y_1)$, which is the probability of reaching infinite values
of $x$ without having been absorbed by the parabola. This is given by
\begin{equation}
\transit(x_1,y_1)=\lim_{x\to\infty}
\int_{-\infty}^{x^2}\trans(x_1,y_1;x,y^\prime)
dy^\prime.
\label{eq:5.4.4}
\end{equation}
Finally, we shall be interested in the relative probabilities of escape
of particles that start very near to the parabola. To this end we define
\begin{eqnarray}
I(x_1,x_2)&=&\left.
\frac{\partial^2}{\partial y_1\partial y_2}\trans(x_1,y_1;x_2,y_2)
\right|_{y_1=x_1^2;y_2=x_2^2}
\label{eq:5.4.5}\\
J(x_1)&=&\left.
\frac{\partial}{\partial y_1}\transit(x_1,y_1)
\right|_{y_1=x_1^2}
\label{eq:5.4.6}
\end{eqnarray}
The final result can now be stated as follows: the concentrations
$c(\mass, v, t)$ obey the following scaling law
\begin{equation}
c(\mass, v, t)=t^{-1}\Phi(x,\eta).
\label{eq:5.4.7}
\end{equation}
Here $x$ and $\eta$ represent the following scaling variables
\begin{eqnarray}
x&=&\frac{\mass}{(2\sigma t)^{2/3}}\nonumber\\
\eta&=&(2\sigma t)^{1/3}v,
\label{eq:5.4.8}
\end{eqnarray}
where $\sigma$ is the r.m.s. velocity in the initial VDF. One
then finds for $\Phi(x,\eta)$
\begin{equation}
\Phi(x,\eta)=I\left(-\frac{x}{2},\frac{x}{2}\right)
J\left(
\frac{\sigma x-\eta}{2\sigma}
\right)
J\left(
\frac{\sigma x+\eta}{2\sigma}
\right)
\label{eq:5.4.9}
\end{equation}
Several points should be noted about this result: first, the mass and
velocity dependence do not factorize, nor do they show the simple kind of
interdependence we have found in multicomponent aggregation. In fact, it is
explicitly shown in \cite{fra99} that the collision probability between
two aggregates does not factorize. In this way, it is clear that any
attempt to build a mean-field theory is ill-founded, and indeed such
attempts have failed to reproduce known results.

The derivation of this exact solution is very interesting in its basic
ideas, but quite laborious in the details. In Appendix \ref{app:ball-exact},
I therefore attempt to present the concepts to the extent that a sufficiently
enthusiastic reader should be able to fill in the missing algebra.

One can then use (\ref{eq:5.4.2}) to evaluate $\trans$, and hence all other
functions defined above,
in terms of Airy functions. This opens the way to an asymptotic analysis
of $\Phi(x,\eta)$, and of its integral over $\eta$. This is a laborious
undertaking, which was performed in the above references, where
they confirm the scaling results stated in \cite{jia93}. Additionally, an
unexpected large $x$ asymptotic is obtained, namely
\begin{equation}
\int_{-\infty}^\infty \Phi(x,\eta)d\eta=const.\cdot\exp\left(
-x^3/12
\right)\qquad(x\gg1)
\label{eq:5.4.10}
\end{equation}
However, it is not clear whether this might not be an artefact of initial
conditions in which all particles are put on a lattice: it seems incredible
that normal density fluctuations when particles are put at random on the
line should still satisfy (\ref{eq:5.4.10}). In any case, however, this
large $x$ behaviour once more shows how impossible it is to treat this
model in mean-field terms.
\subsubsection{Higher Dimensions}
The case of higher dimensions is both far more complex and, to some
extent, controversial. First, it is obvious that it is important, in this
case, to specify how the particle radii grow upon aggregation. We shall
consider two possibilities:
\begin{enumerate}
\item The particle radius remains fixed when the particles stick together.
This is obviously unrealistic, but it has the merit of greater simplicity.
In particular, there is a real hope that mean-field theory might hold in
$d\geq2$.

\item The particle radius grows so as to maintain the total $d$
dimensional aggregate volume constant. Since the total occupied volume
fraction remains always finite, this is a really difficult model. Thus,
even three-particle collisions cannot be discarded out of hand as  being
asymptotically irrelevant. I shall have little to say on that count.
\end{enumerate}
Consider first the case of fixed radius. We may then attempt to formulate
once more the mean-field theory. In this case expression (\ref{eq:5.3.6})
for the reaction kernel still holds, and leads to
\begin{equation}
z=\frac{2}{3}
\label{eq:5.5.1}
\end{equation}
independently of $d$. The monomer decay predicted by (\ref{eq:5.3.7}),
namely a stretched exponential with exponent $1/3$, still holds
independently of $d$. This is in fact quite natural, since the reaction
kernel defined by (\ref{eq:5.3.6}) only depends on $d$ via the geometric
cross-section, which we have assumed constant. In this sense, kernel
(\ref{eq:5.3.6}) is much more general than the previous analysis indicated.

Is this mean-field theory likely to be correct? Since the total number
concentration decays as $t^{-2/3}$, the mean free path grows as
$t^{2(d-1)/3}$, which is always more than the interparticle distance
$t^{2/(3d)}$ when
$d>1$. This is therefore extremely encouraging, since it shows that
typically colliding particles will be distant on the scale of nearest
neighbour separation. In related work it has
been shown that the Boltzmann equation becomes exact at large times
\cite{pia02}
for the model of $d$ dimensional ballistic
annihilation when $d>1$. A numerical confirmation of all this would be of
considerable interest, but has not, to the best of
my knowledge, been performed to date.

The model in which the radii grow as $\mass^{1/d}$, on the other hand, is
far more complex. The obvious mean-field theory has again a reaction
kernel of the form (\ref{eq:5.3.6}), which has values of $\lambda$ and
$\mu$ given by
\begin{eqnarray}
\lambda&=&\frac{d-2}{2d}\nonumber\\
\mu&=&-\half
\label{eq:5.5.2}
\end{eqnarray}
from which one deduces in the ordinary way that
\begin{eqnarray}
z&=&\frac{2d}{d+2}
\label{eq:5.5.3}\\
c_1(t)&=&const.\cdot\exp\left[
-const.\cdot t^{d/(d+2)}
\right]
\label{eq:5.5.4}
\end{eqnarray}
There is therefore again a stretched exponential decay of the monomers
according to this theory. This is again due to the fact that we do not
take into account the selective elimination of high velocity monomers in
this model. But since, in this model, the mean free path of a monomer
always remains of the order of the interparticle distance (due to the
conservation of the occupied volume fraction), it appears very likely that
such an elimination will in fact take place.

Trizac and Hansen have performed some high quality numerical simulation on
the two dimensional version of
this model \cite{tri95}. These lead apparently to predictions quite at
variance with the above theory. However, it should be borne in mind that
the data were never explicitly analyzed in terms of the theory
described here, but rather compared with a rather
peculiar ``mean-field'' theory \cite{pia92}, which includes, however,
some correlation effects by means of various uncontrolled
approximations. This theory's principal merit was to provide a rationale
for the observation of a ``pure exponential''  cluster size distribution,
together with $\tau=0$, which was then thought to be the correct result in
one dimension \cite{car90}. The agreement of the two-dimensional
simulations performed in \cite{tri95}
with the theory presented here is therefore rather difficult to assess.

Nevertheless, let us us shortly discuss the most salient
differences between the simulation \cite{tri95} and theory. The
measurement of the $z$ exponent\footnote{called $\xi$ in \cite{tri95}}
yields a result which is claimed to be significantly different from the
theoretical value of one. The figure published does not, however, show a
very convincing straight line behaviour over any range of sizes. The best
that can be said is that strong crossover effects are certainly present,
which make it difficult to say with certainty what the actual value of $z$
is. The quoted value of $0.8$ (for the dilute system, which is the one of
greater theoretical interest, as well as the one for which deviations from
theory are largest) is really quite far from the expected value, however,
so I feel strongly that a more careful analysis of the data is required
before any definite statement concerning the validity of mean-field theory
can be made.

No measurements either of the monomer decay or of the $\tau$ exponent were
made. It is therefore not possible
to decide the crucial issue whether the stretched exponential decay predicted
by (\ref{eq:5.5.4}) indeed takes place. The approximately exponential
appearance of the cluster size distribution does, however, make  it highly
unlikely.

On the other hand, the average kinetic energy per particle was measured
and yielded a power law decay $t^{-\delta}$ {\em incompatible} with
the prediction of mean-field theory. Indeed, from the mean-field arguments
given above one obtains
\begin{equation}
\delta=z,
\label{eq:5.5.5}
\end{equation}
whereas the quoted value for the dilute case is $1.12$. This is
certainly a serious disagreement with the measured value of $z$.
If we take this result seriously, it suggests that the cancellation
between velocities of colliding particles leads to velocities far less
than what would be implied by the Central Limit Theorem. The intriguing
suggestion that this might be due to the kinematics of collision was made
in \cite{tri01}. A considerable amount of work clearly remains to be done
to settle these important issues.
\section{Conclusions and Outlook}
I have attempted to give a comprehensive overview of the way in which
scaling concepts can be used in the context of the kinetics of irerversible
aggregation. We have seen that applications are numerous and the method is
quite powerful in predicting, among others, various
qualitative behaviours both for the shape of the cluster size distribution
and for the time dependence of the rescaling factors involved, such as
the typical size or length. There remain nevertheless several shortcomings
to the method, which might perhaps be overcome and lead to a technique
that could be more quantitative even for fairly complicated systems. Here
are some aspects which I believe to be of interest.
\begin{itemize}
\item Once one has an integral equation for the scaling function
$\Phi(x)$, difficulties are by no means over. It is then necessary, in
practical applications, to solve it, generally by some numerical approach.
This turns out to be a difficult problem. The various forms into which the
basic equation (\ref{eq:3.7}) can be cast, due to the presence of an
arbitrary function $f(x)$, might possibly be of use. On the other hand,
for numerical work, the original form of the equation as given in
(\ref{eq:3.13}) is presumably optimal, since it contains the desired function
in the r.h.s. of the equation, whereas the corresponding relations
involving exponentials yield objects such as the Laplace transform of
$\Phi(x)$, which is definitely less direct.

The issue of computing scaling functions becomes particularly acute in
cases such as crossover or inhomogeneous systems, for which the function
$\Phi(x,y)$ has two variables and the integral equation is complicated.
In these cases, even the behaviour of the function at the
origin is not straightforward. In this respect, it would be of interest to
have a theory of comparable simplicity as that of regular variation for
functions of two or more variables. It is not clear that this is an
attainable goal, however, as functions in two variables can exhibit a
strikingly complex set of singular behaviours.
\item I believe crossover may be of considerable use in practical
applications, since a competition of two reaction mechanisms is  a
frequent occurence. The progress reported here, while possibly of interest,
is certainly only a first step: more exactly solved models exhibiting
crossover should be looked into and those for which I have not been able to
give solutions, such as the $q$ sum kernel in the limit $q\to1$, should be
viewed as relevant problems. Further, the integral equation has almost
yielded no results of interest on the form of the two-variable scaling
function $\Phi(x,y)$. To obtain a theory yielding the form of the
singularity at least in some limiting cases is of crucial importance. It
must be admitted, however, that the existing exact solutions do
not hold out much hope for broad ranges of simple behaviour.
\item Exact solutions for models with production and diffusion are, to the
best of my knowledge, absent except for some work relying on the
simplifying features arising from stationarity. I believe any such model
would be of interest, since we have shown that the stationary limit in
this case does not arise in an easy way from the non-stationary case.
\item Are there cases in which the scaling function $\Phi(x)$ can be
obtained exactly, though the kernel itself be not solvable? This would
increase considerably our understanding of aggregation, in particular if we
could obtain rigorous results for some type of kernel of type III or type I,
for which, so far, no analytical results are known.
\item Can one go any further than I have done towards elucidating the
singularity of the scaling function $\Phi(x)$
for the gelling case? It should be possible to
obtain some kind of estimates on $\tau$, perhaps by using methods similar
to those developed by Cueille and Sire for Case II kernels
\cite{cue97,cue98}. From the mathematical viewpoint this is certainly
important. On the other hand, from the point of view of applications, it
may well be enough to compute such exponents to sufficient accuracy. To
this end, however, it would again be essential to have a good way to solve
the equation, in spite of the appearance of the unknown exponent
$\tau$ in it.
\item There also remains a certain mystery around the Case I
non-gelling kernels. These are quite important, yet there remains some
doubt concerning the very existence of a scaling function for this case.
The numerical work of Lee \cite{lee01} certainly makes a very strong case
for existence, but numerical evidence, of course, is never entirely
conclusive on its own. This work seems to point to a highly
anomalous patterns of corrections to the leading power-law behaviour, that
are oscillatory and might, in point of fact, violate the hypothesis of
regular behaviour near the origin. A different mathematical framework
should be designed in order to derive such results as are strongly
suggested by the numerics.
\item
Finally, I believe corrections to scaling may be of interest: by this I do
not mean the corrections to the leading power-law of the scaling function:
this has already been done, at least for Case II in \cite{don88}. There
is, however, another issue, hinted at in \cite{boe98}, namely that scaling
behaviour is not reached immediately. Thus, initial conditions may create
broad transients, which should also be treated at least to some
approximation. In \cite{boe98} the consequences of having an initial
condition containing both monomer and $N$-mer are worked out explicitly.
More work along this direction would be desirable, as this is not at all
explored until now.
\end{itemize}
There also remain several mathematical problems related to the theory of
the Smoluchowski equations. First and foremost, I would mention the
necessity of obtaining optimal results for the system of equations
describing coupled diffusion and aggregation. I believe present results,
which only concern diffusion constants bounded from below for large
masses, are probably not as strong as they could be made. Finally, several
old problems are still there: proving the scaling hypothesis under fairly
general hypotheses is one. An issue which may also be related concerns the
relation between the stochastic models of aggregation and the
deterministic Smoluchowski equations. In the framework of van Kampen's
$\Omega$ expansion, very remarkable results have in fact been obtained.
A corresponding rigorous result was proved in \cite{nor99}, which may well
pave the way to future progress.
\begin{ack}
I would first like to thank Itamar Procaccia for encouraging me to write
this article. I have also greatly benefitted from helpful conversations
with F. Calogero, P.L. Krapivski, H. Larralde, S. Redner and E. Trizac.
The financial support of DGAPA
project IN112200 and CONACyT 32173-E is also gratefully acknowledged.
\end{ack}

\appendix
\section{Derivation of the Scaling Equation for $\Phi(x)$}
\setcounter{equation}0
\label{app:deriv-scaling}
The fact that the $c(\mass,t)$ approach a scaling form, as stated in
(\ref{eq:3.7}), can be reformulated as follows:
\begin{equation}
\lim_{t_1, t_2\to\infty}\int_{t_1}^{t_2}\frac{d}{dt}\left[
\int_0^\infty \mass f[\mass/s(t)]c(\mass,t)
\right]=0
\label{eq:a.1}
\end{equation}
We now rewrite the integrand using (\ref{eq:2.2}) as follows
\begin{eqnarray}
&&\frac{s(t)^{-1}}{2}\int d\mass_1\,d\mass_2\,\mass_1\mass_2 c(\mass_1,t)
c(\mass_2,t)
K(\mass_1, \mass_2)G\left[
\mass_1/s(t),\mass_2/s(t)
\right]-\nonumber\\
&&\qquad-\frac{\dot s(t)}{s(t)}\int d\mass\,\mass \frac{\mass}{s(t)}
f^\prime\left[
\mass/s(t)
\right].
\label{eq:a.2}
\end{eqnarray}
Here $G(x,y)$ is given by
\begin{equation}
G(x,y)=\frac{1}{xy}\left[
(x+y)f(x+y)-xf(x)-yf(y)
\right]
\label{eq:a.3}
\end{equation}
If we now assume that we can replace $K(\mass_1, \mass_2)$ by its
continuous limiting form $s(t)^\lambda\rate[\mass_1/s(t),\mass_2/s(t)]$%
---something we shall justify shortly---then (\ref{eq:a.2}) is of the
same form as the l.h.s of (\ref{eq:3.7}), since $G(x,y)$ is bounded
whenever $f(x)$ is bounded and differentiable.
It therefore follows that (\ref{eq:a.2}) can be recast as
\begin{eqnarray}
&&\frac{\constsep^2s(t)^{\lambda-1}}{2}\int_0^\infty dx\,dy \rate(x, y)
\Phi(x)\Phi(y)\left[
(x+y)f(x+y)-xf(x)-yf(y)
\right]-\nonumber\\
&&\qquad-\constsep\frac{\dot s(t)}{s(t)}\int_0^\infty
dx\,x^2f^\prime(x)\Phi(x).
\label{eq:a.4}
\end{eqnarray}
If this, integrated over $t$, should give zero, as
stated in (\ref{eq:a.1}), then surely one must have
\begin{equation}
\dot s(t)=\constsep_1 s^\lambda,
\label{eq:a.5}
\end{equation}
for some constant $\constsep_1$. This allows, in principle, to
determine $s(t)$ completely, including the prefactor.
We find
\begin{equation}
s(t)=\left[(1-\lambda)\constsep_1t+s(0)^{1-\lambda}\right]^{1/(1-\lambda)}
\label{eq:a.6}
\end{equation}
Note that some difficulties arise when $\lambda=1$, and that the
whole scheme becomes inconsistent when $\lambda>1$. The former
difficulties are somewhat technical and will be treated in another
part of this paper. The second involve, as we shall see, convergence
to a scaling form of the second moment rather than the first, and
therefore require different considerations.

The determination of $\constsep_1$ rests on issues which
strongly depend on conventions and personal preferences, such as the
exact way in which typical size is defined. We therefore assume
in the following that $\constsep_1$ has been somehow determined, and
proceed to set the constant $\constsep$, which we had previously left
free, equal to $\constsep_1$. One can then take $\constsep^2$
out of the whole expression in (\ref{eq:a.4}). One further deduces,
from (\ref{eq:a.1}), that (\ref{eq:a.4}) must vanish, yielding
(\ref{eq:3.10}) after observing that
\begin{eqnarray}
G(x,y)&=&\frac{1}{y}\left[
f(x+y)-f(x)
\right]+
\frac{1}{x}\left[
f(x+y)-f(y)
\right]\nonumber\\
&=&G_1(x,y)+G_1(y,x).
\label{eq:a.7}
\end{eqnarray}
Since the rest of the integral is symmetric, we may limit ourselves,
up to a factor of two, to integrating over $G_1(x,y)$.

The one point that is left concerns the justification of
substituting $K(\mass_1, \mass_2)$ by
\begin{equation}
s(t)^{-\lambda}\rate[\mass_1/s(t), \mass_2/s(t)]
\label{eq:a.8}
\end{equation}
Let us first consider the case in which $f(x)$ vanishes identically
on some interval of the form $[0,\epsilon]$.
In this case, the statement is obvious: as
$s(t)\to\infty$, the masses summed over in \ref{eq:a.2}) become
increasingly large, and the convergence of $K$ to $\rate$ as stated
in (\ref{eq:3.11}) is enough to justify the substitution.

We have therefore shown, that (\ref{eq:3.10}) must hold for all
functions $f(x)$ which vanish  on an interval near the origin.
The general statement is
an immediate consequence of this together with the monotone convergence
theorem.
\section{Derivation of the Scaling Equation in the Three-Body Case}
\label{app:deriv-scaling-3}
\setcounter{equation}0
{}From the point of view of the algebraic manipulations involved, the
three-body case is wholly similar to the corresponding two-body
computation, and the reader is referred to appendix
\ref{app:deriv-scaling} for notation and the basic computations. The
issue we shall be addressing here concerns not so much the nature of
the manipulations involved, but their admissibility.

Up to (\ref{eq:a.2},\ref{eq:a.3}), everything is strictly analogous to
the case of appendix \ref{app:deriv-scaling}. $G$ is now a function
of three variables given by
\begin{equation}
G(x,y,z)=\frac{(x+y+z)f(x+y+z)-xf(x)-yf(y)-zf(z)}{2xyz}.
\label{eq:e.1}
\end{equation}
The crucial difference between the two and three body cases now
apppears in the following fact: $G(x, y, z)$ is not a bounded
function, no matter how $f(x)$ is chosen, unless it be a constant.
Indeed, choose $x$ and $y$ so that
\begin{equation}
(x+y)f(x+y)\neq xf(x)+yf(y).
\label{eq:e.2}
\end{equation}
It then immediately follows that $G(x,y,z)\to\infty$ as $z\to0$. The
only possibility left is, therefore, to {\it assume} that things will
go well. If we then go through the computations that correspond to
those given in appendix \ref{app:deriv-scaling}, we obtain
\begin{equation}
\dot s(t)=\constsep_1s^{\lambda_3-1},
\label{eq:e.3}
\end{equation}
which is the result stated in (\ref{eq:3.1.2}). We may then, as in
Appendix \ref{app:deriv-scaling}, set $\constsep_1=\constsep^2$ and
therefore obtain the result stated in (\ref{eq:3.1.2}).
There is, however, no guarantee that this derivation goes through.
It is, in particular, quite possible that the integral
\begin{equation}
\int dx\,dy\,dz\,xyz\,K(x,y,z)\,G(x,y,z)\,\Phi(x)\Phi(y)\Phi(z)
\label{eq:e.4}
\end{equation}
diverges near the origin. This then indicates, from a physical
viewpoint, that processes in which one very small particle coalesces
with two  particles of typical (but not equal) sizes affect the
scaling of the aggregation process in a singular way. I shall not
attempt to make a scaling theory of such cases, merely limiting
myself to stating when the results obtained by the above theory are
self-consistent, that is, when the above mentioned divergences near
the origin do not occur. It should be clear
that all the results discussed in this appendix carry over
straightforwardly to the general case of a reaction of order
$\order\geq3$.

We shall see in subsection
\ref{subsec:3body} that various exactly solved cases fall under
this category. Further, we shall see that in this case, it is not
always possible to neglect a two-body contribution which shoud be
negligible on naive grounds. Thus, the kernel $\mass_1+\mass_2
+\mass_3$ has a naive growth exponent $z=1$. However, the real
growth exponent in the pure three-body case is $z=2$. Nevertheless,
if we perturb the constant kernel, which also
has $z=1$, using this three-body kernel,
the combined process has a growth exponent of $z=1$, that is,
the expected dominance of the model with the higher value of
$z$ does not take place. For larger values of the perturbation, on
the other hand, the exponents vary continuously.

\section{The Large-$\rho$ Behaviour of $I(\rho)$}
\setcounter{equation}0
\label{app:large-rho}
In the following, we shall always assume that $\Phi(x)$ has regular
behaviour near the origin in the sense of (\ref{eq:3.16}) and that the
corresponding power law is given by $\tau$. We now divide
$I(\rho)$ in two parts:
\begin{eqnarray}
I_1&=&\int_0^\infty dx\,\int_0^{C/\rho} dy\,\rate(x,y)\Phi(x)\Phi(y)x
e^{-\rho x}\left[
1-e^{-\rho y}
\right]\nonumber\\
I_2&=&\int_0^\infty dx\,\int_{C/\rho}^\infty dy\,\rate(x,y)\Phi(x)
\Phi(y)xe^{-\rho x}\left[
1-e^{-\rho y}
\right].
\label{eq:b.1}
\end{eqnarray}
I now treat $I_1$ and $I_2$ separately. For $I_1$, a change of
variables yields
\begin{equation}
I_1=\rho^{-(\lambda+3)}\Phi\left(\frac{1}{\rho}
\right)^2\int_0^\infty dx\, x^{1-\tau}e^{-x}\int_0^C dy\,\rate(x,y)y^{-\tau}
\left[
1-e^{-y}.
\right].
\label{eq:b.2}
\end{equation}
Here the substitution of $\Phi(y/\rho)/\Phi(1/\rho)$ by $y^{-\tau}$
is justified by the fact that the range of integration of $y$ is
finite. A similar justification for $x$ exists always, that is, both
for $I_1$ and $I_2$. Should this yield divergent integrals, one would
have to conclude that the regularity assumption was not
self-consistent.

For $I_2$, we only rescale $x$ by $\rho$ and obtain using
(\ref{eq:3.105}) and (\ref{eq:3.110})
\begin{equation}
I_2=\rho^{-2}\Phi\left(\frac{1}{\rho}
\right)\int_0^\infty dx\,x^{1-\tau}e^{-x}\int_{C/\rho}^\infty
dy\,y^\lambda k\left(\frac{1}{\rho y}\right)\Phi(y).
\label{eq:b.3}
\end{equation}
Assuming further that $k(z)$ behaves regularly (in the sense
of (\ref{eq:3.16}) again) with exponent $\mu$ (see (\ref{eq:3.115}))
one finds the final expression
\begin{equation}
I_2=\rho^{-2}\Phi\left(\frac{1}{\rho}
\right)k\left(\frac{1}{\rho}\right)
\int_0^\infty dx\,x^{1-\tau}e^{-x}\int_{C/\rho}^\infty
dy\,y^{\lambda-\mu}\Phi(y).
\label{eq:b.4}
\end{equation}
Now let us consider the issue of convergence of these various
integrals. It is obvious by the above calculation that the sum
$I_1+I_2$ cannot depend on the constant $C$. One sees, moreover, that
the integral appearing in the expression for $I_1$ converges exactly
when the one appearing in $I_2$ diverges. The limit separating these
cases is
\begin{equation}
1+\lambda-\mu-\tau=0.
\label{eq:b.5}
\end{equation}
Whichever integral diverges is therefore compensated by the
correction to scaling generated by the finite cutoff in the
converging integral. The expression containing the convergent
integral must therefore always dominate. We may therefore write
\begin{equation}
I=\const_1\rho^{-(\lambda+3)}\Phi\left(\frac{1}{\rho}
\right)^2+\const_2\rho^{-2}\Phi\left(\frac{1}{\rho}
\right)k\left(\frac{1}{\rho}\right),
\label{eq:b.6}
\end{equation}
where $\const_1$ and $\const_2$ are given by
\begin{eqnarray}
\const_1&=&\int_0^\infty dx\, x^{1-\tau}e^{-x}\int_0^\infty
dy\,\rate(x,y)y^{-\tau}
\left[
1-e^{-y}.
\right]\nonumber\\
\const_2&=&\int_0^\infty dx\,x^{1-\tau}e^{-x}\int_0^\infty
dy\,y^{\lambda-\mu}\Phi(y)=\Gamma(2-\tau)\int_0^\infty
dy\,y^{\lambda-\mu}\Phi(y).
\label{eq:b.7}
\end{eqnarray}
(\ref{eq:b.6}) must be here understood in the sense that only the
term for which the integral in the corresponding prefactor converges
should be taken into account.

Finally, let us consider the case in which $\Phi(x)$ decays faster than any
power near the origin. In this case, one can neglect the contribution due
to $e^{-\rho y}$ in the expression for $I(\rho)$ given in (\ref{eq:3.12}).
It can therefore be rewritten as
\begin{equation}
\int_0^\infty dx\,x^{1+\lambda}\Phi(x)e^{-\rho x}\int_0^\infty
dy\,k\left(
\frac{y}{x}
\right)\Phi(y).
\label{eq:b.8}
\end{equation}
But one has $y\gg x$ always, since $x$ must be of order $1/\rho$
and $y$ cannot be small, due to the behaviour of $\Phi(y)$ at the origin.
One can thus use the following approximation for $k(z)$
\begin{equation}
k(z)=\pref z^\nu[1+o(1)]\qquad(z\to\infty)
\label{eq:b.9}
\end{equation}
from which the result stated in the text readily follows.
\section{The Large-$\rho$ Behaviour for the Three-body Reaction}
\setcounter{equation}0
\label{app:large-rho3}
One follows exactly the same strategy as in appendix
\ref{app:large-rho} to evaluate the various integrals involved. The
result is, with exactly the same notations, and the various
homogeneity exponents defined as in (\ref{eq:3.1.4}):
\begin{eqnarray}
\const_1&=&\int_0^\infty dx\int_0^Cdy\,dz\,x^{1-\tau}e^{-x}
y^{-\tau}z^{-\tau}
K(x,y,z)\left[
1-e^{-(y+z)}
\right]\nonumber\\
\const_2&=&2\int_0^\infty dx\,x^{1-\tau+\lambda_2}e^{-x}
\left(\int_0^Cdy\,y^{-\tau}k_1(y/x)\right)\left(
\int_{C/\rho}^\infty dz\,z^{\lambda_3-\lambda_2}\Phi(z)\right)\\
\const_3&=&\int_0^\infty dx\,x^{1-\tau}e^{-x}\int_{C/\rho}^{\infty}
dy\,dz\,z^{\lambda_2-\mu}\pref_1(y/z)\Phi(y)\Phi(z)\nonumber
\label{eq:c.1}
\end{eqnarray}
\section{The Distribution of Compositions in Multicomponent
Aggregation}
\setcounter{equation}0
\label{app:multi}
In this appendix we show how the compositions scale in the
multicomponent case. To this end we require first a scaling equation
of the same type as (\ref{eq:3.10}) or (\ref{eq:3.2.4}) for the
function $\Psi(\suma, \difer)$. Again, we use the same approach as in
Appendix \ref{app:deriv-scaling} and obtain from (\ref{eq:3.2.1})
\begin{eqnarray}
&&\int_0^\infty d\suma\int_{-\infty}^\infty d\difer\,\difer\left[
\suma f_{\suma}(\suma, \difer)+\alpha\difer f_{\difer}
\right]\Psi(\suma, \difer)=\nonumber\\
&&\qquad\int_0^\infty d\suma_1\,d\suma_2\int_{-\infty}^\infty
d\difer_1\,d\difer_2\,K(\suma_1, \suma_2)\Psi(\suma_1, \difer_1)
\Psi(\suma_2, \difer_2)\times\nonumber\\
&&\qquad\times\left[
(\difer_1+\difer_2)f(\suma_1+\suma_2,\difer_1+\difer_2)-\difer_1
f(\suma_1,\difer_1)
-\difer_2f(\suma_2,\difer_2)
\right].
\label{eq:k.1}
\end{eqnarray}
I now substitute for $f$
\begin{equation}
f(\suma, \difer)=\frac{\difer}{s^\alpha}
\exp(-\rho\suma-iq\difer/\suma^\alpha).
\label{eq:k.2}
\end{equation}
I now take (\ref{eq:3.2.5}) as an {\em ansatz} to be substituted into
(\ref{eq:k.1}) in order to determine the unknown functions $\Phi_1$
and $\chi_\alpha$. This leads to the equation
\begin{eqnarray}
&&-\rho\int_0^\infty d\suma\,\suma^{\alpha+1}
\Phi_1(\suma)e^{-\rho\suma}\int_{-\infty}^\infty d\difer\,
\chi_\alpha(\difer)e^{-iq\difer}=\nonumber\\
&&\qquad\int_0^\infty d\suma_1\,d\suma_2\,\int_{-\infty}^\infty
d\difer_1\,d\difer_2\,\,K(\suma_1, \suma_2)\Phi_1(\suma_1)
\Phi(\suma_2)\chi_{\alpha}(\difer_1)\chi_{\alpha}(\difer_2)
\times\nonumber\\
&&\qquad\times\bigg[
(\suma_1+\suma_2)^\alpha e^{-\rho(\suma_1+\suma_2)}\exp\left(
-iq\frac{\difer_1\suma_1^\alpha+\difer_2\suma_2^\alpha}{(\suma_1+
\suma_2)^\alpha}
\right)-\nonumber\\
&&\qquad-\suma_1^\alpha e^{-\rho\suma_1-iq\difer_1}
-\suma_2^\alpha e^{-\rho\suma_2-iq\difer_2}
\bigg].
\label{eq:k.3}
\end{eqnarray}
For this equation to hold, we need that the r.h.s. show a
factorization in the dependence on $q$, since the l.h.s. does. For
this to be the case, we need
\begin{equation}
\int_{-\infty}^\infty
d\difer_1d\difer_2\,
\chi_{\alpha}(\difer_1)\chi_{\alpha}(\difer_2)
\exp\left(
-iq\frac{\difer_1\suma_1^\alpha+\difer_2\suma_2^\alpha}{(\suma_1+
\suma_2)^\alpha}
\right)
=\int_{-\infty}^\infty d\difer\,\chi_{\alpha}(\difer)
e^{-iq\difer}.
\label{eq:k.4}
\end{equation}
As is readily seen, however, this equation is satisfied by the stable
symmetric L\'evy distributions and by these only. Indeed, the Fourier
transform $\hat\chi_{\alpha}(q)$ of $\chi_{\alpha}(\difer)$ must satisfy
the functional equation
\begin{equation}
\hat\chi_{\alpha}(q)=\hat\chi_{\alpha}\left(
\frac{\suma_1^\alpha}{(\suma_1+\suma_2)^\alpha}q
\right)
\hat\chi_{\alpha}\left(
\frac{\suma_2^\alpha}{(\suma_1+\suma_2)^\alpha}q
\right),
\label{eq:k.5}
\end{equation}
for which the only solution is $\exp(-|q|^{1/\alpha})$.
\section{The Integral Equation for the Gelling Case}
\setcounter{equation}0
\label{app:gel-equation}
In this Appendix we derive the basic equation to be satisfied by the
scaling function $\Phi(x)$ when convergence to scaling occurs in the sense
of the second moment only, as is, for example, characteristic of the
situation in which gelation takes place. We follow exactly the same steps
as in Appendix \ref{app:deriv-scaling} and refer the reader to this
Appendix for more detailed explanations.

The definition we shall use of convergence to scaling is the one given
in (\ref{eq:3.8.2}). In order to ontain the equation for $\Phi(x)$ we shall
again start from an obvious consequence, namely
\begin{equation}
\lim_{t_1,t_2\to\infty}\int_{t_1}^{t_2}I(t^\prime)=0,
\label{eq:r.1}
\end{equation}
where we have defined
\begin{equation}
I(t)=\frac{d}{dt}\left[
\mom[2](t)^{-1}\int_0^\infty d\mass\,c(\mass, t)f\left(
\frac{\mass}{s(t)}
\right)
\right]
\label{eq:r.2}
\end{equation}
Let us now evaluate $I(t)$:
\begin{equation}
I(t)=\constsep^2A(t)s(t)^{\lambda-2}\mom[2](t)-\constsep B(t)
\frac{\dot{\mom[2]}(t)}{\mom[2](t)}-\constsep C(t)\frac{\dot{s(t)}}{s(t)}.
\label{eq:r.3}
\end{equation}
Here the $A(t)$, $B(t)$ and $C(t)$ are defined by
\begin{eqnarray}
A(t)&=&\frac{1}{2\mom[2]^2}\int_0^\infty
d\mass_1d\mass_2\,\mass_1^2\mass_2^2\,K\left(
\frac{\mass_1}{s(t)},\frac{\mass_2}{s(t)}\right)
\bigg[
\left(\frac{\mass_1+\mass_2}{s(t)}
\right)^2f\left(\frac{\mass_1+\mass_2}{s(t)}
\right)-\nonumber\\
&&-\left(
\frac{\mass_1}{s(t)}
\right)^2f\left(
\frac{\mass_1}{s(t)}\right)-\left(
\frac{\mass_2}{s(t)}
\right)^2f\left(
\frac{\mass_2}{s(t)}
\right)
\bigg]\frac{s(t)^4}{\mass_1^2\mass_2^2}\nonumber\\
B(t)&=&\int_0^\infty\mass^2c(\mass, t)f\left(
\frac{\mass}{s(t)}
\right)
\label{eq:r.4}\\
C(t)&=&\int_0^\infty \mass^2c(\mass, t)f^\prime\left(
\frac{\mass}{s(t)}
\right)\frac{\mass}{s(t)}.\nonumber
\end{eqnarray}
Because of the scaling hypothesis, the quantities $A(t)$, $B(t)$ and $C(t)$
all approach limits which can be expressed as follows in terms of the
scaling function $\Phi(x)$:
\begin{eqnarray}
A&=&\int_0^\infty dx\,dy\,xK(x,y)\Phi(x)\Phi(y)\left[
(x+y)f(x+y)-xf(x)
\right]\nonumber\\
B&=&\int_0^\infty dx\,x^2\Phi(x)f(x)
\label{eq:r.5}\\
C&=&\int_0^\infty dx\,x^3\Phi(x)f^\prime(x).\nonumber
\end{eqnarray}
In order to go further, we must distinguish the gelling case and the
non-gelling case with $\lambda=1$. Let us first consider the former.
It is then reasonable to postulate that the various
quantities involved such as $\mom[2](t)$ diverge as a power-law as the
critical time $t_c$ is reached. One thus finds that
\begin{equation}
\frac{\dot{\mom[2]}(t)}{\mom[2](t)}=\frac{C_1}{t_c-t}\qquad
\frac{\dot{s}(t)}{s(t)}=\frac{C_2}{t_c-t}
\label{eq:r.6}
\end{equation}
Assuming that the remaining term of (\ref{eq:r.3}) (proportional to $A$)
also has the same order of magnitude yields
\begin{equation}
s(t)^{\lambda-2}\mom[2](t)=\frac{C_3}{t_c-t}
\label{eq:r.7}
\end{equation}
We now need to evaluate the different constants. An appropriate choice
of $\constsep$ allows to eliminate everything, save the ratio $C_1/C_2$,
which enters via the relation
\begin{equation}
\mom[2](t)=const.\cdot s(t)^{C_1/C_2}.
\label{eq:r.8}
\end{equation}
We therefore need to evaluate this exponent. To this end we make the
following (not really rigorous) observation: from the scaling hypothesis
follows
\begin{equation}
\int_{\epsilon s(t)}^\infty \mass\,c(\mass, t)d\mass=
\frac{\mom[2](t)}{s(t)}\int_\epsilon^\infty x\Phi(x).
\label{eq:r.9}
\end{equation}
The r.h.s of (\ref{eq:r.9}) varies as $\epsilon^{2-\tau}\mom[2](t)/s(t)$.
If, however, we allow $\epsilon$ to go as $1/s(t)$, then the l.h.s should
be constant. From this we deduce
\begin{equation}
\frac{C_1}{C_2}=3-\tau.
\label{eq:r.10}
\end{equation}
For large times, (\ref{eq:r.1}) reduces to
\begin{equation}
\lim_{t_1,t_2\to\infty}\int_{t_1}^{t_2}\frac{dt^\prime}{t^\prime}\left(
A-(3-\tau)B-C)
\right)=0
\label{eq:r.11}
\end{equation}
from which follows
\begin{equation}
A-(3-\tau)B-C=0
\label{eq:r.12}
\end{equation}
which is exactly (\ref{eq:3.8.5}) as stated in the text.
Further, from (\ref{eq:r.7}), (\ref{eq:r.8}) and (\ref{eq:r.10}) follows
straightforwardly
\begin{equation}
\sigma=1+\lambda-\tau
\label{eq:r.13}
\end{equation}

Now let us go over to the case in which we have $\lambda=1$. In this case
we know that $\tau$ is in a sense $2$, but this is not appropriate for
convergence in first moment, since in that case we must always have
\begin{equation}
\int_0^\infty x\Phi(x)dx<\infty.
\label{eq:r.14}
\end{equation}
So let us consider instead convergence in second moment. Up to
(\ref{eq:r.5}) everything remains strictly as above. Now we must attempt
to compare the different contributions. Let us first find a connection
between $\mom[2](t)$ and $s(t)$. Following the same line of reasoning as
above, we obtain from (\ref{eq:r.9}), which is also valid in this case,
that
\begin{equation}
\mom[2](t)=const.\cdot \frac{s(t)}{\ln s(t)}
\label{eq:r.15}
\end{equation}
{}From this we may now deduce, by identifying the orders of magnitude of the
coefficients multiplying $A$ and $C$, that
\begin{equation}
\dot{s}(t)=const.\cdot\frac{s(t)}{\ln s(t)},
\label{eq:r.16}
\end{equation}
from which finally follows
\begin{equation}
s(t)=const.\cdot\exp\left(
const.\cdot\sqrt t
\right),
\label{eq:r.17}
\end{equation}
which is the result obtained by van Dongen \cite{don88} in a slightly
different fashion. We may now choose $\constsep$ in such a way as to
eliminate the various constants involved, so as finally to obtain
\begin{equation}
A-B-C=0.
\label{eq:r.18}
\end{equation}
For this function we only require that $x^2\Phi(x)$ be integrable at the
origin, so that a $\tau=2$ power-law poses no problems. If we wish to cast
the definition of convergence to scaling in a form that is as similar as
possible to the usual one of the non-gelling case, we may reduce it to
\begin{equation}
\lim_{t\to\infty}\left[
\sqrt t\int_0^\infty \mass c(\mass, t)f\left(
\mass e^{-\sqrt t}
\right)
\right]d\mass=\int_0^\infty x\Phi(x)f(x)dx.
\label{eq:r.19}
\end{equation}
This again is identical to the {\em ansatz} discovered by van Dongen in
\cite{don88}. Note, however, that the above reasoning could require
modifications in the presence of additional logarithmic corrections to the
singular behaviour of $\Phi(x)$ near the origin. Whether this may haoppen
or not is not clear to me at present.
\section{The Sum Kernel: The Technique of Characteristics}
\setcounter{equation}0
\label{app:sum}
In this appendix we present the general solution of (\ref{eq:4.3.14a})
with initial conditions given by (\ref{eq:4.3.14b})
by the method of characteristics: if we define $\lapbar(t)$ and
$\fbar(t)$ through the equations
\begin{eqnarray}
\dot{\fbar}&=&-\fbar
\label{eq:d.1a}
\\
\dot{\lapbar}&=&\sinit e^{-t}-\fbar,
\label{eq:d.1b}
\end{eqnarray}
then it immediately follows that, if $F(\lap,t)$ is any solution of
(\ref{eq:4.3.14a}), then $F\left[\lapbar(t),t\right]$
satisfies (\ref{eq:d.1a}). It therefore suffices to match the initial
conditions. This is done as follows; define
\begin{eqnarray}
\lapbar(0)&=&\lap_0
\label{eq:d.3a}\\
\fbar(0)&=&f(\lap_0)
\label{eq:d.3b}
\end{eqnarray}
where $f(\lap)$ is defined via (\ref{eq:4.3.14b}). The characteristic
equations (\ref{eq:d.1a}), (\ref{eq:d.1b}) can now straightforwardly be
solved together with the initial conditions
(\ref{eq:d.3a}), (\ref{eq:d.3b}) to yield the result given in
(\ref{eq:4.3.15a}), (\ref{eq:4.3.15b}). Nevertheless, we present
the details of a particularly straightforward way to perform this
algebra, since this will be used extensively in the sequel for more
complex cases.

We first display an explicit solution of
(\ref{eq:d.1a},\ref{eq:d.1b}) in the scaling limit. This limit will
turn out to be that in which $t$ goes to infinity while
the following variables
\begin{equation}
-\lapp_0=\lap_0e^t\qquad-\lapp=\lapbar e^{2t}
\label{eq:d.4}
\end{equation}
remain constant.
The solution of (\ref{eq:d.1a},\ref{eq:d.1b}) is given by
\begin{eqnarray}
\fbar&=&f(\lap_0)e^{-t}=-\lapp_0e^{-2t}
\label{eq:d.5a}\\
\lapbar&=&\left[
f(\lap_0-f(0)
\right]e^{-t}+\lap_0-f(\lap_0)+f(0)=-e^{-2t}
[\lapp_{0}+\half f^{\prime\prime}(0)\lapp_0^2].
\label{eq:d.5b}
\end{eqnarray}
where the first equality represents the exact solution and the second
involves the scaling limit defined in (\ref{eq:d.4}). Now let us note
the following fundamental relation: in the scaling limit one has,
because of the defining relation for $\Phi(x)$ (\ref{eq:3.801}):
\begin{equation}
\int_0^\infty d\mass\,\mass c(\mass, t)e^{\mass\lap}
\to\int_0^\infty x\Phi(x)
e^{-\lap x}dx.
\label{eq:d.6}
\end{equation}
But the following relation holds for the l.h.s. of (\ref{eq:d.6}):
\begin{equation}
\int_0^\infty d\mass\,\mass c(\mass, t)e^{\mass\lap}
=\frac{\fbar_{\lap_0}}{\lapbar_{\lap_0}}
=\frac{\fbar_{\lapp_0}}{\lapbar_{\lapp_0}},
\label{eq:d.7}
\end{equation}
where the partial derivatives in the final expressions refer to the
scaling expressions derived in (\ref{eq:d.4}). This now readily yields
\begin{equation}
\laplace[x\Phi(x)](s)=\frac{1}{1+f^{\prime\prime}(0)\lapp_0}.
\label{eq:d.8}
\end{equation}
{}From (\ref{eq:d.4},\ref{eq:d.5b}) one now obtains the necessary
relation between $\lapp_0$ and $\lapp$, namely
\begin{equation}
\lapp=\lapp_{0}+\half f^{\prime\prime}(0)\lapp_0^2,
\label{eq:d.9}
\end{equation}
from which the final result eventually follows, namely
\begin{equation}
\laplace[x\Phi(x)](s)=\frac{1}{\sqrt{1+2f^{\prime\prime}(0)\lapp}}.
\label{eq:d.10}
\end{equation}
Finally, the results stated in the text can be obtained in an
elementary way from this equation and (\ref{eq:d.9}).
\section{The Parity Dependent Kernel: Scaling Theory}
\setcounter{equation}0
\label{app:parity-scal}
In the following, we develop the technical details of the parity-dependent
kernel. The kinetic equations for the
concentrations read
\begin{eqnarray}
\dot c_{2j+1}&=&M\sum_{k=0}^j c_{2k+1}c_{2(j-k)}-Kc_{2j+1}\sum_{k=0}^\infty
c_{2k+1}-Mc_{2j+1}\sum_{k=1}^\infty c_{2k}\nonumber\\
\dot c_{2j}&=&\frac{K}{2}\sum_{k=0}^{j-1} c_{2k+1}c_{2(j-k)-1}+\frac{L}{2}
\sum_{k=1}^{j-1}c_{2k}c_{2(j-k)}-\label{eq:s.1}\\
&&\qquad-Mc_{2j}\sum_{k=0}^\infty c_{2k+1}
-Lc_{2j}\sum_{k=1}^\infty c_{2k}.
\nonumber
\end{eqnarray}
If one introduces the generating functions
\begin{eqnarray}
D(\lap,t)&=&\sum_{j=0}^\infty c_{2j+1}(t)\lap^{2j+1}\nonumber\\
P(\lap,t)&=&\sum_{j=1}^\infty c_{2j}(t)\lap^{2j}
\label{eq:s.2}
\end{eqnarray}
and the corresponding moments
\begin{eqnarray}
d(t)&=&D(1,t)\nonumber\\
p(t)&=&P(1,t),
\label{eq:s.3}
\end{eqnarray}
one eventually finds the following equations:
\begin{eqnarray}
\dot D&=&MDP-[(Mp(t)+Kd(t)]D\nonumber\\
\dot P&=&\frac{K}{2}D^2+\frac{L}{2}P^2-P[Lp(t)+Md(t)].
\label{eq:s.4}
\end{eqnarray}
Note that these are ODE, the only dependence in $\lap$ coming in through
the initial  conditions. Equations (\ref{eq:s.4}) then yield the closed
relations for $d(t)$ and $p(t)$:
\begin{eqnarray}
\dot d&=&-Kd^2\nonumber\\
\dot p&=&\frac{K}{2}d^2-\frac{L}{2}p^2-Mpd.
\label{eq:s.5}
\end{eqnarray}
These can be solved exactly. For our purposes, we shall only need the
following elementary facts:
\begin{eqnarray}
d(t)&=&\frac{\alpha}{1+\alpha Kt}\label{eq:s.6}\\
\lim_{t\to\infty}p(t)/d(t)&=&\tpinf=\frac{K-M}{L}\left[
1+\sqrt{
1+\frac{KL}{(K-M)^2}
}
\right]
\label{eq:s.7}
\end{eqnarray}
Here $\alpha$ is the initial value of $d(t)$, which plays no significant
role in the following, since we shall always be interested in the
large-time limit.

Let us now introduce
\begin{eqnarray}
\tD(\lap,\tres)&=&D(\lap,t)/d(t)\nonumber\\
\tP(\lap,\tres)&=&P(\lap,t)/d(t)
\label{eq:s.8}\\
\tp&=&p(t)/d(t)\nonumber\\
\dot\tres(t)&=&Kd(t)\nonumber
\end{eqnarray}
One then finds:
\begin{eqnarray}
K\frac{d\tD}{d\tres}&=&M\tD(\tP-\tp)\nonumber\\
K\frac{d\tP}{d\tres}&=&\frac{K}{2}\tD^2+\frac{L}{2}\tP^2
+(K-M-L\tp)\tP.
\label{eq:s.9}
\end{eqnarray}
The initial conditions are very close to the solution given
by $\tD=1$ and $\tP(\tres)=\tp(\tres)$. They therefore
follow it for considerable time and eventually can be described as
solutions of the following equations
\begin{eqnarray}
K\frac{d\tD}{d\tres}&=&M\tD(\tP-\tpinf)\nonumber\\
K\frac{d\tP}{d\tres}&=&\frac{K}{2}\tD^2+\frac{L}{2}\tP^2
+(K-M-L\tpinf)\tP,
\label{eq:s.10}
\end{eqnarray}
which is a two-dimensional autonomous system. The functions $\tD$ and
$\tP$ then start near the equilibrium $\tD=1$ and $\tP=\tpinf$. It is
then fairly straightforward to convince oneself that letting $\lap$ tend
to one and simultaneously letting $t$ tend to infinity in such a way
that $t(1-\lap)$ remains constant will give a non-trivial limiting
behaviour for both $\tD$ and $\tP$. This then is easily seen to be
equivalent to the existence of a scaling limit. The evaluation of the
scaling function by these means, however, seems to be a hopeless task

What is possible, however, is the evaluation of the $\tau$ and $\explarge$
exponents for the odd and even concentrations. To this end, we assume that
\begin{eqnarray}
\lim_{t\to\infty}\sum_{j=0}^\infty (2j+1)c_{2j+1}(t)f\left(
\frac{2j+1}{Kt}
\right)&=&\int_0^\infty x\Phi_d(x)f(x),\nonumber\\
\lim_{t\to\infty}\sum_{j=1}^\infty 2jc_{2j}(t)f\left(
\frac{2j}{Kt}
\right)&=&\int_0^\infty x\Phi_p(x)f(x).
\label{eq:s.11}
\end{eqnarray}
We may then proceed once more exactly in Appendix \ref{app:deriv-scaling}
in order to obtain an equation of the type of (\ref{eq:3.10}) or
(\ref{eq:3.12}). The final result is
\begin{eqnarray}
&&\int_0^\infty dx\,x^2\Phi_d(x)f^\prime(x)=
\frac{M}{K}\int_0^\infty dx\, dy\,x\Phi_d(x)\Phi_p(y)f(x+y)-\nonumber\\
&&\qquad-\int_0^\infty dx\,dy\,x\Phi_d(x)\left[
\Phi_d(y)+\frac{M}{K}\Phi_p(y)
\right]f(x)
\label{eq:s.12}\\
&&\int_0^\infty dx\,x^2\Phi_p(x)f^\prime(x)=\half\int_0^\infty
dx\,dy\,x\left[
\Phi_d(x)\Phi_d(y)+\frac{L}{K}\Phi_p(x)\Phi_p(y)
\right]f(x+y)-\nonumber\\
&&\qquad-\int_0^\infty dx\,dy\,x\Phi_p(x)\left[
\frac{M}{K}\Phi_d(y)+\frac{L}{K}\Phi_p(y)
\right]f(x)
\label{eq:s.13}
\end{eqnarray}
for arbitrary continuous functions $f(x)$.
In order to fix the normalizations properly, the simplest way is to note
that, from (\ref{eq:s.11}) using $f(x)=1/x$ one finds
\begin{eqnarray}
\int_0^\infty \Phi_d(x)&=&\lim_{t\to\infty}\left[
Ktd(t)
\right]=1\nonumber\\
\int_0^\infty \Phi_p(x)&=&\lim_{t\to\infty}\left[
Ktp(t)
\right]=\tpinf.
\label{eq:s.14}
\end{eqnarray}
Now, putting $f(x)$ equal to $e^{-\rho x}$ into (\ref{eq:s.12},\ref{eq:s.13})
and using
the standard approach, we find from (\ref{eq:s.12}), that all terms
except the first
on the r.h.s. are of the same order for large $\rho$, namely of order
$\rho^{2-\tau_d}$. The first is
manifestly always subdominant. In order to evaluate $\tau_d$ one then
works out all the prefactors and finds in the end
\begin{equation}
\tau_d=1-\frac{M}{K}\tpinf
\label{eq:s.15}
\end{equation}
The equation (\ref{eq:s.13}) is a bit more complex, since one must distinguish
between the case in which the first term dominates or is subdominant. All
other terms are of order $\rho^{2-\tau_p}$, whereas the first is of order
$\rho^{3-2\tau_d}$. Assuming first that the first term is again subdominant
leads once more, using a comparison of coefficients to
\begin{equation}
\tau_p=2-\frac{M}{K}+\frac{L}{K}\tpinf.
\label{eq: s.16}
\end{equation}
On the other hand, if the first term dominates, a simple power-law
comparison leads to
\begin{equation}
\tau_p=2\tau_d-1.
\label{eq:s.17}
\end{equation}
Finally, in order to evaluate the exponent $\explarge$ for large cluster
sizes we proceed again in the standard manner: we develop
(\ref{eq:s.12},\ref{eq:s.13}) around a singularity at some $\rho_c<0$. This
time, it is straightforward to check that the $\explarge_p$ exponent is zero
in all cases. As for the $\explarge_d$ exponent, one can not easily say
anything, except that it must be positive. If this is so, however, then
the $\explarge$ exponent of the full distribution, given by
(\ref{eq:4.10.150}) is equal to zero.

As a final remark concerning the scaling theory, note that equations
similar to (\ref{eq:s.12},\ref{eq:s.13}) can easily be derived in the
much more general case
\begin{equation}
K(k,l)=\left\{
\begin{array}{ll}
a_1K_0(k,l)&\qquad\mbox{for $k$ and $l$ odd}\\
a_2K_0(k,l)&\qquad\mbox{for $k$ and $l$ even}\\
a_3K_0(k,l)&\qquad\mbox{otherwise}
\end{array}
\right.,
\label{eq:s.18}
\end{equation}
where $K_0(k,l)$ is  a homogeneous kernel that is otherwise arbitrary. From
a study of these equations, it might perhaps be shown in general under what
circumstances such changes in the scaling function are to be expected. In
particular, it might confirm a suspicion that this is characteristic of
Type II kernels, the scaling theory of which is in any case non-universal.

Finally, let us evaluate the non-scaling asymptotic behaviour. The easiest
approach is to evaluate first the long-time behaviour of $c_1(t)$ and
$c_2(t)$ and then to convince oneself through induction that this is
indeed the large-time behaviour at fixed $j$ for $j$ arbitrary. From
(\ref{eq:s.1}) one finds for monomers and dimers
\begin{eqnarray}
\dot c_1&=&-\left[Kd(t)+Mp(t)\right]c_1\nonumber\\
\dot c_2&=&\frac{L}{2}c_1^2-\left[
Md(t)+Lp(t)
\right]c_2
\label{eq:s.19}
\end{eqnarray}
{}From these and (\ref{eq:s.7}) the relations (\ref{eq:4.10.4})
stated in the text are readily verified.
\section{The $q$-sum Solution: A Sketch}
\setcounter{equation}0
\label{app:q-sum}
Here we shortly sketch the derivation of the solution
(\ref{eq:4.12.6},\ref{eq:4.12.7}) for the $q$-sum kernel. From
(\ref{eq:4.12.2}) the following equation for $H(\lap,\tres)$ follows,
see (\ref{eq:4.12.5}):
\begin{equation}
\frac{\partial H}{\partial\tres}(\lap,\tres)=H(\lap, \tres)\left[
H(\lap,\tres)-H(\lap-1, \tres)
\right]
\label{eq:t.1}
\end{equation}
This equation is considerably simplified by rewriting $H(\lap, \tres)$ in
the following quotient form:
\begin{equation}
H(\lap,\tres)=\frac{h(\lap-1,\tres)}{h(\lap,\tres)}
\label{eq:t.2}
\end{equation}
One then finds for $h(\lap, \tres)$ the {\em linear} equation
\begin{equation}
\frac{\partial h}{\partial\tres}(\lap, \tres)=-h(\lap-1, \tres)
\label{eq:t.3}
\end{equation}
This equation is linear and translation invariant
with respect to $\tres$, so that it can be solved
by Fourier transform. After some work one verifies that
\begin{equation}
h(\lap, \tres)=e^{i\pi\lap}\sum_{m=0}^\infty\frac{\tres^m}{m!}
\prod_{l=m}^\infty\left(
1-e^{b(\lap-l)}
\right)^{-1}
\label{eq:t.4}
\end{equation}
{}From this it is straightforward, though somewhat tedious, to obtain the
final result stated in the text.
\section{Crossover in the $q$-sum kernel}
\setcounter{equation}0
\label{app:q-cross}
In the following I wish to show that $je^{-2t}$ and $bj$ are the correct
scaling variables for the crossover transition for the
modified $q$-sum kernel $(2-q^k-q^l)/b$. I
do not claim any expression, even formal, for the scaling function: the
methods I shall employ will not be precise enough to fix any kind of
functional dependence. They will, however, determine the order of
magnitude of various quantities, and this will be sufficient.

If the above is correct, then we expect from the definition (\ref{eq:3.5.3})
for the scaling in crossover, that
\begin{equation}
\lim_{t\to\infty}\left\{
y\left[
\sum_{k=1}^\infty c_k(t)
\right]H(\lap, \tres;ye^{-2t})
\right\}=\int_0^\infty \Phi(x,y)\left(
e^{\lap xy}-1
\right)\,dx,
\label{eq:u.1}
\end{equation}
where we have singled out the dependence of $H(\lap, \tres;b)$ on the
argument $b$. To show this, I shall prove the following: first I shall
identify the crossover regime in time
as given by finite values of the parameter
\begin{equation}
\xi=\frac{b\tres-1}{\sqrt b}.
\label{eq:u.2}
\end{equation}
In other words, we require $b\tres$ to be close to one and the difference
of order $b^{1/2}$. Then I show that in this case one has
\begin{equation}
\left[
\sum_{k=1}^\infty c_k(t)
\right]^{-1}=O(b^{-1/2}).
\label{eq:u.3}
\end{equation}
and finally I show that in the crossover regime as defined above one has for
the generating function $H$:
\begin{equation}
H(\lap, \tres)=\sqrt b\phi(\lap, \xi).
\label{eq:u.4}
\end{equation}
Here and throughout this appendix\footnote{but nowhere else in this paper!}
$\phi$ will denote an arbitrary function of
the scaling variables $\lap$ and $\xi$, which may vary from equation to
equation.

Let us first start with the elementary identity:
\begin{equation}
\frac{1}{e_q(q^\alpha)}\sum_{r=0}^\infty\frac{q^{r\alpha}f(q^r)}{(q;q)_r}
=\sum_{m=0}^\infty\frac{f^{(m)}(0)}{m!}(q;q^\alpha)_m
\label{eq:u.5}
\end{equation}
valid for any entire function $f$ for which one of the sides converges.
This is obtained by developing $f$ in a Taylor series, inverting the two
sums and using the two definitions of the $q$-exponential given in
(\ref{eq:4.12.4}).

Using (\ref{eq:u.5}) one obtains from \ref{eq:4.12.8}) the following
expression for $t$
\begin{equation}
t=\sum_{m=0}^\infty\frac{(b\tres)^{m+1}}{m+1}\prod_{l=1}^m\left(
\frac{1-q^l}{
bl
}
\right)
\label{eq:u.6}
\end{equation}
Note the extra factor of $b$ due to the fact that we are working with a
{\em modified} kernel with an additional factor $b^{-1}$ in its
definition. It is now easy to convince oneself that the quantity
\begin{equation}
\prod_{l=1}^m\left(
\frac{1-q^l}{
bl
}
\right)
\label{eq:u.7}
\end{equation}
decays rapidly when $m$ becomes significantly larger than $\sqrt b$ and is
constant before. We may therefore replace, as far as orders of magnitude
go, this prefactor by an appropriate cutoff of the sum, leading to
\begin{equation}
t=\sum_{m=0}^{N(b)}\frac{(b\tres)^{m+1}}{m+1}
\label{eq:u.10}
\end{equation}
where $N(b)=b^{-1/2}$. Using identity (\ref{eq:o.9}) one then finds
\begin{equation}
t=\ln N(b)+\phi(\xi),
\label{eq:u.8}
\end{equation}
from which (\ref{eq:u.2}) follows. An entirely similar evaluation yields
(\ref{eq:u.3}) from the expression
\begin{equation}
\left[
\sum_{k=1}^\infty c_k(t)
\right]^{-1}=\frac{1}{e_q(q)}\sum_{r=1}^\infty\frac{q^re^{\tres q^r}}{
(q;q)_r}.
\label{eq:u.9}
\end{equation}
Again, a similar transformation applied to $S(\lap,\tres)$ yields
\begin{equation}
S(\lap,\tres)=\sum_{m=1}^\infty (b\tres)^m\frac{(q;q^{-\lap})_m}{
(q;q)_m}\prod_{l=1}^m\left(
\frac{1-q^l}{
bl
}
\right).
\label{eq:u.11}
\end{equation}
Once more, we are led to cut the sum off at $N(b)$ leading to
\begin{equation}
S(\lap,\tres)=\sum_{m=1}^{N(b)}(b\tres)^m\frac{(q;q^{-\lap})_m}{
(q;q)_m}.
\label{eq:u.12}
\end{equation}
But, for this range of values of $m$ one has, as $q\to1$,
\begin{equation}
\frac{(q;q^{-\lap})_m}{
(q;q)_m}\to m^{-\lap}.
\label{eq:u.13}
\end{equation}
If one inserts (\ref{eq:u.13}) into (\ref{eq:u.12}) and reduces the sum to
an integral, it follows that
\begin{equation}
S(\lap,\tres)=N(b)^{1-\lap}\phi(\lap,\xi),
\label{eq:u.14}
\end{equation}
from which
(\ref{eq:u.4}) follows straightforwardly. All the results
initially claimed are hence established, and the statements made
in the text follow.
\section{The Simple Product Kernel: Technicalities}
\setcounter{equation}0
\label{app:bilinear}
In the following, we work out the detailed properties of the simple
product kernel, which has all the necessary ingredients to understand the
phenomenon of gelation. It is worthwhile to do this, as the full bilinear
kernel is very much more complex.

We use the generating function $G(\lap,t)$ defined in (\ref{eq:4.5.3}),
which satisfies the PDE (\ref{eq:4.5.4}). The characteristic equations of
this last are given by
\begin{eqnarray}
\frac{d\lapbar}{dt}&=&-[\gbar-\mom[1](t)]=-(\gbar-1)-h(t)\nonumber\\
\frac{d\gbar}{dt}&=&0,
\label{eq:f.1}
\end{eqnarray}
where the second equality of the first equation defines $h(t)$.
{}From this follows immediately that
\begin{equation}
G\left[
\lap_0-[g(\lap_0)-1]t-\int_0^t h(t^\prime)dt^\prime,t
\right]=g(\lap_0).
\label{eq:f.2}
\end{equation}
Setting $h(t)$ identically zero for the time being, one obtains by the
inverse Laplace transform of (\ref{eq:f.2})
\begin{equation}
\mass c(\mass,t)=\frac{1}{2\pi i}\oint_{-i\infty}^{i\infty}
d\lap_0[1-tg^\prime(\lap_0)]g(\lap_0)\exp\left\{
-\mass [\lap_0-(g(\lap_0)-1)t]
\right\}
\label{eq:f.3}
\end{equation}
The case in which $g(\lap_0)$ is $e^{\lap_0}$, which corresponds to
monodisperse initial conditions, is now reduced to an
exercise, which the reader can do in order to verify the result
(\ref{eq:4.5.7}) stated in the text.

We now need to gain some understanding of the failure of the solution
(\ref{eq:f.3}) and of the assumption that $\mom[1](t)$ remains constant.
Equations (\ref{eq:f.1}) specify the surface $G(\lap,t)$ by displaying a
pencil of straight lines as the level curves of this surface. These
straight lines intersect, however, meaning that the surface must be
multivalued. It is important to understand this issue well: indeed, since
we have derived the PDE (\ref{eq:4.5.4}) without giving much thought to
questionns of convergence, we must now take care that the solution we
obtain has the necessary analyticity properties. We may therefore not take
into consideration the second sheet of the solution, but only that sheet
which connects smoothly to $\lap=-\infty$, since there the convergence of
the series is unproblematic.

We therefore need to determine the locus of points on the $(\lap,t)$ plane
for which the function $G(\lap,t)$ becomes singular. This will define a
curve, the position of which we must determine in such a way as to obtain
a meaningful solution for the equations (\ref{eq:2.2}) with the kernel
(\ref{eq:4.5.2}). The determination of the singular curve proceeds as
follows: If one takes two infinitesimally close straight lines from the
pencil defined by the first equation (\ref{eq:f.1}), their
intersection belongs to the singular curve, as is intuitively clear and
readily verified analytically\footnote{Note that a similar pencil exists in
the sum kernel and also has intersections at finite time. The reason this
creates no problem is that, for the sum kernel, all the computations are
carried out using a rescaled time $\tres$, and all such intersections
occur at values of $\tres$ which do not correspond to any physical time.
See Appendix \ref{app:sum} for the detailed definitions.}. Take $\lap_0$
and $\lap_1$ two infinitely close initial values for the first of the
equations (\ref{eq:f.1}). This yields for the time at which these intersect
\begin{equation}
t_c(\lap_0)=1/g^\prime(\lap_0)
\label{eq:f.4}
\end{equation}
and hence
\begin{equation}
\lap_c(\lap_0)=\lap_0-\frac{g(\lap_0)-1}{g^\prime(\lap_0)}
-\int_0^{1/g^\prime(\lap_0)}h(t^\prime)\,dt^\prime.
\label{eq:f.5}
\end{equation}
As long as $t<t_c(0)$, therefore, there is no problem and the solution
(\ref{eq:f.3}) is valid. This is due to the fact that the function
$G(\lap,t)$ is analytic at $\lap=0$ for such times, so that the PDE
(\ref{eq:4.5.4}) can be applied in order to show that $G(0,t)$, which is
the first moment, is indeed constant. This therefore singles out $t_c(0)$,
which is given by
\begin{equation}
t_c\equiv t_c(0)=1/g^\prime(0)=\mom[2](0)^{-1}
\label{eq:f.6}
\end{equation}
as the limiting time beyond which (\ref{eq:f.3}) ceases to hold. It is
known as the gel time, and its determination is in general not a trivial
task; (\ref{eq:f.6}) solves it for arbitrary initial conditions in the
case of the product kernel (\ref{eq:4.5.2}).

Beyond $t_c$, we require a condition that allows to determine the
amount of mass carried to infinity by the run-away aggregation process.
The following remark allows to do this: since $G(\lap,t)$ is a Laplace
transform
of a positive real function, its singularity nearest to the origin
must be on the real $\lap$ axis. We
may therefore identify the singular curve $(\lap_c(\lap_0),t_c(\lap_0))$,
as defined in (\ref{eq:f.5},\ref{eq:f.6}),
as giving the nearest singularity in $\lap$ of $G(\lap,t)$
for fixed $t$. Here $\lap_0$ is simply a parameter for this curve.
But since $c(\mass, t)$ can at most decay as a power-law,
since otherwise $\mom[1](t)$ would remain constant, and since on the
other hand $c(\mass, t)$ cannot grow exponentially, we are left with the
necessary condition that the singular curve coincide with the $\lap=0$ line
on the $(\lap,t)$ plane for $t\geq t_c$. In other words
\begin{equation}
\lap_c\left[\lap_0(t)
\right]=0,
\label{eq:f.7}
\end{equation}
where one defines $\lap_0(t)$ through the relation
\begin{equation}
t=1/\gp[\lap_0(t)].
\label{eq:f.8}
\end{equation}
$h(t)$ is then given by
\begin{equation}
h(t)=\frac{d\lap_0(t)}{dt}-\frac{d}{dt}\left[
t\big(g[\lap_0(t)]-1\big)
\right].
\label{eq:f.9}
\end{equation}
In the particular case of monodisperse initial conditions, this yields
\begin{equation}
h(t)=1-t^{-1},
\label{eq:f.10}
\end{equation}
which corresponds to the results that follows from the solution
(\ref{eq:4.5.7}) stated in the text. To derive the solution itself once
$h(t)$ is known, it is enough to proceed with the expression (\ref{eq:f.2})
in the same way as we obtained (\ref{eq:f.3}) from (\ref{eq:f.2}) under
the assumption that $h(t)$ was identically zero. The final result is given
by
\begin{eqnarray}
\mass c(\mass,t)&=&\frac{1}{2\pi i}\oint_{-i\infty}^{i\infty}
d\lap_0\left[
1-tg^\prime(\lap_0)
\right]
g(\lap_0)\times\nonumber\\
&&\times\exp\left\{
-\mass
\left[
\lap_0-(g(\lap_0)-1)t-\int_0^th(t^\prime)dt^\prime
\right]
\right\}
\label{eq:f.11}
\end{eqnarray}
Again, it is straightforward to verify that the Stockmayer solution given
in the text follows from (\ref{eq:f.11}) in the case of monodisperse
initial conditions.

Let us now finish the subject by deriving the scaling theory for the
product kernel. We shall here derive it in some detail for this simpler
case and later, in Appendix \ref{app:bilinear1}, show tersely that the
general case brings nothing new. The
reader who has well absorbed the following will fill in the necessary
details.

Since we are dealing with a gelling system, we need to use the definition
(\ref{eq:3.810}) of scaling of the $n$-th moment, for $n>1$. In the
following, we will take $n$ equal to two. In principle, one could look at
two limits: the  large time limit, and the behaviour near $t_c$. In the
following, we will only consider the behaviour near $t_c$. In fact, it
turns out that the large-time behaviour is considerably simpler.

Using the monodisperse solution, the moment equations or otherwise, it is
easy to see that the typical size diverges as $(t_c-t)^{-2}$ near $t_c$. We
therefore wish to establish this rigorously for all initial conditions,
and to obtain the scaling function. To this end we first need to compute
the second moment. Using the moment equations
(\ref{eq:3.10.1}) we find for the times
before $t_c$
\begin{equation}
M_2(t)=\frac{M_2(0)}{1-M_2(0)t}=(t_c-t)^{-1}.
\label{eq:f.12}
\end{equation}
The limiting statement predicted by the definition of scaling in second
moment is therefore given by
\begin{equation}
\lim_{t\to t_c}\left\{
(t_c-t)^{-1}\left[
G(s(t_c-t)^2,t)-1
\right]
\right\}
=\int_0^\infty x\Phi(x)(e^{sx}-1)dx.
\label{eq:f.13}
\end{equation}
To evaluate the function $G$ at this value of its arguments we must solve
\begin{equation}
\lap_0-[g(\lap_0)-1]t=s(t_c-t)^2.
\label{eq:f.14}
\end{equation}
Remembering that $t_c$ is $1/g^\prime(0)$, developing everything
to lowest order and introducing $w$ for $(t_c-t)^{-1}\lap_0$,
we reduce (\ref{eq:f.14}) to
\begin{equation}
\gpp-2\gp^2w+2\gp s=0.
\label{eq:f.15}
\end{equation}
This then leads to
\begin{equation}
\int_0^\infty x\Phi(x)(e^{sx}-1)dx=\gp w=\frac{1}{\beta}\left[
1-\sqrt{
1-2\beta s
}
\right]
\label{eq:f.16}
\end{equation}
where $\beta$ is defined as
\begin{equation}
\beta=\frac{\gpp}{\gp^3}
\label{eq:f.17}
\end{equation}
and from this finally follows the expression (\ref{eq:4.5.11}) given in
the text.
\section{The Product Kernel: Power-law initial Conditions}
\setcounter{equation}0
\label{app:bilinear2}
In the following we treat the effect of power-law initial conditons on
the gelling transition. For simplicity, we limit ourselves to the case of
the product kernel. We shall use all the notation of Appendix
\ref{app:bilinear}. We shall start from the assumption that the large
$\mass$ behaviour of $c(\mass,t)$ is given by
\begin{equation}
c(\mass, t)=A\mass^{-1-\alpha}[1+o(1)],
\label{eq:n.1}
\end{equation}
which implies for the generating funcion $g(\lap_0)$
\begin{equation}
g(\lap_0)=\left\{
\begin{array}{ll}
\left[
1+\lap_0\gp+A\Gamma(1-\alpha)(-\lap_0)^{\alpha-1}
\right][1+o(1)]&\qquad(2<\alpha<3)\\
\left[
1-A\Gamma(1-\alpha)(-\lap_0)^{\alpha-1}
\right][1+o(1)]&\qquad(1<\alpha<2)
\end{array}
\right.
\label{eq:n.2}
\end{equation}
We shall treat the two cases separately: in the first ($2<\alpha<3$) the
third moment is infinite, so that the scaling approach developed in
Appendix \ref{app:bilinear} fails, but the second moment is finite, so
that the derivation of the gel time in (\ref{eq:f.6}) holds good. In the
second case $(1<\alpha<2$), on the other hand, the second moment diverges.
{}From a generalization of (\ref{eq:f.6}) one is therefore tempted to think
that instantaneous gelation takes place, an impression that shall indeed be
confirmed in this Appendix.

In the first case, we must solve the equation
\begin{equation}
\lap_0-[g(\lap_0)-1]t=s(t_c-t)^{1/\sigma},
\label{eq:n.3}
\end{equation}
where $\sigma$ is as yet unknown. Developing to leading order
in $\lap_0$ and $t_c-t$ results in:
\begin{equation}
At_c\Gamma(1-\alpha)(-\lap_0)^{\alpha-1}-\frac{(t_c-t)
\lap_0}{t_c}+s(t_c-t)^{1/\sigma}=0.
\label{eq:n.4}
\end{equation}
The three terms are of the same order if we introduce the rescaled
variable $w$ as $\lap_0(t_c-t)^{1/(\alpha-2)}$ and take
\begin{equation}
\sigma=\frac{\alpha-2}{\alpha-1}.
\label{eq:n.5}
\end{equation}
The scaled version of (\ref{eq:n.4}) then reads
\begin{equation}
At_c\Gamma(1-\alpha)(-w)^{\alpha-1}-\frac{w}{t_c}+s=0.
\label{eq:n.6}
\end{equation}
{}From the definition of scaling, one obtains using the same reasoning as in
the derivation of (\ref{eq:f.16})
\begin{equation}
\int_0^\infty x\Phi(x)(e^{-sx}-1)dx=\gp w
\label{eq:n.7}
\end{equation}
{}From this and (\ref{eq:n.6}) one could presumably obtain the exact form
of $\Phi(x)$. This task is left to the enthusiastic reader. On the other
hand, it follows from (\ref{eq:n.6}) that $-w$ goes as $s^{1/(\alpha-1)}$
as $s\to\infty$, so that the $\tau$ exponent can be found easily, and is
given by
\begin{equation}
\tau=\frac{2\alpha-1}{\alpha-1}.
\label{eq:n.8}
\end{equation}
Note that these results are in agreement with the scaling law
(\ref{eq:scaling-rel})
connecting $\tau$ and $\sigma$ in the gelling case, which in this case
is given by
\begin{equation}
3-\tau=\sigma.
\label{eq:n.9}
\end{equation}
In order to evaluate the correction to scaling exponent $\Delta$, we
remark that the subdominant behavior of $-w$ as a function of $\lapp$ for
$\lapp\to\infty$ is linear in $\lapp$, so that
\begin{equation}
\Delta=\frac{\alpha-2}{\alpha-1}
\label{eq:n.905}
\end{equation}
{}From (\ref{eq:n.905}) follows that $\Delta$ and $\sigma$ are identical, as
predicted by the scaling theory developed in the text.

Now let us evaluate the behaviour of $h(t)$ shortly beyond the
gel time. To this end we use (\ref{eq:f.9}), where we first determine the
behaviour of $\lap_0(t)$ in this region. Since we are near the gel time,
$\lap_0(t)$ is small, so that we may use the asymptotic expressions
(\ref{eq:n.2}) for $g(\lap_0)$. Using these yields
\begin{equation}
h(t)=const.\cdot (t-t_c)^{1/(\alpha-2)}
\label{eq:n.10}
\end{equation}
This implies that the flow of mass to infinity is {\em slower}
than in the case where the initial conditions are short-range,
which does seem rather unexpected. In particular, this is an
instance in which the flow rate of sol to gel is zero at the gel
time, therefore strongly
suggesting that $\tau>5/2$, as is indeed borne out by (\ref{eq:n.8}).
On the other hand, it is of course readily verified that, past the gel
time, the singularity of $G(\lap, t)$ at $\lap=0$ is always of square root
type, so that the large $j$ behaviour of the cluster size distribution for
fixed times larger than $t_c$ always has an asymptotic power law decay as
$j^{-5/2}$.

Let us now turn to the second case. There it follows from (\ref{eq:n.2})
that the generalization of (\ref{eq:f.4}) is
\begin{equation}
t_c(\lap_0)=\frac{(-\lap_0)^{2-\alpha}}{(\alpha-1)\Gamma(2-\alpha)}
\label{eq:n.11}
\end{equation}
{}From this it now follows along the same lines as those presented above or
in Appendix \ref{app:bilinear}, that
\begin{equation}
h(t)=const.\cdot t^{1/(2-\alpha)}
\label{eq:n12}
\end{equation}
for small values of $t$, and that $t_c$ is indeed zero. Again this implies
that the flow rate of sol to gel is initially zero, in remarkable contrast
to the case in which the initial conditions are short range. This
underlines the arbitrariness of the assumption (often made in the literature,
see in particular \cite{ley81}) that the flow rate {\em at} $t_c$ is of
order unity.
\section{The Full Bilinear Kernel: A Sketch of the Scaling Theory}
\setcounter{equation}0
\label{app:bilinear1}%
In this Appendix, we give a very short argument showing that the full
bilinear kernel scales in the same way as the product kernel. As the
calculations become quite involved I shall be brief.

If we define as usual
\begin{equation}
G(\lap,t)=\int_0^\infty c(\mass, t)\left(
e^{\lap\mass}-1
\right),
\label{eq:m.1}
\end{equation}
one obtains the following PDE gor $G(\lap,t)$
\begin{equation}
G_t=\half G^2-BG\left(
G_\lap-1
\right)-\half CG_\lap^2.
\label{eq:m.2}
\end{equation}
The characteristic equations for this PDE can be written as follows:
\begin{eqnarray}
\frac{d\lapbar}{dt}&=&-\gbar-C(q-1)\nonumber\\
\frac{dp}{dt}&=&Ap\gbar+Bp(q-1)\nonumber\\
\frac{dq}{dt}&=&AqG+Bq(q-1)
\label{eq:m.3}\\
\frac{dG}{dt}&=&p-q\left[
BG+C(q-1)
\right]\nonumber
\end{eqnarray}
The curves defined by the above system of ODE's has properties similar to
those of the characteristic equations in the quasilinear case.
Specifically, one has
\begin{eqnarray}
G\left[
\lapbar(t), t
\right]&=&\gbar(t)\nonumber\\
G_t\left[
\lapbar(t), t
\right]&=&p(t)
\label{eq:m.205}\\
G_\lap\left[
\lapbar(t), t
\right]&=&q(t)\nonumber
\end{eqnarray}
In order to prove
that gelation occurs at finite time, it is sufficient to look at the
moment equations (\ref{eq:3.10.1}). In order to show that the scaling
theory is the same, it is enough to show that the surface described by
$q$ (which corresponds to the function $G$ of Appendix \ref{app:bilinear}
has the same geometrical structure as the function $G$ described in
Appendix \ref{app:bilinear}.

In order to see this, one  makes the following elementary remarks: for the
usual initial conditions with $\lapbar_0$ negative but very small, both
$q$ and $\gbar$ are small and negative. It therefore follows that $dq/dt$
is also small and negative. Since the gel time is finite, this means that
$q$ is still of order $\lapbar_0$ at the gel time. Of $\gbar$ we can at
least say that it is no larger than of order $\lapbar_0$. The behaviour
of $\lapbar(t)$ is therefore qualitatively similar to the one found for
the product kernel in Appendix \ref{app:bilinear}, that is, the integral
curves for $\lapbar(t)$ will cross at some finite time, which becomes
the gel time as $\lap_0\to0$. We will therefore
observe a similar backbending of the surface described by $q$ and hence a
similar square root singularity, which will therefore lead to the same
scaling function.
\section{The $B\to0$ Crossover When $C=0$}
\setcounter{equation}0
\label{app:crossover2}
In this appendix we study the crossover behaviour of the kernel
\begin{equation}
K(\mass, \massp)=1+\epsilon(\mass+\massp).
\label{eq:g.1}
\end{equation}
For this we use the generating function
\begin{equation}
G(\lap,t;\epsilon)=\int_0^\infty d\massp c(\massp,t)\left[
e^{\massp\lap}-1
\right],
\label{eq:g.2}
\end{equation}
where the $\epsilon$ dependence as been explicitly put in, as we shall
have occasion to single it out in the sequel.
$G(\lap,t)$ then satisfies the following PDE
\begin{equation}
G_t-\epsilon G_\lap=\frac{G^2}{2}-\epsilon G
\label{eq:g.3}
\end{equation}
The characteristic equations are given by
\begin{eqnarray}
\frac{d\lapbar}{dt}&=&-\epsilon\gbar\nonumber\\
\frac{d\gbar}{dt}&=&\frac{\gbar^2}{2}-\epsilon\gbar.
\label{eq:g.4}
\end{eqnarray}
These equations have, as always, the property that
\begin{equation}
G(\lapbar(t),t;\epsilon)=\gbar(t:\epsilon)
\label{eq:g.5}
\end{equation}
We wish to show that scaling holds with $s(t)$ equal to $t$. If this is
true, it follows, using the definition (\ref{eq:3.5.3}) of the scaling
limit in crossover, that
\begin{equation}
\lim_{t\to\infty}\left[
tG(s/t;y/t)
\right]
=\int_0^\infty dx\,\Phi(x,y)\left(e^{sx}-1\right)
\label{eq:g.6}
\end{equation}
Note that this is a legitimate use of (\ref{eq:3.5.3}), since it involves
using as a function $f(x)$ the expression $(e^x-1)/x$, which is indeed
bounded at the origin, as required.

{}From (\ref{eq:g.5}) and (\ref{eq:g.6}) it follows that we need to
evaluate $t\gbar(t)$ as $t\to\infty$ while $\epsilon t$ is kept equal to $y$
and $t\lapbar(t)$ is equal to $s$. This means that the initial conditions
$\lapbar_0$ and $\gbar_0$ are varied appropriately. On the other hand, we
have, as always,
\begin{equation}
G(\lapbar_0, 0:\epsilon)=\gbar(0:\epsilon)=\int d\mass
c(\mass,0)\left(
e^{\lapbar_0\mass}-1
\right)=\lapbar_0+O(\lapbar_0^2)
\label{eq:g.7}
\end{equation}
We now solve (\ref{eq:g.4}) explicitly with the initial conditions given
by (\ref{eq:g.7}), where we shall neglect the higher order terms:
\begin{eqnarray}
\gbar(t;\epsilon)&=&\frac{2\epsilon\lapbar_0e^{-\epsilon t}}
{\lapbar_0e^{-\epsilon t}-(\lapbar_0-2\epsilon)}\nonumber\\
\lapbar(t;\epsilon)&=&\lapbar_0+2\epsilon\ln\left(
\frac{
\lapbar_0e^{-\epsilon t}-(\lapbar_0-2\epsilon)
}{
2\epsilon
}
\right)
\label{eq:g.8}
\end{eqnarray}
{}From these results one finds that in the scaling limit it is necessary to
take $\lapbar_0$ of order $1/t$ (note here in passing the considerable
difference to the similar computations when the ordinary scaling limit is
taken in the sum kernel: in that case one finds a quadratic relationship
between $\lapbar_0$ and $\lapbar(t)$. Again this  shows that we are in two
quite different sorts of asymptotic regimes). If we now define $w$ as the
value at which we keep $t\lapbar_0$ fixed, we obtain in the scaling limit
\begin{eqnarray}
&s&=t\lapbar(t)=w+2y
\ln\left(
\frac{
we^{-y}-(w-2y)
}{
2y
}
\right)\nonumber\\
&t&\gbar(t;\epsilon)=\frac{2ywe^{-y}}
{we^{-y}-(w-2y)}
\label{eq:g.9}
\end{eqnarray}
{}From (\ref{eq:g.6}) and (\ref{eq:g.9}) one finally obtains, by
substracting from the r.h.s. of (\ref{eq:g.6}) its limit as $s\to\infty$
\begin{equation}
\int_0^\infty dx\,\Phi(x,y)e^{sx}=\frac{
4y^2e^{-y}
}{
(1-e^{-y})[2y-w(1-e^{-y})]
}
\label{eq:g.10}
\end{equation}
This together with the first of the relations (\ref{eq:g.9}) yields the
result stated in the text, by using the following inverse transform
\begin{equation}
\Phi(x,y)=\frac{1}{2\pi i}\left(
\frac{4y^2}{1-e^{-y}}
\right)
\oint_{c-i\infty}^{c+i\infty}dw \frac{ds}{dw}\frac{
e^{-s(w)x}
}{
2y-w(1-e^{-y})
}
\label{eq:g.11}
\end{equation}
This integral can be evaluated and yields the result (\ref{eq:4.6.5}) as
stated in the text.
\section{The $C\to0$ Crossover When $B=0$}
\setcounter{equation}0
\label{app:crossover3}
Here we treat the crossover from constant to product kernel defined by the
kernel (\ref{eq:4.6.2}). We start by defining the following generating
function:
\begin{equation}
G(\lap,t)=\int_0^\infty d\mass\,c(\mass, t)\left[
e^{\mass t}-1
\right].
\label{eq:i.1}
\end{equation}
For this function one finds, in the pre-gel regime, which is the only one
we shall be concerned with here:
\begin{equation}
G_t=\half\left[
G^2+\epsilon^2(G_\lap-1)^2
\right]
\label{eq:i.2}
\end{equation}
We may now write the characteristic equations for this system, which is
now quasilinear, in contrast to the others treated thus far.
One finds
\begin{eqnarray}
\dot{p}&=&p\gbar
\label{eq:i.3}\\
\dot{q}&=&qG\label{eq:i.4}\\
\dot{\lapbar}&=&-\epsilon^2\phi\label{eq:i.6}\\
\dot{\gbar}&=&p-\epsilon^2\phi(1+\phi).\label{eq:i.7}
\end{eqnarray}
Here $p$ and $q$ denote $G_t$ and $G_\lap$ respectively. Solving these
equations with the appropriate initial conditions yields the usual
property for the characteristic curves namely that they generate a
solution surface:
\begin{equation}
G\left[
\lapbar(t), t
\right]=\gbar(t)
\label{eq:i.8}
\end{equation}
We also introduce the important auxiliary variable $\phi=q-1$, which
satisfies
\begin{equation}
\dot{\phi}=\gbar(1+\phi)\label{eq:i.5}
\end{equation}

As a first step, we want to scale away the parameter $\epsilon$ from the
equations. This can be done via the following transformations
\begin{equation}
y=\epsilon t\qquad\ptil=\frac{p}{\epsilon^2}\qquad\gtil=
\frac{\gbar}{\epsilon^2}\qquad\laptil=\frac{\lap}{\epsilon}
\label{eq:i.9}
\end{equation}
After these substitutions, the rescaled quantities satisfy
(\ref{eq:i.3}-\ref{eq:i.7}) with $\epsilon$ equal to one. The initial
conditions are now given by
\begin{eqnarray}
\gtil_0&=&\laptil_0\nonumber\\
\phi_0&=&\mu_2\laptil_0\epsilon
\label{eq:i.10}\\
\ptil_0&=&\half\laptil_0^2,
\nonumber
\end{eqnarray}
where $\mu_2$ denotes the second moment of the initial conditions.
It readily follows from (\ref{eq:i.3}) and (\ref{eq:i.4}) that $q$ and $p$
are proportional. If one now substitutes
\begin{equation}
\frac{d}{dt}\ln\ptil=\gtil
\label{eq:i.11}
\end{equation}
into (\ref{eq:i.7}) one obtains using the proportionality of $p$ and $q$
\begin{equation}
\frac{d^2}{dt^2}\ln p=p-\frac{2}{\laptil_0^2}p\left[
\frac{2}{\laptil_0^2}p-1
\right].
\label{eq:i.12}
\end{equation}
This can be straightforwardly solved by noticing the analogy to a Newtonian
equation of motion and using conservation of energy. This eventually leads
to
\begin{equation}
\ptil=\frac{\laptil_0^2}{
2+\laptil_0^2-\left(\laptil_0^2\cos y+2\laptil_0\sin y\right)
},
\label{eq:i.13}
\end{equation}
from which the result stated in the text readily follows upon noticing
that, from the definition of the crossover limit
\begin{equation}
\int_0^\infty dx\,x\Phi(x,y)e^{-sx}=q(-s,y),
\label{eq:i.14}
\end{equation}
where $q(\laptil,y)$ is precisely the function defined above in the
scaling described in (\ref{eq:i.9}).
\section{The Solution of the Finite Constant Kernel}
\setcounter{equation}0
\label{app:const-fin}
The equations for the finite constant kernel may be written as follows:
\begin{eqnarray}
\dot{c_j}(t)&=&\half\sum_{k=1}^{j-1}
c_kc_{j-k}-c_j(t)S_N(t)\qquad(1\leq j\leq N)
\label{eq:o.1a}\\
S_N(t)&=&\sum_{k=1}^Nc_k(t).
\label{eq:o.1b}
\end{eqnarray}
The strategy, which in fact is the one that generalizes straightforwardly
to all other $\Kfin_\alpha(k,l)$ is the following: first one evaluates
$c_j(t)$ for an arbitrary function $S_N(t)$ using only (\ref{eq:o.1a}).
One then considers (\ref{eq:o.1b}) as a self-consistency equation
for $S_N(t)$. After
having solved for it, one has an explicit expression for $S_N(t)$ and
hence for $c_j(t)$.

In the case of (\ref{eq:o.1a}) the first step is reasonably easy: one
transforms to new dependent and independent variables
as follows
\begin{eqnarray}
&&\phi_j(\tres{})=c_j(t)\exp\left(
\int_0^tS_N(t^\prime)\,dt^\prime
\right)\nonumber\\
&&d\tres{}=\half\exp\left(
-\int_0^tS_N(t^\prime)\,dt^\prime
\right)dt.
\label{eq:o.2}
\end{eqnarray}
Equations (\ref{eq:o.1a}) then become
\begin{equation}
\frac{d\phi_j}{d\tres{}}=\sum_{k=0}^{j-1}\phi_k\phi_{j-k},
\label{eq:o.3}
\end{equation}
which, for monodisperse initial conditions, have the solution
\begin{equation}
\phi_j(\tres{})=\tres{}^{j-1}.
\label{eq:o.4}
\end{equation}
The self-consistency condition (\ref{eq:o.1b}) can now be written as
\begin{equation}
S_N(\tres{})=2\frac{d\tres{}}{dt}\sum_{k=1}^N\tres^{k-1}.
\label{eq:o.5}
\end{equation}
One now notes that
\begin{equation}
S_N(\tres{})=-\frac{d}{dt}\ln\frac{d\tres{}}{dt}=-\left(
\frac{d\tres{}}{dt}
\right)^{-1}\frac{d^2\tres{}}{dt^2}
\label{eq:o.6}
\end{equation}
If we now substitute (\ref{eq:o.6}) into (\ref{eq:o.5}) one obtains, after
transforming to the variables $p=d\tres{}/dt$ and $\tres{}$
\begin{equation}
\frac{dp}{d\tres{}}=-2p\sum_{k=1}^N\tres^{k-1},
\label{eq:o.7}
\end{equation}
from which the formulae given in the text follow immediately.

We now proceed to give the asymptotic expressions for the sum
\begin{equation}
\chi(\tres{})=\sum_{k=1}^N\frac{\tres{}^k}{k}-\ln N
\label{eq:o.8}
\end{equation}
in the scaling regime, that is, when $\xi$ defined as $N(1-\tres{})$ is of
order
one and $N \to\infty$. We first observe that
\begin{eqnarray}
\sum_{k=1}^N\frac{\tres{}^k-1}{k}&=&\int_1^\tres{}\frac{y^N-1}{y-1}\nonumber\\
&=&\int_0^{-\xi}\left[
\left(
1+\frac{y}{N}
\right)^N-1
\right]\frac{dy}{y}\\
&=&\int_0^{-\xi}\frac{e^y-1}{y}dy+O(N^{-1})\nonumber,
\label{eq:o.9}
\end{eqnarray}
where in the last equality one uses the scaling limit. The identity stated
in the text then follows straightforwardly.
\section{The Finite-sum Kernel}
\setcounter{equation}0
\label{app:fin-sum}
The general equations for the finite sum kernel read as follows
\begin{equation}
\dot c_j=\sum_{k=1}^{j-1} kc_kc_{j-k}-jc_jM_0(t)-c_jM_1(t),
\label{eq:v.1}
\end{equation}
where, originally, $M_0(t)$ and $M_1(t)$ are arbitrary functions. We shall
then impose the self-consistency conditiions
\begin{eqnarray}
M_0(t)&=&\sum_{j=1}^N c_j(t)\nonumber\\
M_1(t)&=&\sum_{j=1}^N jc_j(t).
\label{eq:v.2}
\end{eqnarray}
The first step is accomplished as follows: introduce the new variables
\begin{eqnarray}
\phi_j(\tres)&=&c_j(t)\exp\left\{
\int_0^t dt^\prime\left[jM_0(t^\prime)+M_1(t^\prime)\right]
\right\}\label{eq:v.3a}\\
d\tres&=&\exp\left[
-\int_0^t dt^\prime M_1(t^\prime)
\right].
\label{eq:v.3b}
\end{eqnarray}
One then finds, denoting the derivative with respect to $\tres$ by a prime
\begin{eqnarray}
\phi_j^\prime&=&\sum_{k=1}^{j-1}k\phi_k\phi_{j-k}\nonumber\\
\phi_j(0)&=&\delta_{j,1},
\label{eq:v.4}
\end{eqnarray}
from which one immediately obtains the solution by induction
\begin{equation}
\phi_j(\tres)=a_j\tres^{j-1},
\label{eq:v.5}
\end{equation}
where the $a_j$ satisfy the relation
\begin{equation}
(j-1)a_j=\sum_{k=1}^{j-1}ka_ka_{j-k}.
\label{eq:v.6}
\end{equation}
If one defines
\begin{equation}
F(\lap)=\sum_{j=1}^\infty a_je^{j\lap},
\label{eq:v.7}
\end{equation}
then one finds immediately that
\begin{equation}
Fe^{-F}=e^\lap,
\label{eq:v.8}
\end{equation}
from which both the explicit expressions for the $a_j$ cited in the text
as well as the various other statements made there are easily derived.

Let us now define
\begin{eqnarray}
X&=&\exp\left\{
\int_0^t dt^\prime\left[M_1(t^\prime)\right]
\right\}
\nonumber\\
Y&=&\exp\left\{
\int_0^t dt^\prime\left[M_0(t^\prime)\right]
\right\}
\label{eq:v.9}
\end{eqnarray}
{}From this we derive the equations
\begin{eqnarray}
X^\prime&=&X\sum_{j=1}^Nja_j\tres^{j-1}Y^{-j}\nonumber\\
Y^\prime&=&\sum_{j=1}^Na_j\tres^{j-1}Y^{1-j}.
\label{eq:v.10}
\end{eqnarray}
If we now introduce $\Lambda=Y/\tres$, we obtain the solution for $\Lambda$
described in the text from the second equation (\ref{eq:v.10}). From the
first equation (\ref{eq:v.10}) one obtains
\begin{equation}
d\ln\tres=\frac{d\ln X}{\sum_{j=1}^N ja_j\Lambda^{-j}}.
\label{eq:v.11}
\end{equation}
Identifying this with the correponding expression of $d\ln\tres$ in terms
of $\Lambda$ one eventually obtains
\begin{equation}
X=\frac{1}{1-\sum_{j=1}^Na_j\Lambda^{-j}},
\label{eq:v.12}
\end{equation}
from which all results quoted in the text follow.
\section{The Three-body Kernel $\mass_1+\mass_2+\mass_3$}
\setcounter{equation}0
\label{app:3body1}
If we define $G$ to be $F(\lap,t)-S(t)$, with the definitions given
in the text, we find that the
characteristic equations for the PDE satisfied by $G$ are as
follows
\begin{eqnarray}
\dot{\lapbar}&=&-\half\gbar(\gbar+2S)\nonumber\\
\dot{\gbar}&=&-S\gbar\label{eq:l.1}\\
\dot{S}&=&-S^2.\nonumber
\end{eqnarray}
Here $S$ is defined as in (\ref{eq:4.1.202b}). Let us introduce $\tres$
through the equation
\begin{equation}
d\tres=S(t)dt.
\label{eq:l.2}
\end{equation}
{}From this the general solution follows straightforwardly:
\begin{eqnarray}
\lapbar&=&\lap_0-\half\frac{f(\lap_0)^2-f(0)^2}{f(0)}(1-e^{-\tres})
\label{eq:l.3a}\\
\gbar&=&[f(\lap_0)-f(0)]e^{-\tres}.
\label{eq:l.3}
\end{eqnarray}
In order to compute the exact solution for the monodisperse initial
condition, one first sets $f(\lap_0)$ equal to $e^{\lap_0}$ in
(\ref{eq:l.3}) and then uses the following representation for the
concentrations $c_j(t)$:
\begin{equation}
c_j(t)=\frac{1}{2\pi i}\oint_C G(\lap, t)e^{-j\lap}d\lap,
\label{eq:l.4}
\end{equation}
where the contour $C$ runs from $-a-\pi i$ to $-a+\pi i$, where $a$ is
an arbitrary real number, which should be large enough for all sums to
converge. We then use the basic properties of characteristic
equations, that is,
\begin{equation}
G[\lap(\tres,\lap_0), \tres]=\gbar(\tres,\lap_0).
\label{eq:l.5}
\end{equation}
Here, both $\lap(\tres,\lap_0)$ and $\gbar(\tres,\lap_0)$ are
explicitly known functions of $\lap_0$ via (\ref{eq:l.3}). It is
therefore possible to rewrite the integral (\ref{eq:l.4}) in the
$\lap_0$ variables. After some fairly tedious but straightforward
evaluations of contour integrals, the result (\ref{eq:4.1.3}) is
obtained. This method is quite general and can be used to evaluate
virtually all kernels for which the generating function is obtained
via characteristics. It is, however, almost invariaby quite cumbersome,
as this example shows.

In order to prove scaling, we proceed entirely as in the sum kernel
case (see appendix \ref{app:sum} for the details of the approach
followed, as well as the notation.) The scaling limit is defined by
\begin{equation}
\lapp=-\lap e^{2\tres}
\qquad \lapp_0=-\lap_0 e^\tres.
\label{eq:l.6}
\end{equation}
In this limit (\ref{eq:l.3a}) yields the following relation
\begin{equation}
\lapp=\lapp_0+\frac{f^{\prime\prime}(0)}{2}\lapp_0^2.
\label{eq:l.7}
\end{equation}
By the usual techniques this yields the results stated in the text.

If, on the other hand, we start with initial conditions of
the type
\begin{equation}
c(\mass, 0)=A\mass^{-\alpha}\qquad(2<\alpha<3),
\label{eq:l.8}
\end{equation}
one finds for large $t$ and hence large $\tres$, that
\begin{equation}
A\Gamma(1-\alpha)(-\lap_0)^{\alpha-1}+\lap_0e^{-\tres}=\lap.
\label{eq:l.9}
\end{equation}
Hence, in order for all the terms in (\ref{eq:l.9}) to scale similarly, one
must introduce the scaling variables
\begin{equation}
\lapp=-\lap e^{\tres(\alpha-1)/(\alpha-2)}
\qquad \lapp_0=-\lap_0 e^{\tres/(\alpha-2)}.
\label{eq:l.10}
\end{equation}
{}From this immediately follows that
\begin{equation}
z=\frac{\alpha-1}{\alpha-2}\qquad\tau=\frac{\alpha}{\alpha-1}.
\label{eq:l.11}
\end{equation}
\section{The Kernel $\mass_1+\mass_2+\mass_3$ Perturbing
the Constant Kernel}
\setcounter{equation}0
\label{app:3body2}
We wish to show here the scaling behaviour of the mixed two- and
three-body system described in (\ref{eq:4.4.1}) in the specific case
given by the first equation (\ref{eq:4.4.2}). For convenience I limit
myself to the case of monodisperse initial conditions, but the general
case  presents no difficulties of principle. In this case, if one
defines
\begin{eqnarray}
F(\lap,t)&=&\sum_{j=1}^\infty c_j(t)e^{j\lap}\nonumber\\
S(t)&=&\sum_{j=1}^\infty c_j(t)\\
G(\lap,t)&=&F(\lap,t)-S(t),\nonumber
\label{eq:j.1}
\end{eqnarray}
one obtains the following set of equations
\begin{eqnarray}
G_t-\half\alpha G(G+2S)G_z&=&\frac{G^2}{2}-\alpha SG
\label{eq:j.2}\\
\dot{S}&=&-(\half+\alpha)S^2.
\label{eq:j.3}
\end{eqnarray}
Applying the method of characteristics to (\ref{eq:j.2}) one gets
\begin{eqnarray}
\dot{\lapbar}&=&-\half\alpha G(G+2S)
\label{eq:j.4}\\
\dot{\gbar}&=&\frac{G^2}{2}-\alpha S G
\label{eq:j.5}
\end{eqnarray}
If we now introduce the rescaled time $\tres$ defined by
\begin{equation}
d\tres=S(t)dt
\label{eq:j.6}
\end{equation}
and define the initial conditions for $\gbar$ and $\lapbar$ to be $G_0$
and $\lap_0$ respectively, one obtains for $\gbar$
\begin{eqnarray}
\gbar(\lapbar, \tres)&=&\frac{G_0e^{\tres/2}}{1+G_0-G_0e^{\tres/2}}
e^{-(1/2+\alpha)\tres}\nonumber\\
&\simeq&\frac{\lap_0e^{\tres/2}}{1-\lap_0e^{\tres/2}}e^{-(1/2+\alpha)\tres},
\label{eq:j.7}
\end{eqnarray}
where in the second line, the smallness of $\lap_0$ has been used
as well as the following relation
\begin{equation}
G(\lap_0,0)=e^{\lap_0}-1=\lap_0+\half\lap_0^2+O(\lap_0^3).
\label{eq:j.8}
\end{equation}
We now want to obtain scaling from this solution of (\ref{eq:j.2}).
This is not quite a trivial task, so we carry it out in some detail.
As was stated in (\ref{eq:3.801}), it is sufficient to show that
there exists a function $s(t)$ such that
\begin{equation}
\lim_{t\to\infty}\left.G_\lap(\lap)\right|_{\lap=-\lapp/s(t)}
=\laplace\left[
x\Phi(x)
\right](\lapp)
\label{eq:j.9}
\end{equation}
We therefore wish to express the l.h.s. of (\ref{eq:j.9}) in terms
of a function of a scaling quantity alone. To do this, we first relate it
to the solutions of the characteristic equations (\ref{eq:j.4},
\ref{eq:j.5}). Since, because of the definition of the characteristic
equations, we have
\begin{equation}
G[\lapbar(t),t]=\gbar(t),
\label{eq:j.10}
\end{equation}
one verifies using L'Hospital's rule, that
\begin{equation}
\lim_{\lap_0\to0}\frac{\gbar(t)}{\lapbar(t)}=G_\lap(\lapbar(t),t),
\label{eq:j.11}
\end{equation}
at least whenever $\lapbar(t)\to0$ when $\lap_0\to0$ for times in
the scaling limit, which is a condition we shall verify shortly.

We therefore need an expression for $\lapbar(t)$. In terms of $\tres$, the
following expression follows from (\ref{eq:j.4},\ref{eq:j.7}) after
some routine algebra:
\begin{equation}
\lapbar=\lap_0\left[
\alpha(-\lap_0)^{2\alpha}\int_{-\lap_0}^{-\lap_0e^{\tres/2}}
x^{-2\alpha}\frac{2x+3}{(1+x)^2}dx+e^{-\alpha \tres}
\right].
\label{eq:j.12}
\end{equation}
{}From now on, we restrict ourselves explicitly to the case
$\alpha<1/2$. Then the integral in the r.h.s. of (\ref{eq:j.12})
converges at the lower end. From this further follows that we
may take $\lap_0e^{\tres/2}$ as the scaling variable.
It follows that
\begin{equation}
\lapbar(\tres)e^{(1/2+\alpha)\tres}=Z(-\lap_0e^{\tres/2}),
\label{eq:j.13}
\end{equation}
where $Z(x)$ is a function which is expressed in terms
of quadratures as follows
\begin{equation}
Z(x)=-x\left[
1+\alpha x^{2\alpha}\int_0^x dy\frac{
y^{-2\alpha}(2y+3)
}{
(1+y)^2
}
\right].
\label{eq:j.14}
\end{equation}
Note that $\lapbar(\tres)\to0$ as $\lap_0\to 0$ at fixed value of
$\lapp_0=\lap_0e^{\tres/2}$. Let us now define $\lapp$ as $\lapbar(\tres)
e^{(1/2+\alpha)\tres}$. We then find that these two quantities
are connected by
\begin{equation}
\lapp=Z(\lapp_0).
\label{eq:j.14.5}
\end{equation}
We may therefore use the relation
(\ref{eq:j.11}) to evaluate the r.h.s. of (\ref{eq:j.9}).
This leads to
\begin{equation}
\laplace\left[
x\Phi(x)
\right](\lapp)=
\frac{\lapp_0}{Z(\lapp_0)(1-\lapp_0)}
=\frac{\lapp_0}{\lapp(1-\lapp_0)}.
\label{eq:j.15}
\end{equation}
This gives the scaling function explicitly, although a more detailed
evaluation would require considerable algebra. Since this is not our
purpose here, however, we merely conclude these remarks by
evaluating the large-$\lapp$ asymptotics of $\laplace[x\Phi(x)]$,
which immediately yields the value of the exponent $\tau$. But,
in the limit of large $\lapp$, one also has $\lapp_0\to\infty$, so that
the asymptotics of the r.h.s. of (\ref{eq:j.15}) is of order
$\lapp^{-1}$, which corresponds to the value of one for the exponent
$\tau$. On the other hand, since the scaling variable is $\lap t$,
as readily follows from (\ref{eq:j.3},\ref{eq:j.6}), it follows that
the typical size grows as $t$ and hence the growth exponent is
one, as stated in the text.

We now turn to the case $\alpha>1/2$. Since the results are
less important and can be obtained by a straightforward extension of
the previous approach I shall be brief. In the limit $\lap_0\to0$ the
relation (\ref{eq:j.12}) reduces to
\begin{equation}
\frac{3\alpha}{1-2\alpha}\lap_0^2+\lap_0e^{-\alpha\tres}=\lap
\label{eq:j.16}
\end{equation}
by extending the integral to infinity.
If we therefore now define
\begin{equation}
\lap_0=\lapp_0 e^{-\alpha\tres}\qquad\lap=\lapp e^{-2\alpha\tres},
\label{eq:j.17}
\end{equation}
we obtain the following relation
\begin{equation}
\frac{3\alpha}{1-2\alpha}\lapp_0^2+\lapp_0e^{-\alpha\tres}=\lapp.
\label{eq:j.18}
\end{equation}
If we use this in connection with (\ref{eq:j.7}), for which we did
not require the hypothesis $\alpha<1/2$, we obtain in the scaling limit
\begin{equation}
\laplace\left[
x\Phi(x)
\right](\lapp)=\frac{\lapp_0(\lapp)}{\lapp},
\label{eq:j.19}
\end{equation}
where $\lapp_0$ is computed from $\lapp$ via (\ref{eq:j.18}). From
this follows the result that $\tau=3/2$, as stated in the text.
\section{The Scaling Function for the One-dimensional PCM}
\setcounter{equation}0
\label{app:scaling-pcm}
Let us show how to prove that (\ref{eq:5.2.7}) holds in general. In this
case, the first moment is equal to $\dens$, so that we should properly
first divide by it to obtain the correctly normalized scaling function.
Let us compute the generaing function $G(\lap,t)$ given by
\begin{equation}
G(\lap,t)=\dens^{-1}\sum_{k=1}^0kc(k,t)e^{k\lap}.
\label{eq:p.1}
\end{equation}
After some straightforward algebra one obtains
\begin{equation}
G(\lap,t)=\frac{4e^\lap}{\sqrt\pi}\int_0^\infty y^2\exp\left[
2\dens\sqrt{Dt}(e^\lap-1)
\right]e^{-y^2}dy.
\label{eq:p.2}
\end{equation}
If one now defines
\begin{equation}
s(t)=2\dens\sqrt{Dt}\qquad \sigma=\lap s(t)
\label{eq:p.3}
\end{equation}
and noting that, as follows from the definition of scaling through
generating functions stated in (\ref{eq:3.801}), one has
\begin{equation}
\frac{4}{\sqrt\pi}
\int_0^\infty y^2 e^{-y^2}e^{\sigma y}=\int_0^\infty x\Phi(x)e^{\sigma x}dx,
\label{eq:p.4}
\end{equation}
from which the result stated in the text follows immediately through
inverse Laplace transform.
\section{Ballistic Aggregation: Sketch of Exact Solution}%
\setcounter{equation}0
\label{app:ball-exact}
The analysis starts by the following basic fact: once the initial condition
is fully known, the cluster size distribution function is known for all
times. In the model considered in \cite{mar94,fra99,fra00,fra01}, the
particle positions are fixed (they occupy a lattice, which we take to have
unit spacing), so that everything is reduced to averaging over the initial
momenta. This is in turn considerably simplified by the assumption that the
initial momenta are independent Gaussian random variables.

One must therefore express the event that, say, particles $1$ to $\mass$
form one cluster at time $t$ in terms of the initial momenta of all
particles. After this, one evaluates the probability of this event using
the independence of the initial momenta. In the following, we shall make
heavy use of the fact that, if a cluster has formed at time $t$, the center
of mass of the particles that constitute it has been moving on a straight
line with  the cluster's velocity ever since the beginning.

There are three conditions (or
rather, classes of conditions) which must all be satisfied for this event
to occur:
\begin{enumerate}
\item A set of conditions guaranteeing that the particles $1$ to $\mass$
do in fact come to form a cluster. One sees that it is necessary and
sufficient that, for all $\mass_1$,
the particle cluster from $1$ to $\mass_1$ have
sufficient velocity to collide with the corresponding
cluster from $\mass_1+1$ to $\mass$ before time $t$. This is
easily seen to be equivalent to the following set of conditions on the
velocities $v(\mass_2)$  of the particles at site $\mass_2$ with respect
to the center of mass of the cluster
\begin{equation}
-2t\sum_{\mass_2=1}^{\mass_1}v(\mass_2)\leq\left(\mass_1-\half\mass
\right)^2-\frac{\mass^2}{4}.
\label{eq:q.1}
\end{equation}
for all $\mass_1$. Now we see a connection with the functions defined in
the text: due to the independence of the $v(\mass_2)$, the partial sums on
the l.h.s. of (\ref{eq:q.1}) perform a random walk
starting from the  origin and returning to it. Conditions
(\ref{eq:q.1}) then state that this random walk must always remain below
the parabolic profile given by the r.h.s. of (\ref{eq:q.1}).

\item A set of conditions guaranteeing that the particles to the right of
the cluster will not themselves coalesce with the cluster. If we again
measure all velocities with respect to that of the center of mass of the
forming cluster, one obtains the following condition for all velocities of
particles to the right of the cluster:
\begin{equation}
-2t\sum_{\mass_2=\mass+1}^{\mass_1}v(\mass_2)\leq\left[
\mass_1+\left(
\half\mass+vt
\right)
\right]^2-
\left(
\half\mass+vt
\right)^2,
\label{eq:q.2}
\end{equation}
where $v$ denotes the velocity of the forming cluster in the rest frame.
These conditions state that no cluster formed by particles $\mass+1$ to
$\mass_1$ can collide before time $t$ with the center of mass of the
forming cluster which, by assumption, is at rest in the frame
we consider. Again, we see that a given random walk is constrained to stay
below a certain parabolic profile. In this case, this must occur for all
values of the upper limit of the sum, so we are dealing with events of the
type described by $\transit$.

\item An entirely similar set of conditions for particles to the left of
the forming cluster. The equations are, of course, entirely similar to
(\ref{eq:q.2}), with $v$ replacd by $-v$.
\end{enumerate}
{}From these considerations the results stated in the text can be obtained
more or less straightforwardly. The crucial remark is the following: in the
scaling limit, that is, when the cluster masses $\mass$ are large, and
$t\gg1$ as well, the random walks will tend, under an appropriate scaling
of the variables, to a continuous diffusion. The fact that all random walks
start on the absorbing parabola causes technical problems, which are
excellently discussed in the various references given above.

As a final excercise for the interested reader, consider the following
problem: if the initial velocities are independent, but distributed
according to some L\'evy stable law, what can we say about the solution?
Everything up to (\ref{eq:q.1}) and (\ref{eq:q.2}) still holds. The real
difficulty consists in building up a continuuum model, since there is, to
my knowledge, no L\'evy analogue to Brownian motion. The discrete problem,
on the other hand, is quite formidable. The numerical results found in
\cite{jia93} make it probable that a scaling picture
does indeed exist when the average velocity is still finite, whereas the
other case is quite pathological.

\end{document}